\documentclass[12pt]{article}%
\usepackage{amsmath}
\usepackage{graphicx}
\usepackage{amsfonts}
\usepackage{amssymb}%
\numberwithin{equation}{section}
\begin{document}
\title{About dual two-dimensional oscillator and Coulomb-like theories on a plane}
\author{G.V. Grigoryan\thanks{%
Yerevan Physics Institute. Yerevan. Armenia; e-mail:gagri@yerphi.am}, 
R.P. Grigoryan\thanks{%
Yerevan Physics Institute. Yerevan. Armenia; e-mail:rogri@yerphi.am}
I.V. Tyutin\thanks{%
Lebedev Physical Institute. Moscow. Russia; e-mail:tyutin@lpi.ru}}
\date{ }
\maketitle

\begin{abstract}
We present a mathematically rigorous quantum-mechanical treatment of a
two-dimensional nonrelativistic quantum dual theories (with oscillator and
Coulomb like potentials) on a plane and compare their spectra and the sets of eigenfunctions.
All self-adjoint Schrodinger operators for these theories are constructed and rigorous
solutions of the corresponding spectral problems are presented. The first part of
the problem is solved by using a method of specifying s.a. extensions by (asymptotic)
s.a. boundary conditions. Solving spectral problems, we follow the Krein's
method of guiding functionals. We show, that there is one to one
correspondence between the spectral points of dual theories in the planes 
energy-coupling constants not only for discrete, but also for continuous spectra.
\end{abstract}

\section{Introduction}

It is well known \cite{Ter-Ant}, that if one introduces in a radial part of
the $D$ dimensional oscillator ($D > 2$)

\begin{equation}  \label{eq1}
{\frac{{d^{2}R}}{{du^{2}}}} + {\frac{{D - 1}}{{u}}}{\frac{{dR}}{{du}}} - {%
\frac{{L\left( {L + D - 2} \right)}}{{u^{2}}}}R + {\frac{{2\mu} }{{\hbar ^{2}%
}}}\left( {E - {\frac{{\mu \omega ^{2}u^{2}}}{{2}}}} \right)R = 0
\end{equation}
(here $R$ is the radial part of the wave function for the $D$ dimensional
oscillator ($D>2$) and $L=0,\mathrm{1},\mathrm{2},...$ are the eigenvalues
of the global angular momentum ) $r=u^{2}$ then equation (\ref{eq1})
transforms into

\begin{equation}  \label{eq2}
{\frac{{d^{2}R}}{{dr^{2}}}} + {\frac{{d - 1}}{{r}}}{\frac{{dR}}{{dr}}} - {%
\frac{{l\left( {l + d - 2} \right)}}{{r^{2}}}}R + {\frac{{2\mu} }{{\hbar ^{2}%
}}}\left( {\mathcal{E} + {\frac{{\alpha} }{{r}}}} \right)R = 0
\end{equation}

\noindent where $d=D/2+1\quad l=L/2\quad \mathcal{E}=-{\frac{{\mu \omega ^{2}%
}}{{8}}}\quad \alpha =E/4$, which formally is identical to the radial
equation for $d$-dimensional hydrogen atom.

Equations (\ref{eq1}) and (\ref{eq2}) are dual to each other and the duality
transformation is $r=u^{2}$. For discreet spectrum of these equations
 (and wave fuctions regular at the origin) it was
proved, that to each state of equation (\ref{eq1}) corresponds a state in (%
\ref{eq2}), and visa versa \cite{{Ners-Ter-Ant},{Hak-Ter-Ant}}. However the
correspondence of the states in general (for discrete, as well as continuous
spectra and for all values of the parameters of the theory) the problems was
not considered.

In \cite{TGG} we constructed all self-adjoint  Schrodinger operators for nonrelativistic 
one-dimen\-sional  quantum dual theories and represented rigorous solutions  of the 
corresponding spectral problems. We have shown that there is one to one correspondence between 
the spectra of dual theories for discreet , as well as continuous spectra.

In this paper will solve the quantum problem of two dimensional quantum dual theories (with oscillator and
Coulomb like potentials ) on a plane and compare their spectra and the sets of eigenfunctions. 
As it was in one dimensional case, we again have  a
correspondence of the states for all values of the parameters ${E}%
_{O}$, ${\lambda }$, $E_{C}$, and $g$, except when the angular momentum $m=1$, 
when the duality is one-to-one only in the case of parameter of s.a. extension $\zeta=\pi/2$ (see below in coulomb case). 
We have similar situation for dual models on two dimensional pseudoshpere.
The interest to these models was stimulated also by the fact that among the theorists dealing with 
similar problems exists a notion, that the "Hamiltonian isn't self adjoit at high energies" \cite{Ners}. 
 In section 2 we will consider the
quantum problem for the oscillator, will find solutions of the equation for
all values of the variable and parameters. In Section 3 we will consider the
quantum problem for Coulomb-like system. The results will be compared in
section 4, where we will show the one-to one correspondense of the spectra
and proper functions of the Hamiltonians of both problems.

\section{Quantum two-dimentional oscillator-like interaction on a plane}
\subsection{Reduction to radial problem}

In the case under consideration, the initial differential operation is
$\check{H}_{O}\equiv\check{H}$,%
\begin{align*}
&  \check{H}=-\Delta+\lambda\ \mathbf{u}^{2},\;\mathbf{u}=(u_{x}%
,u_{y}),\;\Delta=\partial_{u_{x}}^{2}+\partial_{u_{y}}^{2}=\partial_{u}%
^{2}+\frac{1}{u}\partial_{u}+\frac{1}{u^{2}}\partial_{\varphi_{u}}^{2},\\
&  u=\sqrt{u_{x}^{2}+u_{y}^{2}},
\end{align*}
the space of particle quantum states is the Hilbert space $\mathfrak{H}%
_{O}\equiv\mathfrak{H}=L^{2}\left(  \mathbb{R}^{2}\right)  $ of s.-integrable
functions $\psi(\mathbf{u}),\ $, with the scalar product%
\[
(\psi_{1},\psi_{2})=\int\overline{\psi_{1}(\mathbf{u})}\psi_{2}(\mathbf{u}%
)d\mathbf{\rho},\,d\mathbf{u}=du_{x}du_{y}=udud\varphi_{u},\;u\in
\mathbb{R}_{+},\;\varphi_{u}\in\lbrack0,2\pi].
\]

A quantum Hamiltonian should be defined as a s.a. operator in this Hilbert space.

The construction is essentially based on the requirement of rotation symmetry
which certainly holds in a classical description of the system. This
requirement is formulated as the requirement of the invariance of a s.a.
Hamiltonian under rotations around the origin. As in classical mechanics, the
rotation symmetry allows separating the polar coordinates $\rho$ and $\varphi$
and reducing the two-dimensional problem to a one-dimensional radial problem.

The group of rotations $SO(2)$ in $\mathbb{R}^{2}$ naturally acts in the
Hilbert space $\mathfrak{H}$ by unitary operators: if $S\in SO(2)$, then the
corresponding operator $\hat{U}_{S}$ is defined by the relation $(\hat{U}_{S}%
\psi)(\mathbf{u})=\psi(S^{-1}\mathbf{u})$, $\psi\in\mathfrak{H}$.

The Hilbert space $\mathfrak{H}$ is a direct orthogonal sum of subspaces
$\mathfrak{H}_{m}$, that are the eigenspaces of the representation
$\hat{U}_{S}$,
\[
\mathfrak{H}=\sideset{}{^{\,\lower1mm\hbox{$\oplus$}}}  \sum_{m\in\mathbb{Z}%
}\mathfrak{H}_{m},\;\hat{U}_{S}\mathfrak{H}_{m}=e^{-im\theta}\mathfrak{H}%
_{m},\;\mathfrak{H}_{m}=\hat{P}_{m}\mathfrak{H}%
\]
where $\theta$ is the rotation angle corresponding to $S$ and .$\hat{P}_{m}$
is an orthohonal projector on subspace $\mathfrak{H}_{m}$. $\mathfrak{H}_{m}$
consists of eigenfunctions $\psi_{m}(\mathbf{u})$ for angular momentum
operator $\hat{L}_{z}=-i\hbar\partial/\partial\varphi_{u}$, $\psi
_{m}(\mathbf{u})=\frac{1}{\sqrt{u}}\frac{1}{\sqrt{2\pi}}\mathrm{e}%
^{im\varphi_{u}}f_{m}\left(  u\right)  $, where $f_{m}\left(  u\right)
\in\mathfrak{h}_{Om}\equiv\mathfrak{h}_{m}=L^{2}(\mathbb{R}_{+})$,
$L^{2}(\mathbb{R}_{+})$ is the Hilbert space of s.-integrable functions on the
semi-axis $\mathbb{R}_{+}$ with scalar product
\[
\left(  f,g\right)  =\int_{\mathbb{R}_{+}}\overline{f\left(  u\right)
}g\left(  u\right)  du.
\]

We define an initial symmetric operator $\hat{H}_{O}\equiv\hat{H}$ associated
with $\check{H}$\ as follows:%
\[
:\left\{
\begin{array}
[c]{l}%
D_{H}=\{\psi(\mathbf{u}):\ \psi\in\mathcal{D}(\mathbb{R}^{2}\backslash
\{0\})\}\\
\hat{H}\psi=\check{H}\psi,\ \forall\psi\in D_{H}\
\end{array}
\right.  ,
\]
where $\mathcal{D}(\mathbb{R}^{2}\backslash\{0\})$ is the space of smooth and
compactly supported functions vanishing in a neighborhood of the point
$\mathbf{u=}0$. The domain $D_{H}$ is dense in $\mathfrak{H}$ and the
symmetricity of $\hat{H}$ is obvious. It is also obvious that the operator
$\hat{H}$ commutes\footnote{We remind the reader of the notion of
commutativity in this case (where one of the operators, $U_{S}$, is bounded
and defined everywhere): we say that the operators $\hat{H}$ and $U_{S}$
commute if $U_{S}\hat{H}$ $\subseteq\hat{H}U_{S}$, i.e., if $\psi\in D_{H}$,
then also $U_{S}\psi\in D_{H}$ and $U_{S}\hat{H}\psi=\hat{H}U_{S}\psi$} with
the unitary operators $\hat{U}_{S}$,.%
\[
\hat{H}=\sideset{}{^{\,\lower1mm\hbox{$\oplus$}}}  \sum_{m\in\mathbb{Z}%
}\hat{H}_{m},\;\hat{H}_{m}=\hat{P}_{m}\hat{H}.
\]
Operators $\hat{f}$ which commute with the operators $\hat{U}_{S}$, we will
call rotationally-invariant. Such operators can be represented in the form
\[
\hat{f}=\sideset{}{^{\,\lower1mm\hbox{$\oplus$}}}  \sum_{m\in\mathbb{Z}}%
\hat{f}_{m},\;\hat{f}_{m}=\hat{P}_{m}\hat{f},
\]
and $\hat{f}_{m}$ are s.a. operators in sufspaces $\mathfrak{H}_{m}$.

In the polar coordinates $u$, $\varphi_{u}$ the operation $\check{H}$ becomes%
\[
\check{H}=-\partial_{u}^{2}-u^{-1}\partial_{u}+-u^{-2}\partial_{\varphi_{u}%
}^{2}+\lambda u^{2},
\]

Represent $\psi(\mathbf{u})\in\mathfrak{H}$ in the form%
\[
\psi(\mathbf{u})=\sum_{m\in\mathbb{Z}}\psi_{m}(\mathbf{u}),\;\psi
_{m}(\mathbf{u})=\frac{1}{\sqrt{u}}\frac{1}{\sqrt{2\pi}}\mathrm{e}%
^{im\varphi_{u}}f_{m}\left(  u\right)  ,\;f_{m}\left(  u\right)  =\sqrt{u}%
\int_{0}^{2\pi}d\varphi_{u}\frac{1}{\sqrt{2\pi}}\mathrm{e}^{-im\varphi_{u}%
}\psi(\mathbf{u}).
\]
Then we have%
\[
\hat{H}\psi_{m}(\mathbf{u})=\hat{H}_{m}\psi_{m}(\mathbf{u})=\frac{1}{\sqrt{u}%
}\frac{1}{\sqrt{2\pi}}\mathrm{e}^{im\varphi_{u}}\hat{h}_{Om}f_{m}\left(
u\right)  ,
\]
where $\hat{h}_{Om}\equiv\hat{h}_{m}$ is a symmetric operator defined in the
Hilbert space $\mathfrak{h}_{m}$:%
\[
\hat{h}_{m}:\left\{
\begin{array}
[c]{c}%
D_{h_{m}}=\{f\left(  u\right)  :f\left(  u\right)  \in\mathcal{D}%
(\mathbb{R}_{+})\}\\
\hat{h}_{m}f=\check{h}_{m}f,\;\forall f\in D_{h_{m}},\;\check{h}_{m}%
=-\partial_{u}^{2}+u^{-2}(m^{2}-1/4)+\lambda u^{2}%
\end{array}
\right.  .
\]

Let $\hat{h}_{\mathfrak{e}m}$ is a s.a$.$operator associated with the
differential operation $\check{h}_{m}$.in the Hilbert space $\mathfrak{h}_{m}$
Then the operator $\hat{H}_{\mathfrak{e}m}$,%
\begin{equation}
\hat{H}_{\mathfrak{e}m}\psi_{m}(\mathbf{u})=\frac{1}{\sqrt{u}}\frac{1}%
{\sqrt{2\pi}}\mathrm{e}^{im\varphi_{u}}\hat{h}_{\mathfrak{e}m}f_{m}\left(
u\right)  ,\;f_{m}\left(  u\right)  \in D_{h_{\mathfrak{e}m}},
\label{Osc2.1.0}%
\end{equation}
is a s.a. operator operator associated with $\check{H}_{m}$.in the Hilbert
space $\mathfrak{H}_{m}$ and operator $\hat{H}_{\mathfrak{e}}$,%
\begin{equation}
\hat{H}_{\mathfrak{e}}=\sideset{}{^{\,\lower1mm\hbox{$\oplus$}}}  \sum
_{m\in\mathbb{Z}}\hat{H}_{\mathfrak{e}m}, \label{Osc2.1.1}%
\end{equation}
is a s.a. operator in the Hilbert space $\mathfrak{H}$.

Conversely, let $\hat{H}_{\mathfrak{e}}$ be a rotationally invariant s.a.
extension of $\hat{H}$. Then it has the form~(\ref{Osc2.1.1}), where
$\hat{H}_{\mathfrak{e}m}$ are s.a. operators in $\mathfrak{H}_{m}$. The
operator $\hat{H}_{\mathfrak{e}m}$ acts in subspace $\mathfrak{H}_{m}$ by the
rule (\ref{Osc2.1.0}) with some operator $\hat{h}_{\mathfrak{e}m}$ which is
obviously a s.a. operator associated with the symmetric operator $\hat{h}_{m}$
in the Hilbert space $\mathfrak{h}_{m}$.

Thus, the problem of constructing a rotationally-invariant s.a. Hamiltonian
$\hat{H}_{\mathfrak{e}}$ is thus reduced to constructing s.a. radial
Hamiltonians $\hat{h}_{\mathfrak{em}}$.

\subsection{$|m|\geq1$, $\lambda>0$}

\subsection{Useful solutions}

We need solutions of an equation%
\begin{equation}
(\check{h}_{m}-W)\psi(u)=0,\;\check{h}_{m}=-\partial_{u}^{2}+u^{-2}%
(m^{2}-1/4)+\lambda u^{2}, \label{Osc2.2.1.1}%
\end{equation}
where $\hbar^{2}W/2m$ is complex energy, $\hbar^{2}\lambda/2m$ is a coupling
constant,%
\[
W=|W|e^{i\varphi_{W}},\;0\leq\varphi_{W}\leq\pi,\;\operatorname{Im}W\geq0,
\]
and for $\lambda$, we will use the representation $\lambda=\varkappa^{4}$.

It is convenient for our aims first to consider solutions more general
equation%
\begin{equation}
\lbrack-\partial_{u}^{2}+u^{-2}((m+\delta)^{2}-1/4)+\lambda u^{2}%
-W]\psi(u)=0,\;|\delta|<1. \label{Osc2.2.1.2}%
\end{equation}

Introduce a new variable $\rho=(\varkappa u)^{2}$, $\partial_{u}%
=2\varkappa\sqrt{\rho}\partial_{\rho}$, $\partial_{u}^{2}=4\varkappa^{2}%
[\rho\partial_{\rho}^{2}+(1/2)\partial_{\rho}]$, and new function $\phi(\rho
)$, $\psi(u)=\rho^{1/4+|m+\delta|/2}e^{-\rho/2}\phi(\rho)$. Then we obtain%
\begin{align}
&  \rho\partial_{\rho}^{2}\phi(\rho)+(\beta_{\delta}-\rho)\partial_{\rho}%
\phi(\rho)-\alpha_{\delta}\phi(\rho)=0,\;\beta_{\delta}=1+|m+\delta
|,\label{Osc2.2.1.3}\\
&  \alpha_{\delta}=1/2+|m+\delta|/2-w,\;w=w_{O}=W/4\varkappa^{2}.\nonumber
\end{align}

Eq. (\ref{Osc2.2.1.3}) is the equation for confluent hypergeometric functions,
in the terms of which we can express solutions of eq. (\ref{Osc2.2.1.2}). We
will use the following solutions%
\begin{align}
&  O_{1,m,\delta}(u;W)=(\kappa_{0}u)^{1/2+|m+\delta|}e^{-\rho/2}\Phi
(\alpha_{\delta},\beta_{\delta};\rho),\label{Osc2.2.1.4a}\\
&  O_{2,m,\delta}(u;W)=\frac{(\kappa_{0}u)^{1/2-|m+\delta|}}{\Gamma
(\beta_{-,\delta})}e^{-\rho/2}\Phi(\alpha_{-,\delta},\beta_{-,\delta}%
;\rho),\label{Osc2.2.1.4b}\\
&  O_{3,m,\delta}(u;W)=(\kappa_{0}u)^{1/2+|m+\delta|}e^{-\rho/2}\Psi
(\alpha_{\delta},\beta_{\delta};\rho)=\nonumber\\
&  =\frac{\Gamma(-|m+\delta|)}{\Gamma(\alpha_{-,\delta})}O_{1,m,\delta
}(u;W)+\frac{(\kappa_{0}/\varkappa)^{2|m+\delta|}\Gamma(|m+\delta
|)\Gamma(\beta_{-,\delta})}{\Gamma(\alpha_{\delta})}O_{2,m,\delta
}(u;W),\nonumber\\
&  \alpha_{-,\delta}=1/2-|m+\delta|/2-w,\;\beta_{-,\delta}=1-|m+\delta
|.\nonumber
\end{align}
Note that $O_{1,m,\delta}(u;W)$ and $O_{2,m,\delta}(u;W)$ are real-entire in
$W$ solutions of eq. (\ref{Osc2.2.1.2}).

Represent $O_{3,m,\delta}$ in the form%
\begin{align}
&  O_{3,m,\delta}(u;W)=B_{m,\delta}(W)O_{1,m,\delta}(u;W)+\frac{(\kappa
_{0}/\varkappa)^{2|m+\delta|}\Gamma(|m+\delta|)}{\Gamma(\alpha_{\delta}%
)}O_{4,m,\delta}(u;W),\label{Osc2.2.1.5a}\\
&  O_{4,m,\delta}(u;W)=\Gamma(\beta_{-,\delta})\left[  O_{2,m,\delta
}(u;W)-A_{m,\delta}(W)O_{1,m,\delta}(u;W)\right]  ,\label{Osc2.2.1.5b}\\
&  A_{m,\delta}(W)=\frac{(\kappa_{0}/\varkappa)^{-2|m|}\Gamma(\alpha)}%
{\Gamma(\beta_{\delta})\Gamma(\alpha_{-})},\;\frac{\Gamma(\alpha)}%
{\Gamma(\alpha_{-})}=(-1)^{|m|}(1-\alpha)_{|m|},\;(1+x)_{|m|}=(1+x)\cdots
(|m|+x),\label{Osc2.2.1.5c}\\
&  B_{m,\delta}(W)=\frac{\Gamma(-|m+\delta|)}{\Gamma(\alpha_{-,\delta}%
)}+\frac{(\kappa_{0}/\varkappa)^{2|m+\delta|}\Gamma(|m+\delta|)\Gamma
(\beta_{-,\delta})}{\Gamma(\alpha_{\delta})}A_{m,\delta}(W)=\nonumber\\
&  \,=\frac{\Gamma(-|m+\delta|)}{\Gamma(\alpha_{-})}\left[  \frac{\Gamma
(\alpha_{-})}{\Gamma(\alpha_{-,\delta})}-\frac{(\kappa_{0}/\varkappa
)^{2\delta_{m}}\Gamma(\alpha)}{\Gamma(\alpha_{\delta})}\right]
,\label{Osc2.2.1.5d}\\
&  \alpha=\alpha_{0}=1/2+|m|/2-w,\;\alpha_{-}=\alpha_{-,0}=1/2-|m|/2-w,\;\beta
=\beta_{0}=1+|m|,\nonumber
\end{align}
$O_{4,m,\delta}(u;W)$ is real-entire in $W$.

We obtain the solution of eq.(\ref{Osc2.2.1.1}) as the limit $\delta
\rightarrow0$ of the solution of (\ref{Osc2.2.1.2}):%
\begin{align*}
&  O_{1,m}(u;W)=O_{1,m,0}(u;W)=(\kappa_{0}u)^{1/2+|m|}e^{-\rho/2}\Phi
(\alpha,\beta;\rho),\\
&  O_{4,m}(u;W)=\lim_{\delta\rightarrow0}O_{4,m,\delta}(u;W),\\
&  O_{3,m}(u;W)=(\kappa_{0}u)^{1/2+|m|}e^{-\rho/2}\Psi(\alpha,\beta;\rho)=\\
&  =B_{m}(W)O_{1,m}(u;W)+C_{m}(W)O_{4,m}(u;W),\;C_{m}(W)=\frac{(\kappa
_{0}/\varkappa)^{2|m|}\Gamma(|m|)}{\Gamma(\alpha)},\\
&  B_{m}(W)=B_{m,0}(W)=\frac{(-1)^{|m|+1}}{2\Gamma(\beta)\Gamma(\alpha_{-}%
)}\left[  \psi(\alpha_{-})+\psi(\alpha)-4\ln(\kappa_{0}/\varkappa)\right]  .
\end{align*}
where we used relations.%
\begin{align*}
&  |m+\delta|=|m|+\delta_{m},\;\delta_{m}=\delta\mathrm{sign}m,\;\Gamma
(-|m+\delta|)=\frac{(-1)^{|m|+1}}{\delta_{m}\Gamma(\beta)},\\
&  \frac{\Gamma(\alpha_{-})}{\Gamma(\alpha_{-,\delta})}=1+\delta_{m}%
\psi(\alpha_{-})/2,\;\frac{\Gamma(\alpha)}{\Gamma(\alpha_{\delta})}%
=1-\delta_{m}\psi(\alpha)/2,\;(\kappa_{0}/\varkappa)^{2\delta_{m}}%
=1+2\delta_{m}\ln(\kappa_{0}/\varkappa).
\end{align*}

\subsubsection{Asymptotics, $u\rightarrow0$ ($\rho\rightarrow0$)}%

\begin{align}
O_{1,m}(u;W)  &  =(\kappa_{0}u)^{1/2+|m|}(1+O(u^{2})),\label{Osc2.2.1.1.1a}\\
O_{4,m}(u;W)  &  =(\kappa_{0}u)^{1/2-|m|}\left(  1+\left\{
\begin{array}
[c]{c}%
O(u^{2}),\;|m|\geq2\\
O(u^{2}\ln u),\;|m|=1
\end{array}
\right.  \right)  ,\nonumber\\
O_{3,m}(u;W)  &  =.C_{m}(W)(\kappa_{0}u)^{1/2-|m|}\left(  1+\left\{
\begin{array}
[c]{c}%
O(u^{2}),\;|m|\geq2\\
O(u^{2}\ln u),\;|m|=1
\end{array}
\right.  \right)  ,\;\operatorname{Im}W>0\;\mathrm{or}\;W=0.
\label{Osc2.2.1.1.1b}%
\end{align}

\subsubsection{Asymptotics, $u\rightarrow\infty$ ($\rho\rightarrow\infty$),
$\operatorname{Im}W>0$ or $W=0$}%

\begin{align*}
O_{1,m}(u;W)  &  =\frac{\kappa_{0}^{1/2+|m|}\varkappa^{2\alpha-2\beta}%
\Gamma(\beta)}{\Gamma(\alpha)}u^{-1/2-2w}e^{\rho/2}(1+O(u^{-2})),\\
O_{3,m}(u;W)  &  =\kappa_{0}^{1/2+|m|}\varkappa^{-2\alpha}u^{-1/2+2w}%
e^{-\rho/2}(1+O(u^{-2})).
\end{align*}

\subsubsection{Wronskian}%

\begin{equation}
\mathrm{Wr}(O_{1,m},O_{3,m})=-2\kappa_{0}|m|C_{m}(W)=-\omega(W).
\label{Osc2.2.1.3.1}%
\end{equation}

\subsection{Symmetric operator $\hat{h}_{m}$}

For given a differential operation $\check{h}_{m}$ we determine the following
symmetric operator $\hat{h}_{m}$,%

\begin{equation}
\hat{h}_{m}:\left\{
\begin{array}
[c]{l}%
D_{h_{m}}=\mathcal{D}(\mathbb{R}_{+}),\\
\hat{h}_{m}\psi(u)=\check{h}_{m}\psi(u),\;\forall\psi\in D_{h_{m}}%
\end{array}
\right.  . \label{Osc2.2.2.1}%
\end{equation}

\subsection{Adjoint operator $\hat{h}_{m}^{+}=\hat{h}_{m}^{\ast}$}

(in this file, the subindex ``$O$'' is omitted)%

\begin{equation}
\hat{h}_{m}^{+}:\left\{
\begin{array}
[c]{l}%
D_{h_{m}^{+}}=D_{\check{h}_{m}}^{\ast}(\mathbb{R}_{+})=\{\psi_{\ast}%
,\psi_{\ast}^{\prime}\;\mathrm{are\;a.c.\;in}\mathcal{\;}\mathbb{R}_{+}%
,\;\psi_{\ast},\hat{h}_{m}^{+}\psi_{\ast}\in L^{2}(\mathbb{R}_{+})\}\\
\hat{h}_{m}^{+}\psi_{\ast}(u)=\check{h}_{m}\psi_{\ast}(u),\;\forall\psi_{\ast
}\in D_{h_{m}^{+}}%
\end{array}
\right.  . \label{Osc2.2.3.1}%
\end{equation}

\subsubsection{Asymptotics}

I) $|u|\rightarrow\infty$

Because $V(u)>-(|\lambda|+1)u^{2}$ for large $u$, we have: $[\psi_{\ast}%
,\chi_{\ast}](u)\rightarrow0$ as $u\rightarrow\infty$, $\forall\psi_{\ast
},\chi_{\ast}\in D_{h_{m}^{+}}$.

II) $u\rightarrow0$

Because $\check{h}_{m}\psi_{\ast}\in L^{2}(\mathbb{R})$, we have%
\[
\check{h}_{m}\psi_{\ast}(u)=(-\partial_{u}^{2}+u^{-2}(m^{2}-1/4)+\lambda
u^{2})\psi_{\ast}(u)=\eta(u),\;\eta\in L^{2}(\mathbb{R}).
\]
General solution of this equation can be represented in the form%
\begin{align*}
&  \psi_{\ast}(u)=a_{1}O_{1,m}(u;0)+a_{2}O_{3,m}(u;0)+I(u),\\
&  \psi_{\ast}^{\prime}(u)=a_{1}O_{1,m}^{\prime}(u;0)+a_{2}O_{3,m}^{\prime
}(u;0)+I^{\prime}(u),
\end{align*}
where%
\begin{align*}
&  I(u)=\frac{O_{3,m}(u;0)}{\omega(0)}\int_{0}^{u}O_{1,m}(v;0)\eta
(v)dv+\frac{O_{1,m}(u;0)}{\omega(0)}\int_{u}^{\infty}O_{3,m}(v;0)\eta(v)dv,\\
&  I^{\prime}(u)=\frac{O_{3,m}^{\prime}(u;0)}{\omega(0)}\int_{0}^{u}%
O_{1,m}(v;0)\eta(v)dv+\frac{O_{1,m}^{\prime}(u;0)}{\omega(0)}\int_{u}^{\infty
}O_{3,m}(v;0)\eta(v)dv
\end{align*}
We obtain with the help of the Cauchy-Bunyakovskii inequality (CB-inequality):
$I(u)$ is bounded at infinity and%
\[
I(u)=\left\{
\begin{array}
[c]{l}%
O(u^{3/2}),\;|m|\geq2\\
O(u^{3/2}\ln u),\;|m|=1
\end{array}
\right.  ,\;I^{\prime}(u)=\left\{
\begin{array}
[c]{l}%
O(u^{1/2}),\;|m|\geq2\\
O(u^{1/2}\ln u),\;|m|=1
\end{array}
\right.  ,\;u\rightarrow0.
\]
The condition $\psi_{\ast}(u)\in L^{2}(\mathbb{R}_{+})$ gives $a_{1}=a_{2}=0$
such that we find%
\begin{equation}
\psi_{\ast}(u)=\left\{
\begin{array}
[c]{l}%
O(u^{3/2}),\;|m|\geq2\\
O(u^{3/2}\ln u),\;|m|=1
\end{array}
\right.  ,\;\psi_{\ast}^{\prime}(u)=\left\{
\begin{array}
[c]{l}%
O(u^{1/2}),\;|m|\geq2\\
O(u^{1/2}\ln u),\;|m|=1
\end{array}
\right.  ,\;u\rightarrow0, \label{Osc2.2.3.1.1}%
\end{equation}
and $\omega_{h_{m}^{+}}(\chi_{\ast},\psi_{\ast})=\Delta_{h_{m}^{+}}(\psi
_{\ast})=0$.

\subsection{Self-adjoint hamiltonian $\hat{h}_{m\mathfrak{e}}$}

Because $\omega_{h_{m}^{+}}(\chi_{\ast},\psi_{\ast})=\Delta_{h_{m}^{+}}%
(\psi_{\ast})=0$ (and also because $O_{1,m}(u;W)$ and $O_{3,m}(u;W)$ and their
any linear combinations are not s.-integrable for $\operatorname{Im}W\neq0$),
the deficiency indices of initial symmetric operator $\hat{h}_{m}$ are zero,
which means that $\hat{h}_{m\mathfrak{e}}=\hat{h}_{m}^{+}$ is a unique s.a.
extension of the initial symmetric operator $\hat{h}_{m}$:%
\begin{equation}
\hat{h}_{m\mathfrak{e}}:\left\{
\begin{array}
[c]{l}%
D_{h_{m\mathfrak{e}}}=D_{\check{h}_{m}}^{\ast}(\mathbb{R}_{+})\\
\hat{h}_{m\mathfrak{e}}\psi_{\ast}(u)=\check{h}_{m}\psi_{\ast}(u),\;\forall
\psi_{\ast}\in D_{h_{m\mathfrak{e}}}%
\end{array}
\right.  . \label{Osc2.2.4.1}%
\end{equation}

\subsection{The guiding functional $\Phi(\xi;W)$}

As a guiding functional $\Phi(\xi;W)$ we choose%
\begin{align}
&  \Phi(\xi;W)=\int_{0}^{\infty}O_{1,m}(u;W)\xi(u)du,\;\xi\in\mathbb{D}%
=D_{r}(\mathbb{R}_{+})\cap D_{h_{m\mathfrak{e}}}.\label{Osc2.2.5.1}\\
&  D_{r}(\mathbb{R}_{+})=\{\xi(u):\;\mathrm{supp}\xi\subseteq\lbrack
0,\beta_{\xi}],\;\beta_{\xi}<\infty\}.\nonumber
\end{align}

The guiding functional $\Phi(\xi;W)$ is simple. It has, obviously, the
properties 1) and 3) and we should prove the properties 2) only (see \cite{Naima},
pages 245-246). Let $\Phi(\xi_{0};E_{0})=0$, $\xi_{0}\in\mathbb{D}$, $E_{0}%
\in\mathbb{R}$. As a solution $\psi(u)$ of equation
\[
(\check{h}_{m}-E_{0})\psi(u)=\xi_{0}(u),
\]
we choose
\[
\psi(u)=O_{1,m}(u;E_{0})\int_{u}^{\infty}U(v)\xi_{0}(v)dv+U(u)\int_{0}%
^{u}O_{1,m}(v;E_{0})\xi_{0}(v)dv,
\]
where $U(u)$ is any solution of eq. $(\check{h}_{m}-E_{0})U(u)=0$ satisfying
the condition $\mathrm{Wr}(O_{1,m},U)=-1$. Because $\xi_{0}\in D_{r}%
(\mathbb{R}_{+})$, the function $\psi(u)$ is well determined. Because $\xi
_{0}\in D_{r}(\mathbb{R}_{+})$ and $\int_{0}^{u}O_{1,m}(v;E_{0})\xi_{0}(v)=0$
for $u>\beta_{\xi_{0}}$, we have $\psi(u)=0$ for $u>\beta_{\xi_{0}}$. Using
the CB-inequality we show that $\psi(u)$ satisfies the boundary condition
(\ref{Osc2.2.3.1.1}), that is, $\psi\in\mathbb{D}$. Thus, the guiding
functional $\Phi(\xi;W)$ is simple and the spectrum of $\hat{h}_{m\mathfrak{e}%
}$ is simple.

\subsection{Green function $G_{m}(u,v;W)$, spectral function $\sigma_{m}(E)$}

We find the Green function $G_{m}(u,v;W)$ as the kernel of the integral
representation
\[
\psi(u)=\int_{0}^{\infty}G_{m}(u,v;W)\eta(v)dv,\;\eta\in L^{2}(\mathbb{R}%
_{+}),
\]
of unique solution of an equation%
\begin{equation}
(\hat{h}_{m\mathfrak{e}}-W)\psi(u)=\eta(u),\;\operatorname{Im}W>0,
\label{Osc2.2.6.1}%
\end{equation}
for $\psi\in D_{h_{m\mathfrak{e}}}$. General solution of eq. (\ref{Osc2.2.6.1}%
) can be represented in the form%
\begin{align*}
\psi(u)  &  =a_{1}O_{1,m}(u;W)+a_{3}O_{3,m}(u;W)+I(u),\\
I(u)  &  =\frac{O_{1,m}(u;W)}{\omega(W)}\int_{u}^{\infty}O_{3,m}%
(v;W)\eta(v)dv+\frac{O_{3,m}(u;W)}{\omega(W)}\int_{0}^{u}O_{1,m}%
(v;W)\eta(v)dv,\\
I(u)  &  =O\left(  u^{-3/2}\right)  ,\;u\rightarrow\infty,\;I(u)=\left\{
\begin{array}
[c]{l}%
O\left(  u^{3/2}\right)  ,\;|m|\geq2\\
O\left(  u^{3/2}\ln u\right)  ,\;|m|=1
\end{array}
\right.  ,\;u\rightarrow0.
\end{align*}
A condition $\psi\in L^{2}(\mathbb{R}_{+})$ gives $a_{1}=a_{3}=0$, such that
we find%
\begin{align}
&  G_{m}(u,v;W)=\frac{1}{\omega(W)}\left\{
\begin{array}
[c]{l}%
O_{3,m}(u;W)O_{1,m}(v;W),\;u>v\\
O_{1,m}(u;W)O_{3,m}(v;W).\;u<v
\end{array}
\right.  =\nonumber\\
&  \,=\Omega_{m}(W)O_{1,m}(u;W)O_{1,m}(v;W)+\frac{1}{2\kappa_{0}|m|}\left\{
\begin{array}
[c]{c}%
O_{4,m}(u;W)O_{1,m}(v;W),\;u>v\\
O_{1,m}(u;W)O_{4,m}(v;W),\;u<v
\end{array}
\right.  ,\label{Osc2.2.6.2}\\
&  \Omega_{m}(W)\equiv\frac{B_{m}(W)}{\omega(W)}=\frac{[4\ln(\kappa
_{0}/\varkappa)-\psi(\alpha)-\psi(\alpha_{-})](1-\alpha)_{|m|}}{4\kappa
_{0}(\kappa_{0}/\varkappa)^{2|m|}\Gamma^{2}(\beta)}.\nonumber
\end{align}
Note that the last term in the r.h.s. of eq. (\ref{Osc2.2.6.2}) is real for
$W=E$. From the relation%
\[
O_{1,m}^{2}(u_{0};E)\sigma_{m}^{\prime}(E)=\frac{1}{\pi}\operatorname{Im}%
G_{m}(u_{0}-0,u_{0}+0;E+i0),
\]
where $f(E+i0)\equiv\lim_{\varepsilon\rightarrow+0}f(E+i\varepsilon)$,
$\forall f(W)$, we find%
\begin{equation}
\sigma_{m}^{\prime}(E)=\frac{1}{\pi}\operatorname{Im}\Omega_{m}(E+i0).
\label{Osc2.2.6.3}%
\end{equation}

\subsection{Spectrum}

\subsubsection{$E\geq0$}

It is convenient to represent the function $\Omega_{m}(W)$ in the form%

\begin{align}
&  \Omega_{m}(W)=\Omega_{1m}(W)+\Omega_{2m}(W),\;\Omega_{1m}(W)=-\frac{\psi
(\alpha)(1-\alpha)_{|m|}}{2\kappa_{0}(\kappa_{0}/\varkappa)^{2|m|}\Gamma
^{2}(\beta)},\nonumber\\
&  \Omega_{2m}(W)=\frac{(\kappa_{0}/\varkappa)^{-2|m|}[4\ln(\kappa
_{0}/\varkappa)+\Sigma_{m}(\alpha)](1-\alpha)_{|m|}}{4\kappa_{0}\Gamma
^{2}(\beta)},\;\operatorname{Im}\Omega_{2m}(E)=0,\nonumber
\end{align}

where we used relation%
\[
\psi(\alpha_{-})=\psi(\alpha)-\Sigma_{m}(\alpha),\;\Sigma_{m}(\alpha
)=\sum_{l=1}^{|m|}(\alpha-l)^{-1},
\]
such that we have%
\[
\sigma_{m}^{\prime}(E)=-\frac{(\kappa_{0}/\varkappa)^{-2|m|}}{2\pi\kappa
_{0}\Gamma^{2}(\beta)}\left.  (1-\alpha)_{|m|}\right|  _{W=E}\operatorname{Im}%
\psi(1/2+|m|/2-(E+i0)/4\sqrt{\lambda})
\]
The function $\psi(\alpha)$ is real for $W=E$ where $|\psi(\alpha)|<\infty$.
Therefore, $\sigma_{m}^{\prime}(E)$ can be not equal to zero only in the
points $\psi(\alpha)=\pm\infty$, i. e., in the points $\alpha=-n$,
$n\in\mathbb{Z}_{+}$, or%
\[
E_{n}=2\sqrt{\lambda}[1+|m|+2n].
\]
In the neighborhood of the points $E_{n}$ we have ($W=E_{n}+\Delta$,
$\Delta=E-E_{n}+i\varepsilon$, $\alpha=-n-\Delta/4\sqrt{\lambda}$)%
\[
\operatorname{Im}\psi(-n-\Delta/4\sqrt{\lambda})=-4\pi\sqrt{\lambda}%
\delta(E-E_{n}),\;\left.  (1-\alpha)_{|m|}\right|  _{W=E_{n}}=(1+n)_{|m|},
\]
We thus find%
\begin{align*}
&  \sigma_{m}^{\prime}(E)=\sum_{n\in\mathbb{Z}_{+}}Q_{n}^{2}\delta
(E-E_{n}),\;Q_{n}=\frac{(\kappa_{0}/\varkappa)^{-|m|}}{|m|!}\sqrt
{\frac{2\sqrt{\lambda}(1+n)_{|m|}}{\kappa_{0}}},\\
&  \mathrm{spec}\hat{h}_{m\mathfrak{e}}=\{E_{n},\;n\in\mathbb{Z}_{+}\}.
\end{align*}

\subsubsection{$E<0$}

In this case, we have $\alpha=1/2+|m|/2+|E|$, the function $\Omega_{1m}(E)$ is
real and $\sigma_{m}^{\prime}(E)=0$.

Finally: the spectrum of $\hat{h}_{m\mathfrak{e}}$ is simple and discrete,
$\mathrm{spec}\hat{h}_{m\mathfrak{e}}=\{E_{n}>0,\;n\in\mathbb{Z}_{+}\}$, and
the set of functions $\{U_{mn}(u)=Q_{n}O_{1,m}(u;E_{n}),\;n\in\mathbb{Z}%
_{+}\}$ forms a complete orthonormalized system in $L^{2}(\mathbb{R}_{+})$.

\subsection{$|m|\geq1$, $\lambda<0$}

\subsection{Useful solutions}

Here we again will begin with the solutions $O_{1,m,\delta}(u;W)$,
$O_{2,m,\delta}(u;W)$, $O_{3,m,\delta}(u;W)$, and $O_{4,m,\delta}(u;W)$ which
are given by eq. (\ref{Osc2.2.1.4a}), (\ref{Osc2.2.1.4b}) (\ref{Osc2.2.1.5a}%
)-(\ref{Osc2.2.1.5d}), where%
\begin{align*}
&  \varkappa^{2}=-i\sqrt{|\lambda|}=e^{-i\pi/2}\sqrt{|\lambda|},\;\varkappa
^{-1}=e^{i\pi/4}|\lambda|^{-1/4},\;\rho=-i\sqrt{|\lambda|}u^{2}=\\
&  =e^{-i\pi/2}\sqrt{|\lambda|}u^{2},\;\alpha=1/2+|m|/2-i\tilde{w},\;\tilde
{w}\equiv\tilde{w}_{O}=W/(4\sqrt{|\lambda|}),\\
&  \tilde{\alpha}=1/2+|m|/2+i\tilde{w}=\overline{\alpha(\overline{\tilde{w}}%
)},
\end{align*}

The functions $O_{1,m,\delta}(u;W)$, $O_{2,m,\delta}(u;W)$ and $O_{4,m,\delta
}(u;W)$ are real-entire in $W$. Indeed, we have%
\begin{align*}
&  \overline{O_{1,m,\delta}(u;W)}=(\kappa_{0}u)^{1/2+|m+\delta|}e^{\rho/2}%
\Phi(\left.  \alpha_{\delta}\right|  _{W\rightarrow-\overline{W}}%
,\beta_{\delta};-\rho)=\\
&  \,=(\kappa_{0}u)^{1/2+|m+\delta|}e^{-\rho/2}\Phi(\left.  \alpha_{\delta
}\right|  _{W\rightarrow\overline{W}},\beta_{\delta};\rho)=O_{1,m,\delta
}(u;\overline{W}),
\end{align*}
where we used the identity GradRy9.212.1 Analogously, we find%
\[
\overline{O_{2,m,\delta}(u;W)}=O_{2,m,\delta}(u;\overline{W}).
\]
The real-entireness of $O_{4,m,\delta}(u;W)$ follows from the fact that
$O_{1,m,\delta}(u;W)$, $O_{2,m,\delta}(u;W)$, and $A_{m,\delta}(W)$ are
real-entire in $W$.

\subsubsection{Asymptotics, $u\rightarrow0$ ($\rho\rightarrow0$)}

Asymptotics of $O_{1,m,\delta}(u;W)$ and $O_{3,m,\delta}(u;W)$ are given by
eqs. (\ref{Osc2.2.1.1.1a}) and (\ref{Osc2.2.1.1.1b}).

\subsubsection{Asymptotics, $u\rightarrow\infty$ ($\rho\rightarrow-i\infty$),
$\operatorname{Im}W>0$ or $W=0$}%

\begin{align*}
&  O_{1,m}(u;W)=\frac{\kappa_{0}^{1/2+|m|}\varkappa^{2\alpha-2\beta}%
\Gamma(\beta)}{\Gamma(\alpha)}u^{-1/2-2w}e^{\rho/2}(1+O(u^{-2}))=\\
&  =O(u^{-1/2+\operatorname{Im}W/2\sqrt{|\lambda|}}),\\
&  O_{3,m}(u;W)=\kappa_{0}^{1/2+|m|}\varkappa^{-2\alpha}u^{-1/2+2w}e^{-\rho
/2}(1+O(u^{-2}))=\\
&  =O(u^{-1/2-\operatorname{Im}W/2\sqrt{|\lambda|}}).
\end{align*}
-

\subsubsection{Wronskian}

The Wronskian of $O_{1,m,\delta}(u;W)$ and $O_{3,m,\delta}(u;W)$ is given by
eq. (\ref{Osc2.2.1.3.1}).

\subsection{Symmetric operator $\hat{h}_{m}$, adjoint operator $\hat{h}_{m}%
^{+}=\hat{h}_{m}^{\ast}$, s.a. hamiltonian $\hat{h}_{m\mathfrak{e}}$, guiding
functional $\Phi(\xi;W)$, Green function  $G_{m}(u,v;W)$, spectral function $\sigma_{m}(E)$}

These quantities are given exactly by the same formulas (\ref{Osc2.2.2.1}),
(\ref{Osc2.2.3.1}), (\ref{Osc2.2.4.1}), (\ref{Osc2.2.5.1}), (\ref{Osc2.2.6.2}%
), and (\ref{Osc2.2.6.3}) for the corresponding quantities of previous section.

Let us study the structure of exp.(\ref{Osc2.2.6.3}) for $\sigma_{m}^{\prime}(E).$

Let $|m|=2n+1$, $n\in\mathbb{Z}_{+}$. We have%
\begin{align*}
&  \,(\kappa_{0}/\varkappa)^{-2|m|}=(-i)^{|m|}(\sqrt{|\lambda|}/\kappa_{0}%
^{2})^{|m|},\;(1-\alpha)_{|m|}=i^{|m|}\tilde{e}\prod_{l=1}^{n}(l^{2}+\tilde
{e}^{2}),\\
&  \,(\kappa_{0}/\varkappa)^{-2|m|}(1-\alpha)_{|m|}=\tilde{e}q_{1n}%
(E),\;q_{1n}(E)=(\sqrt{|\lambda|}/\kappa_{0}^{2})^{|m|}\prod_{l=1}^{n}%
(l^{2}+\tilde{e}^{2})>0,\\
&  \operatorname{Im}\psi(\alpha)+\operatorname{Im}\psi(\alpha_{-}%
)=\operatorname{Im}\psi(1+n-i\tilde{e})+\operatorname{Im}\psi(-n-i\tilde
{e})=-\pi\coth(\pi\tilde{e}),\\
&  \ln(\kappa_{0}/\varkappa)=i\pi/4+\ln(\kappa_{0}|\lambda|^{-1/4}%
),\;\tilde{e}=E/4\sqrt{|\lambda|},
\end{align*}
and%
\[
\sigma_{m}^{\prime}(E)\equiv\rho_{m}^{2}(E)=\frac{\tilde{e}q_{1n}(E)}%
{4\kappa_{0}\Gamma^{2}(\beta)}[\coth(\pi\tilde{e})+1]>0.
\]
Let $|m|=2n$, $n\in\mathbb{N}$. We have%
\begin{align*}
&  \,(\kappa_{0}/\varkappa)^{-2|m|}=(-i)^{|m|}(\sqrt{|\lambda|}/\kappa_{0}%
^{2})^{|m|},\;(1-\alpha)_{|m|}=i^{|m|}\prod_{l=0}^{n-1}[(l+1/2)^{2}+\tilde
{e}^{2}],\\
&  \,(\kappa_{0}/\varkappa)^{-2|m|}(1-\alpha)_{|m|}=q_{2n}(E),\;q_{2n}%
(E)=(\sqrt{|\lambda|}/\kappa_{0}^{2})^{|m|}\prod_{l=0}^{n-1}[(l+1/2)^{2}%
+\tilde{e}^{2}]>0,\\
&  \operatorname{Im}\psi(\alpha)+\operatorname{Im}\psi(\alpha_{-}%
)=\operatorname{Im}\psi(1/2+n-i\tilde{e})+\operatorname{Im}\psi(1/2-n-i\tilde
{e})=-\pi\tanh(\pi\tilde{e}),\\
&  \ln(\kappa_{0}/\varkappa)=i\pi/4+\ln(\kappa_{0}|\lambda|^{-1/4}),
\end{align*}
and%
\[
\sigma_{m}^{\prime}(E)\equiv\rho_{m}^{2}(E)=\frac{q_{2n}(E)}{4\kappa_{0}%
\Gamma^{2}(\beta)}[1+\tanh(\pi\tilde{e})]>0.
\]

The function $\sigma_{m}^{\prime}(E)$ is absolutely continuous for any
$E\in\mathbb{R}$, such that we obtain: the spectrum of $\hat{h}_{m\mathfrak{e}%
}$ is simple, continuous and it fills out real axis, $\mathrm{spec}%
\hat{h}_{m\mathfrak{e}}=\mathbb{R}.$The set of (generalized) eigenfunctions
$\{U_{mE}(u)=\rho_{m}(E)O_{1,m}(u;E),\;E\in\mathbb{R}\}$ of
$\hat{h}_{m\mathfrak{e}}$ forms a complete orthonormalized system.

\subsection{$|m|\geq1$, $\lambda=0$}

\subsection{Useful solutions}

Here we set $\delta=0$\ at once. Eq. (\ref{Osc2.2.1.1}) is reduced to the form%
\begin{equation}
\lbrack-\partial_{u}^{2}+u^{-2}(m^{2}-1/4)-W]\psi(u)=0. \label{Osc2.4.1.1}%
\end{equation}
The solutions of eq. (\ref{Osc2.4.1.1}) are expressed in the terms of the
Bessel functions. We use the following solutions:%
\begin{align*}
&  O_{1,m}(u;W)=D_{1,m}(W)u^{1/2}J_{|m|}(Ku),\;D_{1,m}(W)=\kappa_{0}%
^{1/2}\Gamma(\beta)(K/2\kappa_{0})^{-|m|},\\
&  O_{3,m}(u;W)=iD_{3,m}(W)u^{1/2}H_{|m|}^{(1)}(Ku),\;D_{3,m}(W)=\pi\kappa
_{0}^{1/2}\Gamma^{-1}(\beta-1)(K/2\kappa_{0})^{|m|},\\
&  O_{4,m}(u;W)=D_{3,m}(W)u^{1/2}\left[  N_{|m|}(Ku)-\frac{2}{\pi}%
J_{|m|}(Ku)\ln(K/\kappa_{0})\right]  ,\\
&  O_{3,m}(u;W)=\pi|m|\omega_{m}(W)\left[  i-\frac{2}{\pi}\ln(K/\kappa
_{0})\right]  O_{1,m}(u;W)-O_{4,m}(u;W),\\
&  K=W^{1/2}=\sqrt{|W|}e^{i\varphi_{W}/2},\;\omega_{m}(W)=\frac{D_{3,m}%
(W)}{\pi|m|D_{1,m}(W)}=\frac{(W/4\kappa_{0}^{2})^{|m|}}{\Gamma^{2}(\beta)}.
\end{align*}

\subsubsection{Asymptotics, $u\rightarrow0$ ($z\rightarrow0$)}%

\begin{align*}
O_{1,m}(u;W)  &  =(\kappa_{0}u)^{1/2+|m|}(1+O(u^{2})),\\
O_{4,m}(u;W)  &  =-(\kappa_{0}u)^{1/2-|m|}\left(  1+\left\{
\begin{array}
[c]{c}%
O(u^{2}),\;|m|\geq2\\
O(u^{2}\ln u),\;|m|=1
\end{array}
\right.  \right)  ,\\
O_{3,m}(u;W)  &  =(\kappa_{0}u)^{1/2-|m|}\left(  1+\left\{
\begin{array}
[c]{c}%
O(u^{2}),\;|m|\geq2\\
O(u^{2}\ln u),\;|m|=1
\end{array}
\right.  \right)  .
\end{align*}

\subsubsection{Asymptotics, $u\rightarrow\infty$ ($z\rightarrow\infty$),
$\operatorname{Im}W>0$ ($\operatorname{Im}K>0$)}%

\begin{align*}
O_{1,m}(u;W)  &  =D_{1,m}(W)\sqrt{\frac{1}{2\pi K}}e^{i\pi(|m|/2+1/4)}%
e^{-iKu}(1+O(u^{-1}))=O(e^{u\operatorname{Im}K}),\\
O_{3,m}(u;W)  &  =iD_{3,m}(W)\sqrt{\frac{2}{\pi K}}e^{-i\pi(|m|/2+1/4)}%
e^{iKu}(1+O(u^{-1})).=O(e^{-u\operatorname{Im}K}).
\end{align*}

\subsubsection{Wronskian}%

\[
\mathrm{Wr}(O_{1,m},O_{3,m})=-2\kappa_{0}|m|.
\]

\subsection{Symmetric operator $\hat{h}_{m}$}

For given a differential operation $\check{h}_{m}=-\partial_{u}^{2}%
+u^{-2}(m^{2}-1/4)$ we determine the following symmetric operator
$\hat{h}_{m}$,%

\[
\hat{h}_{m}:\left\{
\begin{array}
[c]{l}%
D_{h_{m}}=\mathcal{D}(\mathbb{R}_{+}),\\
\hat{h}_{m}\psi(u)=\check{h}_{m}\psi(u),\;\forall\psi\in D_{h_{m}}%
\end{array}
\right.  .
\]

\subsection{Adjoint operator $\hat{h}_{m}^{+}=\hat{h}_{m}^{\ast}$}%

\[
\hat{h}_{m}^{+}:\left\{
\begin{array}
[c]{l}%
D_{h_{m}^{+}}=D_{\check{h}_{m}}^{\ast}(\mathbb{R}_{+})=\{\psi_{\ast}%
,\psi_{\ast}^{\prime}\;\mathrm{are\;a.c.\;in}\mathcal{\;}\mathbb{R}_{+}%
,\;\psi_{\ast},\hat{h}_{m}^{+}\psi_{\ast}\in L^{2}(\mathbb{R}_{+})\}\\
\hat{h}_{m}^{+}\psi_{\ast}(u)=\check{h}_{m}\psi_{\ast}(u),\;\forall\psi_{\ast
}\in D_{h_{m}^{+}}%
\end{array}
\right.  .
\]

\subsubsection{Asymptotics}

I) $|u|\rightarrow\infty$

Because $V(u)>-(|\lambda|+1)u^{2}$ for large $u$, we have: $[\psi_{\ast}%
,\chi_{\ast}](u)\rightarrow0$ as $u\rightarrow\infty$, $\forall\psi_{\ast
},\chi_{\ast}\in D_{H^{+}}$.

II) $u\rightarrow0$

Because $\check{h}_{m}\psi_{\ast}\in L^{2}(\mathbb{R})$, we have%
\[
\check{h}_{m}\psi_{\ast}(u)=[-\partial_{u}^{2}+u^{-2}(m^{2}-1/4)]\psi_{\ast
}(u)=\eta(u),\;\eta\in L^{2}(\mathbb{R}_{+}).
\]
General solution of this equation can be represented in the form%
\begin{align*}
&  \psi_{\ast}(u)=a_{1}u^{1/2+|m|}+a_{2}u^{1/2-|m|}+I(u),\\
&  \psi_{\ast}^{\prime}(u)=(a_{1}u^{1/2+|m|}+a_{2}u^{1/2-|m|})^{\prime
}+I^{\prime}(u),
\end{align*}
where%
\begin{align*}
&  I(u)=\frac{u^{1/2-|m|}}{2|m|}\int_{0}^{u}v^{1/2+|m|}\eta
(v)dv+\frac{u^{1/2+|m|}}{2|m|}\int_{u}^{x_{0}}v^{1/2-|m|}\eta(v)dv,\\
&  I^{\prime}(u)=\frac{(u^{1/2-|m|})^{\prime}}{2|m|}\int_{0}^{u}%
v^{1/2+|m|}\eta(v)dv+\frac{(u^{1/2+|m|})^{\prime}}{2|m|}\int_{u}^{x_{0}%
}v^{1/2-|m|}\eta(v)dv
\end{align*}
We obtain with the help of the Cauchy-Bunyakovskii inequality (CB-inequality):
$I(u)$ is bounded at infinity and%
\[
I(u)=\left\{
\begin{array}
[c]{l}%
O(u^{3/2}),\;|m|\geq2\\
O(u^{3/2}\ln u),\;|m|=1
\end{array}
\right.  ,\;I^{\prime}(u)=\left\{
\begin{array}
[c]{l}%
O(u^{1/2}),\;|m|\geq2\\
O(u^{1/2}\ln u),\;|m|=1
\end{array}
\right.  ,\;u\rightarrow0.
\]
The condition $\psi_{\ast}(u)\in L^{2}(\mathbb{R}_{+})$ gives $a_{1}=0$, such
that we find%
\[
\psi_{\ast}(u)=\left\{
\begin{array}
[c]{l}%
O(u^{3/2}),\;|m|\geq2\\
O(u^{3/2}\ln u),\;|m|=1
\end{array}
\right.  ,\;\psi_{\ast}^{\prime}(u)=\left\{
\begin{array}
[c]{l}%
O(u^{1/2}),\;|m|\geq2\\
O(u^{1/2}\ln u),\;|m|=1
\end{array}
\right.  ,\;u\rightarrow0,
\]
and $\omega_{h_{m}^{+}}(\chi_{\ast},\psi_{\ast})=\Delta_{h_{m}^{+}}(\psi
_{\ast})=0$

\subsection{Self-adjoint hamiltonian $\hat{h}_{m\mathfrak{e}}$}

Because $\omega_{h_{m}^{+}}(\chi_{\ast},\psi_{\ast})=\Delta_{h_{m}^{+}}%
(\psi_{\ast})=0$ (and also because $O_{1,m}(u;W)$ and $O_{3,m}(u;W)$ and their
linear combinations are not s.-integrable for $\operatorname{Im}W\neq0$), the
deficiency indices of initial symmetric operator $\hat{h}_{m}$ are zero, which
means that $\hat{h}_{m\mathfrak{e}}=\hat{h}_{m}^{+}$ is a unique s.a.
extension of the initial symmetric operator $\hat{h}_{m}$:%
\[
\hat{h}_{m\mathfrak{e}}:\left\{
\begin{array}
[c]{l}%
D_{h_{m\mathfrak{e}}}=D_{\check{h}_{m}}^{\ast}(\mathbb{R}_{+})\\
\hat{h}_{m\mathfrak{e}}\psi_{\ast}(u)=\check{h}_{m}\psi_{\ast}(u),\;\forall
\psi_{\ast}\in D_{h_{m\mathfrak{e}}}%
\end{array}
\right.  .
\]

\subsection{The guiding functional $\Phi(\xi;W)$}

As a guiding functional $\Phi(\xi;W)$ we choose%
\begin{align*}
&  \Phi(\xi;W)=\int_{0}^{\infty}O_{1,m}(u;W)\xi(u)du,\;\xi\in\mathbb{D}%
_{\zeta}=D_{r}(\mathbb{R}_{+})\cap D_{h_{m\mathfrak{e}}}.\\
&  D_{r}(a,b)=\{\psi(u):\;\mathrm{supp}\psi\subseteq\lbrack a,\beta_{\psi
}],\;\beta_{\psi}<b.
\end{align*}

The guiding functional $\Phi_{\zeta}(\xi;W)$ is simple. and the spectrum of
$\hat{h}_{m\mathfrak{e}}$ is simple.

\subsection{Green function $G_{m}(u,v;W)$, spectral function $\sigma_{m}(E)$}

We find the Green function $G_{m}(u,v;W)$ as the kernel of the integral
representation
\[
\psi(u)=\int_{0}^{\infty}G_{m}(u,v;W)\eta(v)dv,\;\eta\in L^{2}(\mathbb{R}%
_{+}),
\]
of unique solution of an equation%
\begin{equation}
(\hat{h}_{m\mathfrak{e}}-W)\psi(u)=\eta(u),\;\operatorname{Im}W>0,
\label{Osc2.4.6.1}%
\end{equation}
for $\psi\in D_{h_{m\mathfrak{e}}}$. General solution of eq. (\ref{Osc2.4.6.1}%
) can be represented in the form%
\begin{align*}
\psi(u)  &  =a_{1}O_{1,m}(u;W)+a_{3}O_{3,m}(u;W)+I(u),\\
I(u)  &  =\frac{O_{1,m}(u;W)}{2\kappa_{0}|m|}\int_{u}^{\infty}O_{3,m}%
(v;W)\eta(v)dv+\frac{O_{3,m}(u;W)}{2\kappa_{0}|m|}\int_{0}^{u}O_{1,m}%
(v;W)\eta(v)dv,\\
I(u)  &  =O\left(  u^{-3/2}\right)  ,\;u\rightarrow\infty,\;I(u)=\left\{
\begin{array}
[c]{l}%
O\left(  u^{3/2}\right)  ,\;|m|\geq2\\
O\left(  u^{3/2}\ln u\right)  ,\;|m|=1
\end{array}
\right.  ,\;u\rightarrow0.
\end{align*}
A condition $\psi\in L^{2}(\mathbb{R}_{+})$ gives $a_{1}=a_{3}=0$, such that
we find%
\begin{align}
&  G_{m}(u,v;W)=\frac{1}{2\kappa_{0}|m|}\left\{
\begin{array}
[c]{l}%
O_{3,m}(u;W)O_{1,m}(v;W),\;u>v\\
O_{1,m}(u;W)O_{3,m}(v;W).\;u<v
\end{array}
\right.  =\nonumber\\
&  \,=\Omega_{m}(W)O_{1,m}(u;W)O_{1,m}(v;W)-\frac{1}{2\kappa_{0}|m|}\left\{
\begin{array}
[c]{c}%
O_{4,m}(u;W)O_{1,m}(v;W),\;u>v\\
O_{1,m}(u;W)O_{4,m}(v;W),\;u<v
\end{array}
\right.  ,\label{Osc2.4.6.2}\\
&  \Omega_{m}(W)\equiv\frac{\pi}{2\kappa_{0}}\omega_{m}(W)[i-(2/\pi
)\ln(K/\kappa_{0})]=\frac{\pi(W/4\kappa_{0}^{2})^{|m|}}{2\kappa_{0}\Gamma
^{2}(\beta)}[i-(2/\pi)\ln(K/\kappa_{0})].\nonumber
\end{align}
Note that the last term in the r.h.s. of eq. (\ref{Osc2.4.6.2}) is real for
$W=E$. We find
\[
\sigma_{m}^{\prime}(E)=\frac{1}{\pi}\operatorname{Im}\Omega_{m}(E+i0).
\]

\subsection{Spectrum}

\subsubsection{$E=p^{2}\geq0$ ($K=p$)}%

\[
\sigma_{m}^{\prime}(E)=\rho_{m}^{2}(E),\;\rho_{m}(E)=\frac{(p/2\kappa
_{0})^{|m|}}{\sqrt{2\kappa_{0}}|m|!}%
\]

The spectrum is simple and continuous on whole nonnegative $E$-semiaxis.

\subsubsection{$E=-\tau^{2}<0$}

In this case, we have $K=e^{i\pi/2}\tau$, the function $\Omega_{m}(E)$ is real
and $\sigma_{m}^{\prime}(E)=0$.

Finally: the spectrum $\hat{h}_{m\mathfrak{e}}$ is simple and continuous,
$\mathrm{spec}\hat{h}_{m\mathfrak{e}}=\{E\geq0\}$, and the set of functions
$\{U_{mE}(u),\;E\geq0\}$,%
\[
U_{mE}(u)=\rho_{m}(E)O_{1,m}(u;E)=\sqrt{u/2}J_{|m|}(pu),
\]
forms a complete orthonormalized system in $L^{2}(\mathbb{R}_{+})$.

Note that the results of this section can be obtained as limit $\lambda
\rightarrow0$ of the corresponding results of previous sec.3.

Indeed, we have%
\begin{align*}
&  \lim_{\lambda\rightarrow0}\left.  O_{1,m}(u;W)\right|  _{\lambda\neq
0}=(\kappa_{0}u)^{1/2+|m|}\sum_{k=0}^{\infty}\frac{(-W/4\varkappa^{2}%
)^{k}\Gamma(\beta)}{\Gamma(\beta+k)}\frac{\varkappa^{2k}u^{2k}}{k!}=\\
&  \,=\kappa_{0}^{1/2}(K/2\kappa_{0})^{-|m|}\Gamma(\beta)\left[
(Ku/2)^{|m|}\sum_{k=0}^{\infty}\frac{(-1)^{k}(Ku/2)^{2k}}{\Gamma(\beta
+k)k!}\right]  =\\
&  =D_{1,m}(W)u^{1/2}J_{|m|}(Ku)=\left.  O_{1,m}(u;W)\right|  _{\lambda=0}.
\end{align*}
Further, we have for $\sigma_{m}^{\prime}(E)$

i) $E>0$%
\[
\lim_{\lambda\rightarrow-0}\left.  \sigma_{m}^{\prime}(E)\right|  _{\lambda
<0}=\frac{1}{2\kappa_{0}\Gamma^{2}(\beta)}\left(  \frac{E}{4\kappa_{0}^{2}%
}\right)  ^{|m|}%
\]

ii) $E<0$%
\[
\lim_{\lambda\rightarrow-0}\left.  \sigma_{m}^{\prime}(E)\right|  _{\lambda
<0}=0.
\]

iii) $E=0$%
\[
\left.  \sigma_{m}^{\prime}(0)\right|  _{\lambda<0}=0.
\]

\subsection{$m=0$}

\subsection{Useful solutions}

For $m=0$, eqs. (\ref{Osc2.2.1.1}) and (\ref{Osc2.2.1.2}) are redused
respectively to%
\begin{equation}
(-\partial_{u}^{2}-u^{-2}/4+\lambda u^{2}-W)\psi(u)=0\label{Osc2.5.1.1}%
\end{equation}
and%
\[
\lbrack-\partial_{u}^{2}+u^{-2}(\delta^{2}-1/4)+\lambda u^{2}-W]\psi(u)=0.
\]

We will use the following solutions of eq. (\ref{Osc2.5.1.1})%
\begin{align*}
&  O_{1,0}(u;W)=O_{1,0,0}(u;W)=(\kappa_{0}u)^{1/2}e^{-\rho/2}\Phi
(\alpha,1;\rho),\\
&  O_{2,0}(u;W)=\left.  \partial_{\delta}O_{1,0,\delta}(u;W)\right|
_{\delta=+0}=\\
&  \,=(\kappa_{0}u)^{1/2}e^{-\rho/2}\left.  \partial_{\delta}\Phi
(\alpha_{\delta},\beta_{\delta};\rho)\right|  _{\delta=+0}+O_{1,0}%
(u;W)\ln(\kappa_{0}u),\\
&  O_{3,0}(u;W)=O_{3,0,0}(u;W)=(\kappa_{0}u)^{1/2}e^{-\rho/2}\Psi
(\alpha,1;\rho)=\\
&  \,=\frac{\omega_{0}(W)}{\Gamma(\alpha)}O_{1,0}(u;W)-\frac{2}{\Gamma
(\alpha)}O_{2,0}(u;W),\\
&  \alpha=1/2-w,\;\omega_{0}(W)=2\ln(\kappa_{0}/\varkappa)+2\psi
(1)-\psi(\alpha).
\end{align*}

\subsubsection{Asymptotics, $u\rightarrow0$}

We have%
\begin{align*}
O_{1,0}(u;W)  &  =(\kappa_{0}u)^{1/2}(1+O(u^{2})),\;O_{2,0}(u;W)=(\kappa
_{0}u)^{1/2}\ln(\kappa_{0}u)(1+O(u^{2})),\\
O_{3,0}(u;W)  &  =\left[  \frac{\omega_{0}(W)}{\Gamma(\alpha)}(\kappa
_{0}u)^{1/2}-\frac{2}{\Gamma(\alpha)}(\kappa_{0}u)^{1/2}\ln(\kappa
_{0}u)\right]  (1+O(u^{2})).
\end{align*}

\subsubsection{Wronskian}%

\[
\mathrm{Wr}(O_{1,0},O_{3,0})=-\frac{2\kappa_{0}}{\Gamma(\alpha)}%
\]

\subsection{$\lambda>0$ ($\varkappa=\lambda^{1/4}$)}

\subsubsection{Asymptotics,\ $u\rightarrow\infty$, $\operatorname{Im}W>0$ or
$W=0$}

We have%
\begin{align*}
O_{1,0}(u;W)  &  =\frac{\kappa_{0}^{1/2}\varkappa^{-1-2w}}{\Gamma(\alpha
)}u^{-1/2-2w}e^{\rho/2}(1+O(u^{-2})),\\
O_{3,0}(u;W)  &  =\kappa_{0}^{1/2}\varkappa^{-1+2w}u^{-1/2+2w}e^{-\rho
/2}(1+O(u^{-2})).
\end{align*}

\subsubsection{Symmetric operator $\hat{h}_{0}$}

For given a differential operation $\check{h}_{0}=-\partial_{u}^{2}%
-u^{-2}/4+\lambda u^{2}$ we determine the following symmetric operator
$\hat{h}_{0}$,%

\begin{equation}
\hat{h}_{0}:\left\{
\begin{array}
[c]{l}%
D_{h_{0}}=\mathcal{D}(\mathbb{R}_{+}),\\
\hat{h}_{0}\psi(u)=\check{h}_{0}\psi(u),\;\forall\psi\in D_{h_{0}}%
\end{array}
\right.  . \label{Osc2.5.2.2.1}%
\end{equation}

\subsubsection{Adjoint operator $\hat{h}_{0}^{+}=\hat{h}_{0}^{\ast}$}%

\begin{equation}
\hat{h}_{0}^{+}:\left\{
\begin{array}
[c]{l}%
D_{h_{0}^{+}}=D_{\check{h}_{0}}^{\ast}(\mathbb{R}_{+})=\{\psi_{\ast}%
,\psi_{\ast}^{\prime}\;\mathrm{are\;a.c.\;in}\mathcal{\;}\mathbb{R}_{+}%
,\;\psi_{\ast},\hat{h}_{0}^{+}\psi_{\ast}\in L^{2}(\mathbb{R}_{+})\}\\
\hat{h}_{0}^{+}\psi_{\ast}(u)=\check{h}_{0}\psi_{\ast}(u),\;\forall\psi_{\ast
}\in D_{h_{0}^{+}}%
\end{array}
\right.  . \label{Osc2.5.2.3.1}%
\end{equation}

\paragraph{Asymptotics}

I) $u`\rightarrow\infty$

Because $V(u)>-(|\lambda|+1)u^{2}$ for large $u$, we have: $[\psi_{\ast}%
,\chi_{\ast}](u)\rightarrow0$ as $u\rightarrow\infty$, $\forall\psi_{\ast
},\chi_{\ast}\in D_{H^{+}}$.

II) $u\rightarrow0$

By the standard way, we obtain%
\begin{align}
&  \psi_{\ast}(u)=c_{1}u_{1\mathrm{as}}(u)+c_{2}u_{2\mathrm{as}}%
(u)+O(u^{3/2}\ln u),\nonumber\\
&  \psi_{\ast}^{\prime}(u)=c_{1}u_{1\mathrm{as}}^{\prime}(u)+c_{2}%
u_{2\mathrm{as}}^{\prime}(u)+O(u^{1/2}\ln u),\label{Osc2.5.2.3.1.1}\\
&  u_{1\mathrm{as}}(u)=(\kappa_{0}u)^{1/2},\;u_{2\mathrm{as}}(u)=(\kappa
_{0}u)^{1/2}\ln(\kappa_{0}u).\nonumber
\end{align}

For the asymmetry form $\Delta_{h_{0}^{+}}(\psi_{\ast})$, we find%
\begin{align}
&  \Delta_{h_{0}^{+}}(\psi_{\ast})=\kappa_{0}(\overline{c_{2}}c_{1}%
-\overline{c_{1}}c_{2})=i\kappa_{0}(\overline{c_{+}}c_{+}-\overline{c_{-}%
}c_{-}),\label{Osc2.5.2.3.1.2}\\
&  c_{\pm}=\frac{1}{\sqrt{2}}(c_{1}\pm ic_{2}).\nonumber
\end{align}

\subsubsection{Self-adjoint hamiltonians}

The condition $\Delta_{h_{0}^{+}}(\psi)=0$ gives%
\begin{align*}
&  c_{-}=e^{2i\theta}c_{+},\;0\leq\theta\leq\pi,\;\theta=0\sim\theta
=\pi,\;\Longrightarrow\;\\
&  c_{1}\cos\zeta=c_{2}\sin\zeta,\;\zeta=\theta-\pi/2,\;|\zeta|\leq
\pi/2,\;\zeta=-\pi/2\sim\zeta=\pi/2,
\end{align*}
or%
\begin{align}
&  \psi(u)=C\psi_{\mathrm{as}}(u)+O(u^{3/2}\ln u),\;\psi^{\prime}%
(u)=C\psi_{\mathrm{as}}^{\prime}(u)+O(u^{1/2}\ln u),\label{Osc2.5.2.4.1}\\
&  \psi_{\mathrm{as}}(u)=u_{1\mathrm{as}}(u)\sin\zeta+u_{2\mathrm{as}}%
(u)\cos\zeta.\nonumber
\end{align}
We thus have a family of s.a. $\hat{h}_{0\zeta}$,%
\begin{equation}
\hat{h}_{0\zeta}:\left\{
\begin{array}
[c]{l}%
D_{h_{0\zeta}}-\{\psi\in D_{h_{0}^{+}},\;\psi
\;\mathrm{satisfy\;the\;boundary\;condition\;(\ref{Osc2.5.2.4.1})}\\
\hat{h}_{0\zeta}\psi=\check{h}_{0}\psi,\;\forall\psi\in D_{h_{0\zeta}}%
\end{array}
\right.  . \label{Osc2.5.2.4.2}%
\end{equation}

\subsubsection{The guiding functional}

As a guiding functional $\Phi_{\zeta}(\xi;W)$ we choose%
\begin{align}
&  \Phi_{\zeta}(\xi;W)=\int_{0}^{\infty}U_{\zeta}(u;W)\xi(u)du,\;\xi
\in\mathbb{D}_{\zeta}=D_{r}(\mathbb{R}_{+})\cap D_{h_{0\zeta}}%
.\label{Osc2.5.2.5.1}\\
&  .U_{\zeta}(u;W)=O_{1,0}(u;W)\sin\zeta+O_{2,0}(u;W)\cos\zeta,\nonumber
\end{align}
$U_{\zeta}(u;W)$ is real-entire solution of eq. (\ref{Osc2.5.1.1}) satisfying
the boundary condition (\ref{Osc2.5.2.4.1}).

The guiding functional $\Phi_{\zeta}(\xi;W)$ is simple and the spectrum of
$\hat{h}_{0\zeta}$ is simple.

\subsubsection{Green function $G_{\zeta}(u,v;W)$, spectral function
$\sigma_{\zeta}(E)$.}

We find the Green function $G_{\zeta}(u,v;W)$ as the kernel of the integral
representation
\[
\psi(u)=\int_{0}^{\infty}G_{\zeta}(u,v;W)\eta(v)dv,\;\eta\in L^{2}%
(\mathbb{R}_{+}),
\]
of unique solution of an equation%
\begin{equation}
(\hat{h}_{0\zeta}-W)\psi(u)=\eta(u),\;\operatorname{Im}W>0,
\label{Osc2.5.2.6.1}%
\end{equation}
for $\psi\in D_{h_{0\zeta}}$. General solution of eq. (\ref{Osc2.5.2.6.1})
(under condition $\psi\in L^{2}(\mathbb{R}_{+})$) can be represented in the
form%
\begin{align*}
\psi(u)  &  =aO_{3,0}(u;W)+\frac{\Gamma(\alpha)}{2\kappa_{0}}O_{1,0}%
(u;W)\eta_{3}(W)+\frac{\Gamma(\alpha)}{2\kappa_{0}}I(u),\;\eta_{3}(W)=\int
_{0}^{\infty}O_{3,0}(v;W)\eta(v)dv,\\
I(u)  &  =O_{3,0}(u;W)\int_{0}^{u}O_{1,0}(v;W)\eta(v)dv-O_{1,0}(u;W)\int
_{0}^{u}O_{3,0}(v;W)\eta(v)dv,\\
I(u)  &  =O\left(  u^{3/2}\ln u\right)  ,\;u\rightarrow0.
\end{align*}
A condition $\psi\in D_{h_{0\zeta}\text{ , }}$(i.e.$\psi$ satisfies the
boundary condition (\ref{Osc2.5.2.4.1})) gives%
\[
a=-\frac{\Gamma^{2}(\alpha)\cos\zeta}{4\kappa_{0}\omega_{\zeta}(W)}\eta
_{3}(W),\;\omega_{\zeta}(W)=\frac{1}{2}\omega_{0}(W)\cos\zeta+\sin\zeta,
\]%
\begin{align}
&  G_{\zeta}(u,v;W)=\Omega_{\zeta}(W)U_{\zeta}(u;W)U_{\zeta}(v;W)+\nonumber\\
&  \,+\frac{1}{\kappa_{0}}\left\{
\begin{array}
[c]{c}%
\tilde{U}_{\zeta}(u;W)U_{\zeta}(v;W),\;u>v\\
U_{\zeta}(u;W)\tilde{U}_{\zeta}(v;W),\;u<v
\end{array}
\right.  ,\label{Osc2.5.2.6.2}\\
&  \Omega_{\zeta}(W)\equiv\frac{\tilde{\omega}_{\zeta}(W)}{\kappa_{0}%
\omega_{\zeta}(W)},\;\tilde{\omega}_{\zeta}(W)=\frac{1}{2}\omega_{0}%
(W)\sin\zeta-\cos\zeta,\nonumber\\
&  \tilde{U}_{\zeta}(u;W)=O_{1,0}(u;W)\cos\zeta-O_{2,0}(u;W)\sin
\zeta,\nonumber
\end{align}
where we used an equality%
\[
\Gamma(\alpha)O_{3,0}(u;W)=2\tilde{\omega}_{\zeta}(W)U_{\zeta}(u;W)+2\omega
_{\zeta}(W)\tilde{U}_{\zeta}(u;W).
\]
Note that the function $\tilde{U}_{\zeta}(u;W)$ is real-entire in $W$ and the
last term in the r.h.s. of eq. (\ref{Osc2.5.2.6.2}) is real for $W=E$. For
$\sigma_{\zeta}^{\prime}(E)$, we find%
\begin{equation}
\sigma_{\zeta}^{\prime}(E)=\frac{1}{\pi}\operatorname{Im}\Omega_{\zeta}(E+i0).
\label{Osc2.5.2.6.3}%
\end{equation}

\subsubsection{Spectrum}

\paragraph{$\zeta=\pi/2$}

First we consider the case $\zeta=\pi/2$.

In this case, we have $U_{\pi/2}(u;W)=O_{1,0}(u;W)$ and%
\[
\sigma_{\pi/2}^{\prime}(E)=-\frac{1}{2\pi\kappa_{0}}\operatorname{Im}%
\psi(1/2-(E+i0)/4\sqrt{\lambda}),
\]

\subparagraph{$E\geq0$}

In this case, we find%
\begin{align*}
&  \sigma_{\pi/2}^{\prime}(E)=\sum_{n=0}^{\infty}\frac{2\sqrt{\lambda}}%
{\kappa_{0}}\delta(E-\mathcal{E}_{n}),\;\mathcal{E}_{n}=2\sqrt{\lambda
}(1+2n),\\
&  \mathrm{spec}\hat{h}_{0\pi/2}=\{\mathcal{E}_{n},\;n\in\mathbb{Z}_{+}\}..
\end{align*}

\subparagraph{$E<0$}

In this case, we have $\alpha=1/2+|E|/4\sqrt{\lambda}$, the function
$\omega_{0}(E)$ is real and $\sigma_{0}^{\prime}(E)=0$.

Finally: the spectrum $\hat{h}_{\pi/2}$ is simple and discrete, $\mathrm{spec}%
\hat{h}_{\pi/2}=\{\mathcal{E}_{n}>0,\;n\in\mathbb{Z}_{+}\}$, and the set of
eigenfunctions $\{U_{n}(u)=(2\sqrt{\lambda}/\kappa_{0})^{1/2}O_{1,0}%
(u;\mathcal{E}_{n}),\;n\in\mathbb{Z}_{+}\}$ forms a complete orthonormalized
system in $L^{2}(\mathbb{R}_{+})$.

Note that these results for spectrum and the set of eigenfunctions can be
obtained from the corresponding results of sec. 2 by formal substitution
$|m|\rightarrow0$.

The same results we obtain for the case $\zeta=-\pi/2$.

\paragraph{$|\zeta|<\pi/2$}

Now we consider the case $|\zeta|<\pi/2$.

In this case, we can represent $\sigma_{\zeta}^{\prime}(E)$ in the form%
\begin{align*}
\sigma_{\zeta}^{\prime}(E)  &  =-\frac{1}{\pi\kappa_{0}\cos^{2}\zeta
}\operatorname{Im}\frac{1}{f_{\zeta}(E+i0)},\;f_{\zeta}(W)=f(W)+\tan\zeta,\\
f(W)  &  =\omega_{0}(W)/2,\;f^{\prime}(E)=4\sqrt{\lambda}\partial_{\alpha}%
\psi(\alpha)/8\sqrt{\lambda}>0,\;\alpha\neq-n,\;n\in\mathbb{Z}_{+}.
\end{align*}
The function $f(E)$ has the properties: $f(E)\rightarrow-\infty$ as
$E\rightarrow\mathcal{E}_{-1}\equiv-\infty$; $f(\mathcal{E}_{n}\pm0)=\mp
\infty$, $n\in\mathbb{Z}_{+}$; in any interval $(\mathcal{E}_{n-1}%
,\mathcal{E}_{n})$, $n\in\mathbb{Z}_{+}$, $f(E)$ increases monotonically from
$-\infty$ to $\infty$ as $E$ run from $\mathcal{E}_{n-1}$ to $\mathcal{E}_{n}$.

The function $f_{\zeta}^{-1}(E)$ is real in the points where $f_{\zeta}%
(E)\neq0$ therefore $\sigma_{\zeta}^{\prime}(E)$ can be not equal to zero only
in the points $E_{n}(\zeta)$ satisfying the equation $f_{\zeta}(E)=0$ or%
\begin{equation}
f(E_{n}(\zeta))=-\tan\zeta,\;\partial_{\zeta}E_{n}(\zeta)=-\frac{1}{f^{\prime
}(E_{n}(\zeta))\cos^{2}\zeta}<0.. \label{Osc2.5.2.7.2.1}%
\end{equation}

The described above properties of function $f(E)$ imply the following
structure of discrete spectrum: in each interval energy $(\mathcal{E}%
_{n-1},\mathcal{E}_{n})$, $n\in\mathbb{Z}_{+}$, for fixed $\zeta\in(-\pi
/2,\pi/2)$, exists only one solution $E_{n}(\zeta)$ of eq.
(\ref{Osc2.5.2.7.2.1}) monotonically increasing from $\mathcal{E}_{n-1}+0$ to
$\mathcal{E}_{n}-0$ as $\zeta$ run from $\pi/2$ to $-\pi/2$. Note the relation%
\[
\lim_{\zeta\rightarrow-\pi/2}E_{n}(\zeta)=\lim_{\zeta\rightarrow\pi/2}%
E_{n+1}(\zeta)=\mathcal{E}_{n},\;n\in\mathbb{Z}_{+}.
\]
Finally we obtain:%
\[
\sigma_{\zeta}^{\prime}(E)=\sum_{n\in\mathbb{Z}_{+}}Q_{n}^{2}\delta
(E-E_{n}(\zeta)),\;Q_{n}=\frac{1}{\cos\zeta}\frac{1}{\kappa_{0}\sqrt
{\kappa_{0}f^{\prime}(E_{n}(\zeta))}}>0,
\]
the spectrum of $\hat{h}_{0\zeta}$ is simple and discrete, $\mathrm{spec}%
\hat{h}_{0\zeta}=\{E_{n}(\zeta),\;n\in\mathbb{Z}_{+}\}$, the set
$\{U_{n}(u)=Q_{n}U_{\zeta}(u;E_{n}(\zeta)\},\;n\in\mathbb{Z}_{+}\}$ of
eigenfunctions of $\hat{h}_{0\zeta}$ forms the complete orthohonalized system
in $L^{2}(\mathbb{R}_{+})$.

\subsection{$\lambda<0$ ($\varkappa=e^{-i\pi/4}|\lambda|^{1/4}$,
$\rho=e^{-i\pi/2}|\lambda|^{1/2}u^{2}$, $w=i\tilde{e}$)}

\subsubsection{Asymptotics,\ $u\rightarrow\infty$, $\operatorname{Im}W>0$ or
$W=0$}

We have%
\begin{align*}
O_{1,0}(u;W)  &  =\frac{\kappa_{0}^{1/2}\varkappa^{-1-2i\tilde{e}}}%
{\Gamma(\alpha)}u^{-1/2-2i\tilde{e}}e^{i|\lambda|^{1/2}u^{2}/2}(1+O(u^{-2}%
))=O(u^{-1/2+\operatorname{Im}W/2\sqrt{|\lambda|}},\\
O_{3,0}(u;W)  &  =\kappa_{0}^{1/2}\varkappa^{-1+2i\tilde{e}}u^{-1/2+2i\tilde
{e}}e^{i|\lambda|^{1/2}u^{2}/2}(1+O(u^{-2}))=O(u^{-1/2-\operatorname{Im}%
W/2\sqrt{|\lambda|}}).
\end{align*}

\subsubsection{Symmetric operator $\hat{h}_{0}$, adjoint operator
$\hat{h}_{0}^{+}=\hat{h}_{0}^{\ast}$}

Symmetric operator $\hat{h}_{0}$ is given by eq. (\ref{Osc2.5.2.2.1}), adjoint
operator $\hat{h}_{0}^{+}=\hat{h}_{0}^{\ast}$ is given by eqs.
(\ref{Osc2.5.2.3.1}) -- (\ref{Osc2.5.2.3.1.2})

\subsubsection{Self-adjoint hamiltonians $\hat{h}_{0\vartheta}$, guiding functional}

There are a family of s.a. hamiltonians $\hat{h}_{0\vartheta}$ given by eqs.
(\ref{Osc2.5.2.4.2}), (\ref{Osc2.5.2.4.1}) (with substitution $\zeta
\rightarrow\vartheta$), a simple guiding functional $\Phi_{\vartheta}(\xi;W)$
is given by eq. (\ref{Osc2.5.2.5.1}) (with substitution $\zeta\rightarrow
\vartheta$). The spectra of the s.a. hamiltonians $\hat{h}_{0\zeta}$ are simple.

\subsubsection{Green function $G_{\vartheta}(u,v;W)$, spectral function
$\sigma_{\vartheta}(E)$}

The Green function $G_{\vartheta}(u,v;W)$ is given by eq. (\ref{Osc2.5.2.6.2})
and the derivative of the spectrum function $\sigma_{\vartheta}^{\prime}(E)$
is given by eq. (\ref{Osc2.5.2.6.3}) (with substitution $\zeta\rightarrow
\vartheta$).

\subsubsection{Spectrum}

In this case, $\sigma_{\vartheta}^{\prime}(E)$ can be represented in the form%
\begin{align*}
\sigma_{\vartheta}^{\prime}(E)  &  =\frac{2}{\kappa_{0}}\frac{B(E)}%
{(A(E)\cos\vartheta+2\sin\vartheta)^{2}+\pi^{2}B^{2}(E)\cos^{2}\vartheta
}\equiv\rho_{\vartheta}^{2}(E),\\
A(E)  &  =\operatorname{Re}\omega_{0}(E),\;B(E)=\pi^{-1}\operatorname{Im}%
\omega_{0}(E)=(1/2)[1+\tanh(\pi\tilde{e})]>0.
\end{align*}
We see that $\sigma_{\vartheta}^{\prime}(E)$ is positive a.c. function on
whole $E$-axis, such that we obtain: spectrum of $\hat{h}_{0\vartheta}$ is
simple and continuous, $\mathrm{spec}\hat{h}_{0\vartheta}=\mathbb{R}$. The set
of generalized eigenfunctions $\{U_{\vartheta E}(u)=\rho_{\vartheta
}(E)U_{\zeta}(u;W),\;E\in\mathbb{R}\}$ forms a complete orthonormalized system
in $L^{2}(\mathbb{R}_{+}).$

\subsection{$\lambda=0$}

For $\lambda=0$, eq. (\ref{Osc2.5.1.1}) is reduced to%
\begin{equation}
(\partial_{u}^{2}+u^{-2}/4+W)\psi(u)=0. \label{Osc2.5.4.1}%
\end{equation}

\subsubsection{Useful solutions}

We use the following solutions of eq. (\ref{Osc2.5.4.1}):%
\begin{align*}
&  O_{1,00}(u;W)=(\kappa_{0}u)^{1/2}J_{0}(Ku),\\
&  O_{2,00}(u;W)=\frac{\pi}{2}(\kappa_{0}u)^{1/2}\left\{  N_{0}(Ku)-\frac{2}%
{\pi}\left[  \ln(K/2\kappa_{0})-\psi(1)\right]  J_{0}(Ku)\right\}  =\\
&  =(\kappa_{0}u)^{1/2}J_{0}(Ku)\ln(\kappa_{0}u)+(\kappa_{0}u)^{1/2}%
Q(z),\;z=Wu^{2},\\
&  Q(z)=\sum_{k=1}^{\infty}\frac{(-1)^{k}}{(k!)^{2}}\left(  \frac{z}%
{4}\right)  ^{k}[\psi(1)-\psi(k+1)],\;Q(z)=O(z),\;z\rightarrow0,\\
&  O_{3,00}(u;W)=-\frac{i\pi}{2}(\kappa_{0}u)^{1/2}H_{0}^{(1)}(Ku)=-\omega
_{00}(W)O_{1,00}(u;W)+O_{2,00}(u;W),\\
&  K=W^{1/2}=\sqrt{|W|}e^{i\varphi_{W}/2},\;\omega_{00}(W)=i\pi/2+\psi
(1)-\ln(K/2\kappa_{0}).
\end{align*}

Note that $O_{1,00}(u;W)$ and $O_{2,00}(u;W)$ are real-entire in $W$.

\paragraph{Asymptotics, $u\rightarrow0$}

We have%
\begin{align*}
O_{1,00}(u;W)  &  =(\kappa_{0}u)^{1/2}(1+O(u^{2})),\;O_{2,00}(u;W)=(\kappa
_{0}u)^{1/2}\ln(\kappa_{0}u)(1+O(u^{2})),\\
O_{3,00}(u;W)  &  =-\omega_{00}(W)(\kappa_{0}u)^{1/2}(1+O(u^{2}))+(\kappa
_{0}u)^{1/2}\ln(\kappa_{0}u)(1+O(u^{2})).
\end{align*}

\paragraph{Asymptotics, $u\rightarrow\infty$, $\operatorname{Im}W>0$
($\operatorname{Im}K>0$) or $W=0$}

We have%
\begin{align*}
O_{1,00}(u;W)  &  =\left(  \frac{\kappa_{0}}{2\pi K}\right)  ^{1/2}%
e^{-i(Ku-\pi/4)}(1+O(u^{-1}))=O(e^{u\operatorname{Im}K}),\\
O_{3,00}(u;W)  &  =-i\left(  \frac{\pi\kappa_{0}}{2K}\right)  ^{1/2}%
e^{i(Ku-\pi/4)}(1+O(u^{-1}))=O(e^{-u\operatorname{Im}K}).
\end{align*}

\paragraph{Wronskian}

We have%
\[
\mathrm{Wr}(O_{1,00},O_{3,00})=\kappa_{0}.
\]

\subsubsection{Symmetric operator $\hat{h}_{00}$}

Symmetric operator $\hat{h}_{00}$ associated with s.a. differential operation
$\check{h}_{00}=-\partial_{u}^{2}-u^{2}/4$ is given by eq. (\ref{Osc2.5.2.2.1}).

\subsubsection{Adjoint operator $\hat{h}_{00}^{+}=\hat{h}_{00}^{\ast}$}

Adjoint operator $\hat{h}_{00}^{+}=\hat{h}_{00}^{\ast}$ is given by is given
by eqs. (\ref{Osc2.5.2.3.1}) -- (\ref{Osc2.5.2.3.1.2}) with substitution
$\check{h}_{00}$, $\hat{h}_{00}$, and $\hat{h}_{00}^{\ast}$ for $\check{h}%
_{0}$, $\hat{h}_{0}$, and $\hat{h}_{0}^{\ast}$.

\subsubsection{Self-adjoint hamiltonians $\hat{h}_{0\vartheta}$, guiding functional}

There are a family of s.a. hamiltonians $\hat{h}_{00\theta}$ given by eqs.
(\ref{Osc2.5.2.4.2}), (\ref{Osc2.5.2.4.1}) (with substitution $\zeta
\rightarrow\theta$, $\hat{h}_{0\zeta}\rightarrow\hat{h}_{00\theta}$). A simple
guiding functional $\Phi_{\theta}(\xi;W)$ is given by%
\begin{align*}
&  \Phi_{\theta}(\xi;W)=\int_{0}^{\infty}U_{\theta}(u;W)\xi(u)du,\;\xi
\in\mathbb{D}_{\theta}=D_{r}(\mathbb{R}_{+})\cap D_{h_{00\theta}}.\\
&  .U_{\theta}(u;W)=O_{1,00}(u;W)\sin\theta+O_{2,00}(u;W)\cos\theta.
\end{align*}
$U_{\theta}(u;W)$ is real-entire solution of eq.(\ref{Osc2.5.4.1}) satisfying
the boundary condition (\ref{Osc2.5.2.4.1}).

The spectrum of $\hat{h}_{00\theta}$ is simple.

\subsubsection{Green function $G_{\theta}(u,v;W)$, spectral function
$\sigma_{\theta}(E)$}

We find the Green function $G_{\theta}(u,v;W)$ as the kernel of the integral
representation
\[
\psi(u)=\int_{0}^{\infty}G_{\theta}(u,v;W)\eta(v)dv,\;\eta\in L^{2}%
(\mathbb{R}_{+}),
\]
of unique solution of an equation%
\begin{equation}
(\hat{h}_{00\theta}-W)\psi(u)=\eta(u),\;\operatorname{Im}W>0,
\label{Osc2.5.4.5.1}%
\end{equation}
for $\psi\in D_{h_{00\theta}}$. General solution of eq. (\ref{Osc2.5.4.5.1})
(under condition $\psi\in L^{2}(\mathbb{R}_{+})$) can be represented in the
form%
\begin{align*}
\psi(u)  &  =aO_{3,00}(u;W)-\frac{1}{\kappa_{0}}O_{1,00}(u;W)\eta
_{3}(W)+\frac{1}{\kappa_{0}}I(u),\;\eta_{3}(W)=\int_{0}^{\infty}%
O_{3,00}(v;W)\eta(v)dv,\\
I(u)  &  =O_{1,00}(u;W)\int_{0}^{u}O_{3,00}(v;W)\eta(v)dv-O_{3,00}%
(u;W)\int_{0}^{u}O_{1,00}(v;W)\eta(v)dv,\\
I(u)  &  =O\left(  u^{3/2}\ln u\right)  ,\;u\rightarrow0.
\end{align*}
A condition $\psi\in D_{h_{00\theta}\text{ , }}$(i.e.$\psi$ satisfies the
boundary condition (\ref{Osc2.5.2.4.1})) gives%
\[
a=-\frac{\cos\theta}{\kappa_{0}\omega_{\theta}(W)}\eta_{3}(W),\;\omega
_{\theta}(W)=\omega_{00}(W)\cos\theta+\sin\theta,
\]%
\begin{align}
&  G_{\theta}(u,v;W)=\Omega_{\theta}(W)U_{\theta}(u;W)U_{\theta}%
(v;W)+\nonumber\\
&  \,+\frac{1}{\kappa_{0}}\left\{
\begin{array}
[c]{c}%
\tilde{U}_{\theta}(u;W)U_{\theta}(v;W),\;u>v\\
U_{\theta}(u;W)\tilde{U}_{\theta}(v;W),\;u<v
\end{array}
\right.  ,\label{Osc2.5.4.5.2}\\
&  \Omega_{\theta}(W)\equiv\frac{\tilde{\omega}_{\theta}(W)}{\kappa_{0}%
\omega_{\theta}(W)},\;\tilde{\omega}_{\theta}(W)=\omega_{00}(W)\sin\theta
-\cos\theta,\nonumber\\
&  \tilde{U}_{\theta}(u;W)=O_{1,00}(u;W)\cos\theta-O_{2,00}(u;W)\sin
\theta,\nonumber
\end{align}
where we used an equality%
\[
O_{3,00}(u;W)=-\tilde{\omega}_{\theta}(W)U_{\theta}(u;W)-\omega_{\theta
}(W)\tilde{U}_{\theta}(u;W).
\]
Note that the function $\tilde{U}_{\theta}(u;W)$ is real-entire in $W$ and the
last term in the r.h.s. of eq. (\ref{Osc2.5.4.5.2}) is real for $W=E$. For
$\sigma_{\theta}^{\prime}(E)$, we find%
\[
\sigma_{\theta}^{\prime}(E)=\frac{1}{\pi}\operatorname{Im}\Omega_{\theta
}(E+i0).
\]

\subsubsection{Spectrum}

\paragraph{$E=p^{2}\geq0$, $K=p\geq0$}

In this case, we have $\omega_{00}(E)=i\pi/2+\psi(1)-\ln(p/2\kappa_{0})$, such
that we find%
\begin{align*}
&  \sigma_{\theta}^{\prime}(E)=\frac{2}{\kappa_{0}}\frac{1}{[g(E)\cos
\theta+2\sin\theta]^{2}+\pi^{2}\cos^{2}\theta}\equiv\rho_{\theta}^{2}(E),\\
&  g(E)=2\psi(1)-\ln(E/4\kappa_{0}^{2}),
\end{align*}
the spectrum of $\hat{h}_{00\theta}$ is simple and continuous for all $E\geq
0$, $\mathrm{spec}\hat{h}_{00\theta}=\mathbb{R}_{+}$.

\paragraph{$E=-\tau^{2}<0$, $K=i\tau=e^{i\pi/2}\tau$}

In this case, we have $\omega_{00}(E)=\psi(1)-\ln(\tau/2\kappa_{0})$.

Let $\theta=\pm\pi/2$. In this case we find%
\[
\sigma_{\theta}^{\prime}(E)=\frac{1}{\pi\kappa_{0}}\operatorname{Im}%
\omega_{00}(E+i0)=\frac{1}{\pi\kappa_{0}}\operatorname{Im}\omega_{00}(E)=0,
\]
such that the spectrum points are absent on the negative $E$-semiaxis.

Let $|\theta|<\pi/2$. In this case, an expression for $\sigma_{\theta}%
^{\prime}(E)$ can be represented in the form%
\[
\sigma_{\theta}^{\prime}(E)=-\frac{1}{\pi\kappa_{0}\cos^{2}\theta
}\operatorname{Im}\frac{1}{f_{\theta}(E+i0)},\;.
\]
Because $f_{\theta}(E)=\psi(1)-\ln(\tau/2\kappa_{0})+\tan\theta$ is real,
$\sigma_{\theta}^{\prime}(E)$ can be different from zero only in the point
$E_{(-)}(\theta)$,%
\[
f_{\theta}(E_{(-)}(\theta))=0,\;E_{(-)}(\theta)=-4\kappa_{0}^{2}%
e^{2(\psi(1)+\tan\theta)},
\]
such that we have%
\[
\sigma_{\theta}^{\prime}(E)=\frac{2|E_{(-)}(\theta)|}{\kappa_{0}\cos^{2}%
\theta}\delta(E-E_{(-)}(\theta)),\;.
\]

Finally we obtain:

for $\theta=\pm\pi/2$, $\mathrm{spec}\hat{h}_{00\pm\pi/2}=\mathbb{R}_{+}$, the
set $\{U_{\pm\pi/2E}(u)=\rho_{\pm\pi/2}(E)U_{\pi/2}(u;E)=\sqrt{u/2}%
J_{0}(pu),$ $E\in\mathbb{R}_{+}\}$ of generalized eigenfunctions of
$\hat{h}_{00\pm\pi/2}$ forms the complete orthohonalized system in
$L^{2}(\mathbb{R}_{+})$;

for $\theta\in(-\pi/2,\pi/2)$, $\mathrm{spec}\hat{h}_{00\pm\pi/2}%
=\mathbb{R}_{+}\cup\{E_{(-)}(\theta)\}$, the set $\{U_{\theta E}%
(u)=\rho_{\theta}(E)U_{\theta}(u;E),$ $E\in\mathbb{R}_{+},\;U_{\theta
(-)}(u)=\sqrt{\frac{2|E_{(-)}(\theta)|}{\kappa_{0}\cos^{2}\theta}}U_{\theta
}(u;E_{(-)}(\theta))\}$ of (generalized) eigenfunctions of $\hat{h}_{00\theta
}$ forms the complete orthohonalized system in $L^{2}(\mathbb{R}_{+})$.

Note that the results of this section can be obtained as limit $\lambda
\rightarrow0$ of the corresponding results of previous subsec.3.

Indeed, we have%
\begin{align*}
&  \lim_{\lambda\rightarrow0}\left.  O_{1,0}(u;W)\right|  _{\lambda\neq
0}=(\kappa_{0}u)^{1/2}\sum_{k=0}^{\infty}\frac{(-W/4\varkappa^{2})^{k}}%
{k!}\frac{\varkappa^{2k}u^{2k}}{k!}=\\
&  \,=(\kappa_{0}u)^{1/2}\sum_{k=0}^{\infty}\frac{(-1)^{k}(Ku/2)^{2k}%
}{(k!)^{2}}=(\kappa_{0}u)^{1/2}J_{0}(Ku)=O_{1,00}(u;W).
\end{align*}%
\begin{align*}
&  \lim_{\lambda\rightarrow0}\left.  O_{2,0}(u;W)\right|  _{\lambda\neq
0}=\partial_{\delta}\left[  (\kappa_{0}u)^{1/2+\delta}\sum_{k=0}^{\infty
}\frac{\Gamma(\alpha_{\delta}+k)}{\Gamma(\alpha_{\delta})}\frac{\Gamma
(\beta_{\delta}}{\Gamma(\beta_{\delta}+k)}\frac{\varkappa^{2k}u^{2k}}%
{k!}\right]  _{\lambda,\delta\rightarrow0}=\\
&  \,=(\kappa_{0}u)^{1/2}J_{0}(Ku)\ln(\kappa_{0}u)+(\kappa_{0}u)^{1/2}%
\sum_{k=1}^{\infty}\frac{(-1)^{k}(Ku/2)^{2k}}{(k!)^{2}}\left[  \psi
(1)-\psi(1+k)\right]  =\\
&  =(\kappa_{0}u)^{1/2}\left[  J_{0}(Ku)\ln(\kappa_{0}u)+Q(z)\right]
=O_{2,00}(u;W)..
\end{align*}
Further, we have for $\sigma_{\vartheta}^{\prime}(E)$

i) $E>0$, $\ln(\kappa_{0}/\varkappa)=i\pi/4+\ln(\kappa_{0}|\lambda|^{-1/4})$,
$\alpha=1/2-iE/4\sqrt{|\lambda|}=1/2+e^{-i\pi/2}E/4\sqrt{|\lambda|}%
\equiv1/2+z$, $\psi(\alpha)=\ln\alpha-1/2\alpha+O(\alpha^{-2})=\ln
z+O(z^{-2})=-i\pi/2+\ln(E/4)-(1/2)\ln|\lambda|+O(|\lambda|)$%
\begin{align*}
&  \omega_{0}(E)=i\pi-\ln(E/4\kappa_{0}^{2})+2\psi(1)+O(|\lambda
|)\;\Longrightarrow\\
&  A(E)=2\psi(1)-\ln(E/4\kappa_{0}^{2})+O(|\lambda|)=g(E)+O(|\lambda|),\\
&  B(E)=1+O(|\lambda|),
\end{align*}%
\begin{align*}
&  \lim_{\lambda\rightarrow-0}\left.  \sigma_{\vartheta}^{\prime}(E)\right|
_{\lambda<0}=\frac{2}{\kappa_{0}}\frac{1}{(g(E)\cos\vartheta+2\sin
\vartheta)^{2}+\pi^{2}\cos^{2}\vartheta}=\\
&  \,=\left.  \sigma_{\theta}^{\prime}(E)\right|  _{\theta=\vartheta}.
\end{align*}

ii) $E=-\tau^{2}<0$, $\ln(\kappa_{0}/\varkappa)=i\pi/4+\ln(\kappa_{0}%
|\lambda|^{-1/4})$, $\alpha=1/2+i|E|/4\sqrt{|\lambda|}=1/2+e^{i\pi/2}%
E/4\sqrt{|\lambda|}\equiv1/2+z$,

$\psi(\alpha)=\ln\alpha-1/2\alpha+O(\alpha^{-2})=\ln z+O(z^{-2})=i\pi
/2+\ln(|E|/4)-(1/2)\ln|\lambda|+O(|\lambda|)$, $\operatorname{Im}\psi
(\alpha)=\pi/2-2\varepsilon$, $\varepsilon=(\pi/2)\exp(-\pi|E|/2\sqrt
{|\lambda|})$, $|\lambda|=\frac{\pi^{2}|E|^{2}}{2\ln^{2}(\pi/2\varepsilon)}$%
\begin{align*}
&  \omega_{0}(E)=2\psi(1)-\ln(E/4\kappa_{0}^{2})+O(|\lambda|)+2i\varepsilon,\\
&  A(E)=2\psi(1)-\ln(E/4\kappa_{0}^{2})+O(|\lambda|),\;B(E)=2\varepsilon/\pi,
\end{align*}%
\[
\sigma_{\vartheta}^{\prime}(E)=\frac{1}{\pi\kappa_{0}}\frac{\varepsilon
}{([\psi(1)-\ln(\tau/2\kappa_{0})+O(|\lambda|]\cos\vartheta+\sin\vartheta
)^{2}+\varepsilon^{2}\cos^{2}\vartheta}.
\]

ii$_{a}$) $\vartheta=\pm\pi/2$

We have%
\[
\sigma_{\vartheta=\pm\pi/2}^{\prime}(E)=\frac{\varepsilon}{\pi\kappa_{0}%
}\rightarrow0=\sigma_{\theta=\pm\pi/2}^{\prime}(E).
\]

ii$_{b}$) $|\lambda|<\pi/2$

Represent $\sigma_{\vartheta}^{\prime}(E)$ in the form%
\[
\sigma_{\vartheta}^{\prime}(E)=\frac{1}{\pi\kappa_{0}\cos^{2}\vartheta
}\frac{\varepsilon}{[f_{\vartheta}(E)+O(|\lambda|]^{2}+\varepsilon^{2}}.
\]

If $E\neq E_{(-)}(\vartheta)$, then $\lim_{\lambda\rightarrow-0}\left.
\sigma_{\vartheta}^{\prime}(E)\right|  _{\lambda<0}=0$. Represent
$\sigma_{\vartheta}^{\prime}(E)$ in the form%
\[
\sigma_{\vartheta}^{\prime}(E)=\frac{1}{\pi\kappa_{0}\cos^{2}\vartheta
}\frac{\varepsilon}{[f_{\vartheta}^{\prime}(E_{(-)}(\vartheta))\Delta
+\Delta^{2}b(\Delta)+O(|\lambda|)]^{2}+\varepsilon^{2}},
\]
where $\Delta=E-E_{(-)}(\vartheta)$, $b(\Delta)=\Delta^{-2}[f_{\vartheta
}(E)-f_{\vartheta}^{\prime}(E_{(-)}(\vartheta))\Delta]$. It is easy to obtain
\[
\lim_{\lambda\rightarrow-0}\left.  \sigma_{\vartheta}^{\prime}(E)\right|
_{\lambda<0}=\frac{1}{\kappa_{0}f_{\vartheta}^{\prime}(E_{(-)}(\vartheta
))\cos^{2}\vartheta}\delta(\Delta)=.\frac{2|E_{(-)}(\vartheta)|}{\kappa
_{0}\cos^{2}\vartheta}\delta(\Delta)=\left.  \sigma_{\theta}^{\prime
}(E)\right|  _{\theta=\vartheta}.
\]%
\[
\lim_{\lambda\rightarrow-0}\left.  \sigma_{m}^{\prime}(E)\right|  _{\lambda
<0}=0.
\]

iii) $E=0$

iii$_{a}$) $\vartheta=\pm\pi/2$
\begin{align*}
&  \left.  \sigma_{\vartheta=\pm\pi/2}^{\prime}(+0)\right|  _{\lambda
=-0}=\left.  \left.  \sigma_{\vartheta=\pm\pi/2}^{\prime}(E)\right|
_{\lambda=-0}\right|  _{E\rightarrow+0}=\frac{1}{2\kappa_{0}}=\sigma
_{\theta=\pm\pi/2}^{\prime}(+0),\\
&  \left.  \sigma_{\vartheta=\pm\pi/2}^{\prime}(0)\right|  _{\lambda
=-0}=\frac{1}{4\kappa_{0}},\\
&  \left.  \sigma_{\vartheta=\pm\pi/2}^{\prime}(-0)\right|  _{\lambda
=-0}=\left.  \left.  \sigma_{\vartheta=\pm\pi/2}^{\prime}(E)\right|
_{\lambda=-0}\right|  _{E\rightarrow-0}=0=\sigma_{\theta=\pm\pi/2}^{\prime
}(-0).
\end{align*}
iii$_{b}$) $|\vartheta|<\pi/2$
\begin{align*}
&  \left.  \sigma_{\vartheta}^{\prime}(+0)\right|  _{\lambda=-0}=\left.
\left.  \sigma_{\vartheta}^{\prime}(E)\right|  _{\lambda=-0}\right|
_{E\rightarrow+0}=0=\left.  \sigma_{\theta}^{\prime}(+0)\right|
_{\theta=\vartheta},\\
&  \left.  \sigma_{\vartheta}^{\prime}(0)\right|  _{\lambda=-0}=0,\\
&  \left.  \sigma_{\vartheta}^{\prime}(-0)\right|  _{\lambda=-0}=\left.
\left.  \sigma_{\vartheta}^{\prime}(E)\right|  _{\lambda=-0}\right|
_{E\rightarrow-0}=0=\left.  \sigma_{\theta}^{\prime}(-0)\right|
_{\theta=\vartheta}.
\end{align*}

\section{Quantum two-dimensional Coulomb-like interaction on a plane}

\subsection{Preliminaries}

Let we have%
\[
\Psi(x,\varphi)=x^{-1/2}\sum_{m\in\mathbb{Z}}e^{im\varphi_{x}/2}\psi
_{m}(x),\;x\in\mathbb{R}_{+},\;0\leq\varphi_{x}\leq4\pi.
\]
The Schroedinger equation%
\[
(\check{H}-\mathcal{E})\Psi(x,\varphi)=0,\;\check{H}=-\partial_{x}^{2}%
-x^{-1}\partial_{x}-x^{-2}\partial_{\varphi_{x}}^{2}+gx^{-1},
\]
is reduced to the radial equation%
\begin{equation}
(\check{h}_{m}-\mathcal{E})\psi_{m}(x)=0,\;\check{h}_{m}=-\partial_{x}%
^{2}+(2x)^{-2}(m^{2}-1)+gx^{-1}, \label{Coul2.1.0}%
\end{equation}
or%

\begin{equation}
\partial_{x}^{2}\psi_{m}(x)+(-\frac{m^{2}-1}{4x^{2}}-\frac{g}{x}%
+\mathcal{E})\psi_{m}(x)=0, \label{Coul2.1.1}%
\end{equation}
where $\hbar^{2}\mathcal{E}/2\mu$ is complex energy, $\hbar^{2}g/2\mu$ is the
Coulomb coupling constant, $\check{h}_{m}$ are radial differential
operations,
\[
\mathcal{E}=|\mathcal{E}|e^{i\varphi_{\mathcal{E}}},\;0\leq\varphi
_{\mathcal{E}}\leq\pi,\;\operatorname{Im}\mathcal{E}\geq0.
\]

It is convenient for our aims first to consider solutions more general
equation%
\begin{equation}
\partial_{x}^{2}\psi_{m}(x)+(-\frac{(m+\delta)^{2}-1}{4x^{2}}-\frac{g}%
{x}+\mathcal{E})\psi_{m}(x)=0,\;|\delta|<1. \label{Coul2.1.1a}%
\end{equation}

Introduce a new variable $$z=2Kx, K=\sqrt{-\mathcal{E}}=\sqrt{|\mathcal{E}%
|}e^{i(\varphi_{\mathcal{E}}-\pi)/2}=\sqrt{|\mathcal{E}|}\left[  \sin
(\varphi_{\mathcal{E}}/2)-i\cos(\varphi_{\mathcal{E}}/2)\right],$$
$\partial_{x}=2K\partial_{z}$, $\partial_{x}^{2}=4K^{2}\partial_{z}^{2}$, and
new function $\phi(z)$, $\psi(x)=z^{1/2+|m+\delta|/2}e^{-z/2}\phi(z)$. Then we obtain

\begin{align}
&  z\partial_{z}^{2}\phi(z)+(\beta_{\delta}-z)\partial_{z}\phi(z)-\alpha
_{\delta}\phi(z)=0.\label{Coul2.1.2}\\
&  \alpha_{\delta}=1/2+|m+\delta|/2-w,\;w=-g/2K,\;\beta_{\delta}%
=1+|m+\delta|.\nonumber
\end{align}

Eq. (\ref{Coul2.1.2}) is the equation for confluent hypergeometric functions,
in the terms of which we can express solutions of eq. (\ref{Coul2.1.1a}). We
will use the following solutions:%
\begin{align}
&  C_{1,m,\delta}(x;\mathcal{E})=(\kappa_{0}x)^{1/2+|m+\delta|/2}e^{-z/2}%
\Phi(\alpha_{\delta},\beta_{\delta};z),\nonumber\\
&  C_{2,m,\delta}(x;\mathcal{E})=\Gamma^{-1}(\beta_{-,\delta})(\kappa
_{0}x)^{1/2-|m+\delta|/2}e^{-z/2}\Phi(\alpha_{-,\delta},\beta_{-,\delta
};z),\nonumber\\
&  C_{3,m,\delta}(x;\mathcal{E})=(\kappa_{0}x)^{1/2+|m+\delta|/2}e^{-z/2}%
\Psi(\alpha_{\delta},\beta_{\delta};z)=\nonumber\\
&  \,=\frac{\Gamma(-|m+\delta|)}{\Gamma(\alpha_{-,\delta})}C_{1,m,\delta
}(x;\mathcal{E})+\frac{(\kappa_{0}/2K)^{|m+\delta|}\Gamma(|m+\delta
|)\Gamma(\beta_{-,\delta})}{\Gamma(\alpha_{\delta})}C_{2,m,\delta
}(x;\mathcal{E}),\label{Coul2.1.3}\\
&  \alpha_{-,\delta}=1/2-|m+\delta|/2-w,\;\beta_{-,\delta}=1-|m+\delta
|,\nonumber
\end{align}
where $\kappa_{0}$ is parameter of dimensionality of inverse length introduced
in the file $<$Osc2-...tex$>$.\ Using relation GradRy.9.212.1, we find%
\begin{align*}
&  \left.  C_{1,m,\delta}(x;\mathcal{E})\right|  _{K\rightarrow-K}=(\kappa
_{0}x)^{1/2+|m+\delta|/2}e^{z/2}\Phi(1/2+|m+\delta|/2+w,\beta_{\delta};-z)=\\
&  \,=(\kappa_{0}x)^{1/2+|m+\delta|/2}e^{-z/2}\Phi(\alpha_{\delta}%
,\beta_{\delta};z)=C_{1,m,\delta}(x;\mathcal{E}).
\end{align*}
Similarly, we obtain $\left.  C_{2,m,\delta}(x;\mathcal{E})\right|
_{K\rightarrow-K}=C_{2,m,\delta}(x;\mathcal{E})$. We thus obtain that the
functions $C_{1,m,\delta}(x;\mathcal{E})$ and $C_{2,m,\delta}(x;\mathcal{E})$
are real-entire in $\mathcal{E}$.

\subsection{$|m|\geq2$}

\subsection{Useful solutions}

Represent $C_{3,m,\delta}$ in the form%
\begin{align*}
&  C_{3,m,\delta}(x;\mathcal{E})=B_{m,\delta}(\mathcal{E})C_{1,m,\delta
}(x;\mathcal{E})+\frac{(\kappa_{0}/2K)^{|m+\delta|}\Gamma(|m+\delta|)}%
{\Gamma(\alpha_{\delta})}C_{4,m,\delta}(x;\mathcal{E}),\\
&  C_{4,m,\delta}(x;\mathcal{E})=\Gamma(\beta_{-,\delta})\left[
C_{2,m,\delta}(x;\mathcal{E})-A_{m,\delta}(\mathcal{E})C_{1,m,\delta
}(x;\mathcal{E})\right]  ,\\
&  A_{m,\delta}(\mathcal{E})=\frac{(\kappa_{0}/2K)^{-|m|}\Gamma(\alpha
)}{\Gamma(\beta_{\delta})\Gamma(\alpha_{-})},\;\frac{\Gamma(\alpha)}%
{\Gamma(\alpha_{-})}=(-1)^{|m|}(1-\alpha)_{|m|},\;(1+x)_{|m|}=(1+x)\cdots
(|m|+x),\\
&  B_{m,\delta}(\mathcal{E})=\frac{\Gamma(-|m+\delta|)}{\Gamma(\alpha
_{-,\delta})}+\frac{(\kappa_{0}/2K)^{|m+\delta|}\Gamma(|m+\delta|)\Gamma
(\beta_{-,\delta})}{\Gamma(\alpha_{\delta})}A_{m,\delta}(\mathcal{E})=\\
&  \,=\frac{\Gamma(-|m+\delta|)}{\Gamma(\alpha_{-})}\left[  \frac
{\Gamma(\alpha_{-})}{\Gamma(\alpha_{-,\delta})}-\frac{(\kappa_{0}%
/2K)^{\delta_{m}}\Gamma(\alpha)}{\Gamma(\alpha_{\delta})}\right]  ,\\
&  \alpha=\alpha_{0}=1/2+|m|/2-w,\;\alpha_{-}=\alpha_{-,0}=1/2-|m|/2-w,\;\beta
=\beta_{0}=1+|m|,
\end{align*}
Consider $A_{m,\delta}(\mathcal{E})$ in more details.

Let $|m|=2k-1$, $k\in\mathbb{N}$. We have%
\begin{align*}
&  \,(\kappa_{0}/2K)^{-|m|}=(-4\mathcal{E}/\kappa_{0}^{2})^{k}(\kappa
_{0}/2K),\;(1-\alpha)_{|m|}=-\frac{g}{2K}(-4\mathcal{E})^{1-k}\prod
_{l=1}^{k-1}(g^{2}+4\mathcal{E}l^{2}),\\
&  \,(\kappa_{0}/2K)^{-|m|}(1-\alpha)_{|m|}=-\kappa_{0}^{-|m|}g\prod
_{l=1}^{k-1}(g^{2}+4\mathcal{E}l^{2}).
\end{align*}
Let $|m|=2k$, $k\in\mathbb{N}$. We have%
\begin{align*}
&  \,(\kappa_{0}/2K)^{-|m|}=(-4\mathcal{E}/\kappa_{0}^{2})^{k},\;(1-\alpha
)_{|m|}=(-4\mathcal{E})^{-k}\prod_{l=0}^{k-1}[g^{2}+4\mathcal{E}%
(l+1/2)^{2}],\\
&  \,(\kappa_{0}/2K)^{-|m|}(1-\alpha)_{|m|}=\kappa_{0}^{-|m|}\prod_{l=0}%
^{k-1}[g^{2}+4\mathcal{E}(l+1/2)^{2}],
\end{align*}
such that we obtain that $A_{m,\delta}(\mathcal{E})$ is real-entire polynomial
function of $\mathcal{E}$ and $C_{4,m,\delta}(x;\mathcal{E})$ is real-entire
in $\mathcal{E}$.

We obtain the solutions of eq.(\ref{Coul2.1.1}) as the limit $\delta
\rightarrow0$ of the solutions of eq. (\ref{Coul2.1.3}):%
\begin{align*}
&  C_{1,m}(x;\mathcal{E})=C_{1,m,0}(x;\mathcal{E})=(\kappa_{0}x)^{1/2+|m|/2}%
e^{-z/2}\Phi(\alpha,\beta;z),\\
&  C_{4,m}(x;\mathcal{E})=\lim_{\delta\rightarrow0}C_{4,m,\delta
}(x;\mathcal{E}),\\
&  C_{3,m}(x;\mathcal{E})=(\kappa_{0}x)^{1/2+|m|/2}e^{-z/2}\Psi(\alpha
,\beta;z)=\\
&  =B_{m}(\mathcal{E})C_{1,m}(x;\mathcal{E})+C_{m}(\mathcal{E})C_{4,m}%
(x;\mathcal{E}),\;C_{m}(\mathcal{E})=\frac{(\kappa_{0}/2K)^{|m|}\Gamma
(|m|)}{\Gamma(\alpha)},\\
&  B_{m}(\mathcal{E})=B_{m,0}(\mathcal{E})=\frac{(-1)^{|m|+1}}{2\Gamma
(\beta)\Gamma(\alpha_{-})}\left[  \psi(\alpha_{-})+\psi(\alpha)+2\ln
(2K/\kappa_{0})\right]  .
\end{align*}
where we used relations.%
\begin{align*}
&  |m+\delta|=|m|+\delta_{m},\;\delta_{m}=\delta\mathrm{sign}m,\;\Gamma
(-|m+\delta|)=\frac{(-1)^{|m|+1}}{\delta_{m}\Gamma(\beta)},\\
&  \frac{\Gamma(\alpha_{-})}{\Gamma(\alpha_{-,\delta})}=1+\delta_{m}%
\psi(\alpha_{-})/2,\;\frac{\Gamma(\alpha)}{\Gamma(\alpha_{\delta})}%
=1-\delta_{m}\psi(\alpha)/2,\;(\kappa_{0}/2K)^{\delta_{m}}=1+\delta_{m}%
\ln(\kappa_{0}/2K).
\end{align*}

\subsubsection{Asymptotics, $x\rightarrow0$}

We have%
\begin{align*}
&  C_{1,m}(x;\mathcal{E})=C_{1,m\mathrm{as}}(x)(1+O(x)),\\
&  C_{4,m}(x;\mathcal{E})=C_{4,m\mathrm{as}}(x)\left(  1+O(x)\right)  ,\\
&  C_{3,m}(x;\mathcal{E})=C_{m}(\mathcal{E})C_{4,m\mathrm{as}}(x)\left(
1+O(x)\right)  ,\;\operatorname{Im}\mathcal{E}>0,\\
&  C_{1,m\mathrm{as}}(x)=(\kappa_{0}x)^{1/2+|m|/2},\;C_{4,m\mathrm{as}%
}(x)=(\kappa_{0}x)^{1/2-|m|/2}.
\end{align*}

\subsubsection{Asymptotics, $x\rightarrow\infty$, $\operatorname{Im}%
\mathcal{E}>0$ ($\operatorname{Re}K>0$)}%

\begin{align*}
C_{1,m}(x;\mathcal{E})  &  =\frac{\kappa_{0}^{1/2+|m|/2}(2K)^{\alpha-\beta
}\Gamma(\beta)}{\Gamma(\alpha)}x^{-w}e^{Kx}(1+O(x^{-1}))=O(x^{-w}%
e^{x\operatorname{Re}K}),\\
C_{3,m}(x;\mathcal{E})  &  =\kappa_{0}^{1/2+|m|/2}(2K)^{-\alpha}x^{w}%
e^{-Kx}(1+O(x^{-1}))=O(x^{w}e^{-x\operatorname{Re}K})
\end{align*}

\subsubsection{Wronskians}%

\begin{align*}
&  \mathrm{Wr}(C_{1,m},C_{4,m})=-\kappa_{0}|m|,\\
&  \mathrm{Wr}(C_{1,m},C_{3,m})=-\kappa_{0}|m|C_{m}(\mathcal{E})=-\frac
{\kappa_{0}(\kappa_{0}/2K)^{|m|}\Gamma(\beta)}{\Gamma(\alpha)}=-\omega
_{m}(\mathcal{E})
\end{align*}

\subsection{Symmetric operator $\hat{h}_{m}$}

For given a differential operation $\check{h}_{m}$ (\ref{Coul2.1.0}), we
determine the following symmetric operator $\hat{h}_{m}$,%

\[
\hat{h}_{m}:\left\{
\begin{array}
[c]{l}%
D_{h_{m}}=\mathcal{D}(\mathbb{R}_{+}),\\
\hat{h}_{m}\psi(x)=\check{h}_{m}\psi(x),\;\forall\psi\in D_{h_{m}}%
\end{array}
\right.  .
\]

\subsection{Adjoint operator $\hat{h}_{m}^{+}=\hat{h}_{m}^{\ast}$}%

\[
\hat{h}_{m}^{+}:\left\{
\begin{array}
[c]{l}%
D_{h_{m}^{+}}=D_{\check{h}_{m}}^{\ast}(\mathbb{R}_{+})=\{\psi_{\ast}%
,\psi_{\ast}^{\prime}\;\mathrm{are\;a.c.\;in}\mathcal{\;}\mathbb{R}_{+}%
,\;\psi_{\ast},\hat{h}_{m}^{+}\psi_{\ast}\in L^{2}(\mathbb{R}_{+})\}\\
\hat{h}_{m}^{+}\psi_{\ast}(x)=\check{h}_{m}\psi_{\ast}(x),\;\forall\psi_{\ast
}\in D_{h_{m}^{+}}%
\end{array}
\right.  .
\]

\subsubsection{Asymptotics}

I) $x\rightarrow\infty$

Because $V(x)\rightarrow0$ for large $x$, we have: $[\psi_{\ast},\chi_{\ast
}](x)\rightarrow0$ as $x\rightarrow\infty$, $\forall\psi_{\ast},\chi_{\ast}\in
D_{h_{m}^{+}}$.

II) $x\rightarrow0$

Because $\check{h}_{m}\psi_{\ast}\in L^{2}(\mathbb{R})$, we have%
\[
\check{h}_{m}\psi_{\ast}(x)=[-\partial_{x}^{2}+(2x)^{-2}(m^{2}-1)+gx^{-1}%
]\psi_{\ast}(x)=\eta(x),\;\eta\in L^{2}(\mathbb{R}_{+}).
\]
or%
\[
(\check{h}_{m}-\mathcal{E}_{0})\psi_{\ast}(x)=\tilde{\eta}(x),\;\tilde{\eta
}(x)=\eta(x)-\mathcal{E}_{0}\psi_{\ast}(x),\;\tilde{\eta}\in L^{2}%
(\mathbb{R}_{+}),
\]
where $\mathcal{E}_{0}$ is an arbitrary (but fixed) number with
$\operatorname{Im}\mathcal{E}_{0}>0$. General solution of this equation can be
represented in the form%
\begin{align*}
&  \psi_{\ast}(x)=a_{1}C_{1,m}(x;\mathcal{E}_{0})+a_{2}C_{4,m}(x;\mathcal{E}%
_{0})+I(x),\\
&  \psi_{\ast}^{\prime}(x)=a_{1}C_{1,m}^{\prime}(x;\mathcal{E}_{0}%
)+a_{2}C_{4,m}^{\prime}(x;\mathcal{E}_{0})+I^{\prime}(x),
\end{align*}
where%
\begin{align*}
&  I(x)=\frac{C_{4,m}(x;\mathcal{E}_{0})}{\kappa_{0}|m|}\int_{0}^{x}%
C_{1,m}(y;\mathcal{E}_{0})\tilde{\eta}(y)dy+\frac{C_{1,m}(x;\mathcal{E}_{0}%
)}{\kappa_{0}|m|}\int_{x}^{x_{0}}C_{4,m}(y;\mathcal{E}_{0})\tilde{\eta
}(y)dy,\\
&  I^{\prime}(x)=\frac{C_{4,m}^{\prime}(x;\mathcal{E}_{0})}{\kappa_{0}|m|}%
\int_{0}^{x}C_{1,m}(y;\mathcal{E}_{0})\tilde{\eta}(y)dy+\frac{C_{1,m}^{\prime
}(x;\mathcal{E}_{0})}{\kappa_{0}|m|}\int_{x}^{x_{0}}C_{4,m}(y;\mathcal{E}%
_{0})\tilde{\eta}(y)dy.
\end{align*}
We obtain with the help of the CB-inequality: $I(x)$ is bounded at infinity
and%
\[
I(x)=\left\{
\begin{array}
[c]{l}%
O(x^{3/2}),\;|m|\geq3\\
O(x^{3/2}\sqrt{\ln x}),\;|m|=2
\end{array}
\right.  ,\;I^{\prime}(x)=\left\{
\begin{array}
[c]{l}%
O(x^{1/2}),\;|m|\geq3\\
O(x^{1/2}\sqrt{\ln x}),\;|m|=2
\end{array}
\right.  ,\;x\rightarrow0.
\]
The condition $\psi_{\ast}(x)\in L^{2}(\mathbb{R}_{+})$ gives $a_{1}=a_{2}=0$
for all $|m|\geq2$ such that we find%
\begin{equation}
\psi_{\ast}(x)=\left\{
\begin{array}
[c]{l}%
O(x^{3/2}),\;|m|\geq3\\
O(x^{3/2}\sqrt{\ln x}),\;|m|=2
\end{array}
\right.  ,\;\psi_{\ast}^{\prime}(x)=\left\{
\begin{array}
[c]{l}%
O(x^{1/2}),\;|m|\geq3\\
O(x^{1/2}\sqrt{\ln x}),\;|m|=2
\end{array}
\right.  ,\;x\rightarrow0, \label{Coul2.2.3.1.1}%
\end{equation}
and $\omega_{h_{m}^{+}}(\chi_{\ast},\psi_{\ast})=\Delta_{h_{m}^{+}}(\psi
_{\ast})=0$.

\subsection{Self-adjoint hamiltonian $\hat{h}_{m\mathfrak{e}}$}

Because $\omega_{h_{m}^{+}}(\chi_{\ast},\psi_{\ast})=\Delta_{h_{m}^{+}}%
(\psi_{\ast})=0$ (and also because $C_{1,m}(x;\mathcal{E})$ and $C_{3,m}%
(x;\mathcal{E})$ and their linear combinations are not s.-integrable for
$\operatorname{Im}\mathcal{E}\neq0$), the deficiency indices of initial
symmetric operator $\hat{h}_{m}$ are zero, which means that $\hat
{h}_{m\mathfrak{e}}=\hat{h}_{m}^{+}$ is a unique s.a. extension of the initial
symmetric operator $\hat{h}_{m}$:%
\begin{equation}
\hat{h}_{m\mathfrak{e}}:\left\{
\begin{array}
[c]{l}%
D_{h_{m\mathfrak{e}}}=D_{\check{h}_{m}}^{\ast}(\mathbb{R}_{+})\\
\hat{h}_{m\mathfrak{e}}\psi(x)=\check{h}_{m}\psi(x),\;\forall\psi\in
D_{h_{m\mathfrak{e}}}%
\end{array}
\right.  . \label{Coul2.2.4.1}%
\end{equation}

\subsection{The guiding functional $\Phi(\xi;\mathcal{E})$}

As a guiding functional $\Phi(\xi;\mathcal{E})$ we choose%
\begin{align*}
&  \Phi(\xi;\mathcal{E})=\int_{0}^{\infty}C_{1,m}(x;\mathcal{E})\xi
(x)dx,\;\xi\in\mathbb{D}=D_{r}(\mathbb{R}_{+})\cap D_{h_{m\mathfrak{e}}}.\\
D_{r}(a,b)=\{\psi(x  &  ):\;\mathrm{supp}\psi\subseteq\lbrack a,\beta_{\psi
}],\;\beta_{\psi}<b\}.
\end{align*}

The guiding functional $\Phi(\xi;\mathcal{E})$ is simple and the spectrum of
$\hat{h}_{m\mathfrak{e}}$ is simple.

\subsection{Green function $G_{m}(x,y;\mathcal{E})$, spectral function
$\sigma_{m}(E)$}

We find the Green function $G_{m}(x,y;\mathcal{E})$ as the kernel of the
integral representation
\[
\psi(x)=\int_{0}^{\infty}G_{m}(x,y;\mathcal{E})\eta(y)dy,\;\eta\in
L^{2}(\mathbb{R}_{+}),
\]
of unique solution of an equation%
\begin{equation}
(\hat{h}_{m\mathfrak{e}}-\mathcal{E})\psi(x)=\eta(x),\;\operatorname{Im}%
\mathcal{E}>0, \label{Coul2.2.6.1}%
\end{equation}
for $\psi\in D_{h_{m\mathfrak{e}}}$. General solution of eq.
(\ref{Coul2.2.6.1}) can be represented in the form%
\begin{align*}
\psi(x)  &  =a_{1}C_{1,m}(x;\mathcal{E})+a_{3}C_{3,m}(x;\mathcal{E})+I(x),\\
I(x)  &  =\frac{C_{1,m}(x;\mathcal{E})}{\omega_{m}(\mathcal{E})}\int
_{x}^{\infty}C_{3,m}(y;\mathcal{E})\eta(y)dy+\frac{C_{3,m}(x;\mathcal{E}%
)}{\omega_{m}(\mathcal{E})}\int_{0}^{x}C_{1,m}(y;\mathcal{E})\eta(y)dy,\\
I(x)  &  =\left\{
\begin{array}
[c]{l}%
O\left(  x^{3/2}\right)  ,\;|m|\geq3\\
O\left(  x^{3/2}\sqrt{\ln x}\right)  ,\;|m|=2
\end{array}
\right.  ,\;x\rightarrow0.
\end{align*}
A condition $\psi\in L^{2}(\mathbb{R}_{+})$ gives $a_{1}=a_{3}=0$, such that
we find%
\begin{align}
&  G_{m}(x,y;\mathcal{E})=\frac{1}{\omega_{m}(\mathcal{E})}\left\{
\begin{array}
[c]{l}%
C_{3,m}(x;\mathcal{E})C_{1,m}(y;\mathcal{E}),\;x>y\\
C_{1,m}(x;\mathcal{E})C_{3,m}(y;\mathcal{E}).\;x<y
\end{array}
\right.  =\nonumber\\
&  \,=\Omega_{m}(\mathcal{E})C_{1,m}(x;\mathcal{E})C_{1,m}(y;\mathcal{E}%
)+\frac{1}{\kappa_{0}|m|}\left\{
\begin{array}
[c]{c}%
C_{4,m}(x;\mathcal{E})C_{1,m}(y;\mathcal{E}),\;x>y\\
C_{1,m}(x;\mathcal{E})C_{4,m}(y;\mathcal{E}),\;x<y
\end{array}
\right.  ,\label{Coul2.2.6.2}\\
&  \Omega_{m}(\mathcal{E})\equiv\frac{B_{m}(\mathcal{E})}{\omega
_{m}(\mathcal{E})}=D_{m}(\mathcal{E})[2\ln(\kappa_{0}/2K)-\psi(\alpha
)-\psi(\alpha_{-})],\nonumber\\
&  D_{m}(\mathcal{E})=\frac{(2K/\kappa_{0})^{|m|}(1-\alpha)_{|m|}}{2\kappa
_{0}\Gamma^{2}(\beta)}=\nonumber\\
&  =\frac{1}{2\kappa_{0}^{|m|+1}\Gamma^{2}(\beta)}\times\left\{
\begin{array}
[c]{l}%
-g\prod_{l=1}^{k-1}((g^{2}+4\mathcal{E}l^{2}),\;|m|=2k-1,\\
\prod_{l=0}^{k-1}[g^{2}+4\mathcal{E}(l+1/2)^{2}],\;|m|=2k
\end{array}
\right.  ,\;k\in\mathbb{N},\nonumber\\
&  \operatorname{Im}D_{m}(E)=0,\;\forall E\in\mathbb{R}.\nonumber
\end{align}
Note that the last term in the r.h.s. of eq. (\ref{Coul2.2.6.2}) is real for
$\mathcal{E}=E$. From the relation%
\[
C_{1,m}^{2}(u_{0};E)\sigma_{m}^{\prime}(E)=\frac{1}{\pi}\operatorname{Im}%
G_{m}(u_{0}-0,u_{0}+0;E+i0),
\]
we find%
\begin{equation}
\sigma_{m}^{\prime}(E)=\frac{1}{\pi}\operatorname{Im}\Omega_{m}(E+i0).
\label{Coul2.2.6.3}%
\end{equation}

\subsection{Spectrum}

\subsubsection{$E=p^{2}\geq0$, $p\geq0$, $K=-ip=e^{-i\pi/2}p$}

We have%
\begin{align*}
&  \alpha=1/2+|m|/2-i\tilde{w},\;\alpha_{-}=1/2-|m|/2-i\tilde{w},\;\tilde
{w}=-g/2p,\\
&  \operatorname{Im}\Omega_{m}(E)=D_{m}(E)[\pi-\operatorname{Im}(\psi
(\alpha)+\psi(\alpha_{-}))]=\\
&  \,=\pi D_{m}(E)\times\left\{
\begin{array}
[c]{l}%
\coth(\pi\tilde{w})+1,\;|m|=2n-1\\
1+\tanh(\pi\tilde{w}),\;|m|=2n
\end{array}
\right.  ,\;n\in\mathbb{N.}%
\end{align*}
We thus find%
\begin{align*}
&  \sigma_{m}^{\prime}(E)=\frac{1}{2\kappa_{0}^{|m|+1}\Gamma^{2}(\beta)}%
\times\left\{
\begin{array}
[c]{l}%
g[\coth(\pi g/2p)-1]\prod_{l=1}^{k-1}((g^{2}+4El^{2}),\;|m|=2k-1\\
\lbrack1-\tanh(\pi g/2p)]\prod_{l=0}^{k-1}[g^{2}+4E(l+1/2)^{2}],\;|m|=2k
\end{array}
\right.  =\\
&  =\kappa_{0}^{-1-|m|}\frac{|\Gamma(\alpha)|^{2}(2p)^{|m|}e^{-\pi g/2p}}%
{2\pi\Gamma^{2}(\beta)}\equiv\rho_{m}^{2}(E),\;\mathrm{spec}\hat
{h}_{m\mathfrak{e}}=\mathbb{R}_{+}.
\end{align*}

\subsubsection{$E=-\tau^{2}<0$, $\tau>0$, $K=\tau$}

We have%
\begin{align*}
&  \alpha=1/2+|m|/2+g/2\tau,\;\alpha_{-}=1/2-|m|/2+g/2\tau,\\
&  \operatorname{Im}\Omega_{m}(E+i0)=-2D_{m}(E)\left.  \operatorname{Im}%
(\psi(\alpha)\right|  _{\mathcal{E}=E+i0}.
\end{align*}
The function $\psi(\alpha)$ is real for $\mathcal{E}=E$ where $|\psi
(\alpha)|<\infty$. Therefore, $\sigma_{m}^{\prime}(E)$ can be not equal to
zero only in the points $\psi(\alpha)=\pm\infty$, i. e., in the points
$\alpha=-n$, $n\in\mathbb{Z}_{+}$.

\paragraph{$g\geq0$}

In this case, we have $\alpha>0$ and the equation $\alpha=-n$ has no
solutions, i.e., $\sigma_{m}^{\prime}(E)=0$.

\paragraph{$g<0$}

In this case, the equation $\alpha=-n$ has solutions,%
\begin{align*}
&  \tau_{n}=\frac{|g|}{1+|m|+2n},\;E_{n}=-\frac{g^{2}}{(1+|m|+2n)^{2}}%
,\;n\in\mathbb{Z}_{+},\\
&  \operatorname{Im}\psi(\alpha)=-\sum_{n\in\mathbb{Z}_{+}}\frac{4\pi\tau
_{n}^{3}}{|g|}\delta(E-E_{n}).
\end{align*}

We thus find%
\begin{align*}
&  \sigma_{m}^{\prime}(E)=\sum_{n\in\mathbb{Z}_{+}}Q_{n}^{2}\delta
(E-E_{n}),\;Q_{n}=\sqrt{8|g|^{-1}\tau_{n}^{3}D_{m}(E_{n})}=\\
&  =(2\tau_{n}/\kappa_{0})^{1/2+|m|/2}\sqrt{\frac{2\tau_{n}(1+n)_{|m|}%
}{1+|m|+2n}},\;\mathrm{spec}\hat{h}_{m\mathfrak{e}}=\{E_{n},\;n\in
\mathbb{Z}_{+}\}.
\end{align*}

Finally, we find:

i) $g\geq0$. The spectrum of $\hat{h}_{m\mathfrak{e}}$ is simple and
continuous, $\mathrm{spec}\hat{h}_{m\mathfrak{e}}=\mathbb{R}_{+}.$The set of
the generalized eigenfunctions $\{U_{mE}=\rho_{m}(E)C_{1,m}(x;E),\;E\geq0\}$
forms a complete orthonormalized system in $L^{2}(\mathbb{R}_{+})$.

ii) $g<0$. The spectrum of $\hat{h}_{m\mathfrak{e}}$ is simple and has
additionally the discrete part, $\mathrm{spec}\hat{h}_{m\mathfrak{e}%
}=\mathbb{R}_{+}\cup\{E_{n}<0,\;n\in\mathbb{Z}_{+}\}$. The set of the
generalized eigenfunctions $\{U_{mE}(x)=\rho_{m}(E)C_{1,m}(x;E),\;E\geq0\}$
and the eigenfunctions $\{U_{mn}(x)=Q_{n}C_{1,m}(x;E_{n}),\;n\in\mathbb{Z}%
_{+}\}$forms a complete orthonormalized system in $L^{2}(\mathbb{R}_{+})$.

\subsection{$m=1$}

\subsection{Useful solutions}

We obtain the solutions of eq.(\ref{Coul2.1.1}) with $m=1$ as the limit
$\delta\rightarrow0$ of the solutions of eq. (\ref{Coul2.1.3}) with $m=1$:%
\begin{align*}
&  C_{1,1}(x;\mathcal{E})=C_{1,1,0}(x;\mathcal{E})=\kappa_{0}xe^{-z/2}%
\Phi(\alpha,2;z),\\
&  C_{4,1}(x;\mathcal{E})=\lim_{\delta\rightarrow0}C_{4,1,\delta
}(x;\mathcal{E}),\\
&  C_{3,1}(x;\mathcal{E})=\kappa_{0}xe^{-z/2}\Psi(\alpha,2;z)=\\
&  =B_{1}(\mathcal{E})C_{1,1}(x;\mathcal{E})+C_{1}(\mathcal{E})C_{4,1}%
(x;\mathcal{E}),\;C_{1}(\mathcal{E})=\frac{\kappa_{0}}{2K\Gamma(\alpha)},\\
&  B_{1}(\mathcal{E})=B_{1,0}(\mathcal{E})=\frac{1}{2\Gamma(\alpha_{-}%
)}\left[  \psi(\alpha_{-})+\psi(\alpha)+2\ln(2K/\kappa_{0})\right]  ,\\
\alpha &  =1-w=1+\frac{g}{2K},\;\alpha_{-}=-w=\frac{g}{2K}.
\end{align*}

\subsubsection{Asymptotics, $x\rightarrow0$}

We have%
\begin{align*}
&  C_{1,1}(x;\mathcal{E})=C_{1,1\mathrm{as}}(x)(1+O(x)),\\
&  C_{4,1}(x;\mathcal{E})=C_{4,1\mathrm{as}}(x)\left(  1+O(x^{2}\ln x)\right)
,\\
&  C_{3,m}(x;\mathcal{E})=[B_{1}(\mathcal{E})C_{1,1\mathrm{as}}(x)+C_{1}%
(\mathcal{E})C_{4,1\mathrm{as}}(x)]\left(  1+O(x^{2}\ln x)\right)
,\;\operatorname{Im}\mathcal{E}>0,\\
&  C_{1,1\mathrm{as}}(x)=\kappa_{0}x,\;C_{4,1\mathrm{as}}(x)=1+(\mathbf{C}%
-1)gx+gx\ln(\kappa_{0}x).
\end{align*}

It follows from these results that any solutions are s.-integrable at the origion.

\subsubsection{Asymptotics, $x\rightarrow\infty$, $\operatorname{Im}%
\mathcal{E}>0$ ($\operatorname{Re}K>0$)}%

\begin{align*}
C_{1,1}(x;\mathcal{E})  &  =\frac{\kappa_{0}(2K)^{\alpha-\beta}\Gamma(\beta
)}{\Gamma(\alpha)}x^{-w}e^{Kx}(1+O(x^{-1}))=O(x^{-w}e^{x\operatorname{Re}%
K}),\\
C_{3,1}(x;\mathcal{E})  &  =\kappa_{0}(2K)^{-\alpha}x^{w}e^{-Kx}%
(1+O(x^{-1})).=O(x^{w}e^{-x\operatorname{Re}K})
\end{align*}

We see that the function $C_{3,1}(x;\mathcal{E})$ is s.-integrable for
$\operatorname{Im}\mathcal{E}>0$.

\subsubsection{Wronskians}%

\begin{align*}
&  \mathrm{Wr}(C_{1,m},C_{4,m})=-\kappa_{0},\\
&  \mathrm{Wr}(C_{1,m},C_{3,m})=-\kappa_{0}C_{1}(\mathcal{E})=-\frac
{\kappa_{0}^{2}\Gamma(\beta)}{2K\Gamma(\alpha)}=-\omega_{1}(\mathcal{E})
\end{align*}

\subsection{Symmetric operator $\hat{h}_{1}$}

For given a differential operation $\check{h}_{1}$ (\ref{Coul2.1.0}), we
determine the following symmetric operator $\hat{h}_{1}$,%

\[
\hat{h}_{1}:\left\{
\begin{array}
[c]{l}%
D_{h_{1}}=\mathcal{D}(\mathbb{R}_{+}),\\
\hat{h}_{1}\psi(x)=\check{h}_{1}\psi(x),\;\forall\psi\in D_{h_{1}}%
\end{array}
\right.  .
\]

\subsection{Adjoint operator $\hat{h}_{1}^{+}=\hat{h}_{1}^{\ast}$}%

\[
\hat{h}_{1}^{+}:\left\{
\begin{array}
[c]{l}%
D_{h_{1}^{+}}=D_{\check{h}_{1}}^{\ast}(\mathbb{R}_{+})=\{\psi_{\ast}%
,\psi_{\ast}^{\prime}\;\mathrm{are\;a.c.\;in}\mathcal{\;}\mathbb{R}_{+}%
,\;\psi_{\ast},\hat{h}_{1}^{+}\psi_{\ast}\in L^{2}(\mathbb{R}_{+})\}\\
\hat{h}_{1}^{+}\psi_{\ast}(x)=\check{h}_{1}\psi_{\ast}(x),\;\forall\psi_{\ast
}\in D_{h_{1}^{+}}%
\end{array}
\right.  .
\]

\subsubsection{Asymptotics}

I) $x\rightarrow\infty$

Because $V(x)\rightarrow0$ for large $x$, we have: $[\psi_{\ast},\chi_{\ast
}](x)\rightarrow0$ as $x\rightarrow\infty$, $\forall\psi_{\ast},\chi_{\ast}\in
D_{h_{1}^{+}}$.

II) $x\rightarrow0$

We represent the functions $\psi_{\ast}\in D_{h_{1}^{+}}$ in the form%
\begin{align*}
&  \psi_{\ast}(x)=c_{1}C_{1,1}(x;\mathcal{E}_{0})+c_{2}C_{4,1}(x;\mathcal{E}%
_{0})+I(x),\\
&  \psi_{\ast}^{\prime}(x)=c_{1}C_{1,1}^{\prime}(x;\mathcal{E}_{0}%
)+c_{2}C_{4,1}^{\prime}(x;\mathcal{E}_{0})+I^{\prime}(x),
\end{align*}
where%
\begin{align*}
&  I(x)=\frac{C_{4,1}(x;\mathcal{E}_{0})}{\kappa_{0}}\int_{0}^{x}%
C_{1,m}(y;\mathcal{E}_{0})\tilde{\eta}(y)dy-\frac{C_{1,1}(x;\mathcal{E}_{0}%
)}{\kappa_{0}}\int_{0}^{x}C_{4,m}(y;\mathcal{E}_{0})\tilde{\eta}(y)dy,\\
&  I^{\prime}(x)=\frac{C_{4,1}^{\prime}(x;\mathcal{E}_{0})}{\kappa_{0}}%
\int_{0}^{u}C_{1,1}(y;\mathcal{E}_{0})\tilde{\eta}(y)dy-\frac{C_{1,}^{\prime
}(x;\mathcal{E}_{0})}{\kappa_{0}}\int_{0}^{x}C_{4,1}(y;\mathcal{E}_{0}%
)\tilde{\eta}(y)dy.
\end{align*}
We obtain with the help of the CB-inequality that%
\[
I(x)=O(x^{3/2}),\;I^{\prime}(x)=O(x^{1/2}),\;x\rightarrow0.
\]
such that we find%
\begin{align*}
\psi_{\ast}(x) &  =c_{1}C_{1,1\mathrm{as}}(x)+c_{2}C_{4,1\mathrm{as}%
}(x)+O(x^{3/2}),\\
\psi_{\ast}^{\prime}(x) &  =c_{1}C_{1,1\mathrm{as}}^{\prime}(x)+c_{2}%
C_{4,1\mathrm{as}}^{\prime}(x)+O(x^{1/2}),\;x\rightarrow0,
\end{align*}
and%
\begin{align*}
&  \Delta_{h_{1}^{+}}(\psi_{\ast})=\kappa_{0}(\overline{c_{2}}c_{1}%
-\overline{c_{1}}c_{2})=i\kappa_{0}(\overline{c_{+}}c_{+}-\overline{c_{-}%
}c_{-}),\\
&  c_{\pm}=\frac{1}{\sqrt{2}}(c_{1}\pm ic_{2}).
\end{align*}

\subsection{Self-adjoint hamiltonians $\hat{h}_{1,\zeta}$}

The condition $\Delta_{h_{1}^{+}}(\psi)=0$ gives%
\begin{align*}
&  c_{-}=e^{2i\theta}c_{+},\;0\leq\theta\leq\pi,\;\theta=0\sim\theta
=\pi,\;\Longrightarrow\;\\
&  c_{1}\cos\zeta=c_{2}\sin\zeta,\;\zeta=\theta-\pi/2,\;|\zeta|\leq
\pi/2,\;\zeta=-\pi/2\sim\zeta=\pi/2,
\end{align*}
or%
\begin{align}
&  \psi(x)=C\psi_{\mathrm{as}}(x)+O(x^{3/2}),\;\psi^{\prime}(x)=C\psi
_{\mathrm{as}}^{\prime}(x)+O(x^{1/2}),\label{Coul2.3.4.1}\\
&  \psi_{\mathrm{as}}(x)=C_{1,1\mathrm{as}}(x)\sin\zeta+C_{4,1\mathrm{as}%
}(x)\cos\zeta.\nonumber
\end{align}
We thus have a family of s.a.hamiltonians $\hat{h}_{1,\zeta}$,%
\[
\hat{h}_{1,\zeta}:\left\{
\begin{array}
[c]{l}%
D_{h_{1\zeta}}=\{\psi\in D_{h_{1}^{+}},\;\psi
\;\mathrm{satisfy\;the\;boundary\;condition\;(\ref{Coul2.3.4.1})}\\
\hat{h}_{1,\zeta}\psi=\check{h}_{1}\psi,\;\forall\psi\in D_{h_{1,\zeta}}%
\end{array}
\right.  .
\]

\subsection{The guiding functional $\Phi_{1,\zeta}(\xi;\mathcal{E})$}

As a guiding functional $\Phi_{1,\zeta}(\xi;\mathcal{E})$ we choose%
\begin{align*}
&  \Phi_{1,\zeta}(\xi;\mathcal{E})=\int_{0}^{\infty}U_{1,\zeta}(x;\mathcal{E}%
)\xi(x)dx,\;\xi\in\mathbb{D}_{1,\zeta}=D_{r}(\mathbb{R}_{+})\cap
D_{h_{1,\zeta}}.\\
&  .U_{1,\zeta}(x;\mathcal{E})=C_{1,1}(x;\mathcal{E})\sin\zeta+C_{4,1}%
(x;\mathcal{E})\cos\zeta,
\end{align*}
$U_{1,\zeta}(u;\mathcal{E})$ is real-entire solution of eq. (\ref{Coul2.1.1})
(with $m=1$) satisfying the boundary condition (\ref{Coul2.3.4.1}).

The guiding functional $\Phi_{1,\zeta}(\xi;\mathcal{E}W)$ is simple and the
spectrum of $\hat{h}_{\zeta}$ is simple.

\subsection{Green function $G_{1,\zeta}(x,y;\mathcal{E})$, spectral function
$\sigma_{1,\zeta}(E)$}

We find the Green function $G_{1,\zeta}(x,y;\mathcal{E})$ as the kernel of the
integral representation
\[
\psi(x)=\int_{0}^{\infty}G_{1,\zeta}(x,y;\mathcal{E})\eta(y)dy,\;\eta\in
L^{2}(\mathbb{R}_{+}),
\]
of unique solution of an equation%
\begin{equation}
(\hat{h}_{1,\zeta}-\mathcal{E})\psi(x)=\eta(x),\;\operatorname{Im}%
\mathcal{E}>0,\label{Coul2.3.6.1}%
\end{equation}
for $\psi\in D_{h_{1,\zeta}}$. General solution of eq. (\ref{Coul2.3.6.1}) can
be represented in the form%
\begin{align*}
\psi(x) &  =a_{1}C_{1,1}(x;\mathcal{E})+a_{3}C_{3,1}(x;\mathcal{E})+I(x),\\
I(x) &  =\frac{C_{1,1}(x;\mathcal{E})}{\omega_{1}(\mathcal{E})}\int
_{x}^{\infty}C_{3,1}(y;\mathcal{E})\eta(y)dy+\frac{C_{3,1}(x;\mathcal{E}%
)}{\omega_{1}(\mathcal{E})}\int_{0}^{x}C_{1,1}(y;\mathcal{E})\eta(y)dy,\\
I(x) &  =\frac{C_{1,1as}(x;\mathcal{E})}{\omega_{1}(\mathcal{E})}+O\left(
x^{3/2}\right)  ,\;x\rightarrow0.
\end{align*}
A condition $\psi\in L^{2}(\mathbb{R}_{+})$ gives $a_{1}=0$. The condition
$\psi\in D_{h_{1\zeta}}$, i.e., $\psi$ satisfies the boundary condition
(\ref{Coul2.3.4.1}), gives%
\begin{align*}
&  a_{3}=-\frac{1}{\omega_{1}}\tilde{a}_{3}\int_{0}^{\infty}C_{3,1}%
(y;\mathcal{E})\eta(y)dy,\;\\
&  \tilde{a}_{3}=\frac{\cos\zeta}{C_{1}\omega_{1,\zeta}},\;\omega_{1,\zeta
}=\omega_{1,\zeta}(\mathcal{E})=f_{1}\cos\zeta-\sin\zeta,\\
&  f_{1}=f_{1}(\mathcal{E})=\frac{B_{1}(\mathcal{E})}{C_{1}(\mathcal{E}%
)}.=\frac{g}{2\kappa_{0}}[\psi(\alpha)+\psi(\alpha_{-})+2\ln(2K/\kappa_{0})]
\end{align*}

Using the relations%
\begin{align*}
&  \tilde{a}_{3}C_{3,1}-C_{1,1}=\frac{1}{\omega_{1,\zeta}}C_{1,\zeta},\\
&  C_{3,1}=C_{1}[\tilde{\omega}_{1,\zeta}C_{1,\zeta}+\omega_{1,\zeta}\tilde
{C}_{1,\zeta}],\\
&  C_{1,\zeta}=C_{1,1}\sin\zeta+C_{4,1}\cos\zeta,\\
&  \tilde{C}_{1,\zeta}=C_{1,1}\cos\zeta-C_{4,1}\sin\zeta,\\
&  \tilde{\omega}_{1,\zeta}=\tilde{\omega}_{1,\zeta}(\mathcal{E}%
)=f_{1}(\mathcal{E})\sin\zeta+\cos\zeta,\;\frac{\omega_{1}(\mathcal{E})}%
{C_{1}(\mathcal{E})}=\kappa_{0},
\end{align*}
we find%

\begin{align}
&  G_{1,\zeta}(x,y;\mathcal{E})=\frac{1}{\omega_{1}}\left\{
\begin{array}
[c]{l}%
C_{3,1}(x;\mathcal{E})[C_{1,1}(y;\mathcal{E})-\tilde{a}_{3}C_{3,1}%
(y;\mathcal{E})],\;x>y\\
\lbrack C_{1,m}(x;\mathcal{E})-\tilde{a}_{3}C_{3,1}(x;\mathcal{E}%
)]C_{3,m}(y;\mathcal{E}).\;x<y
\end{array}
\right.  =\nonumber\\
&  \,=-\frac{1}{\kappa_{0}}\Omega_{1,\zeta}(\mathcal{E})C_{1,\zeta
}(x;\mathcal{E})C_{1,\zeta}(y;\mathcal{E})-\frac{1}{\kappa_{0}}\left\{
\begin{array}
[c]{c}%
\tilde{C}_{1,\zeta}(x;\mathcal{E})C_{1,\zeta}(y;\mathcal{E}),\;x>y\\
C_{1,\zeta}(x;\mathcal{E})C_{1,\zeta}(y;\mathcal{E}),\;x<y
\end{array}
\right.  ,\label{Coul2.3.6.2}\\
&  \Omega_{m}(\mathcal{E})\equiv\frac{\tilde{\omega}_{1,\zeta}(\mathcal{E}%
)}{\omega_{1,\zeta}(\mathcal{E})}.\nonumber
\end{align}
Note that the last term in the r.h.s. of eq. (\ref{Coul2.3.6.2}) is real for
$\mathcal{E}=E$. From the relation%
\[
C_{1,1}^{2}(x_{0};E)\sigma_{m}^{\prime}(E)=\frac{1}{\pi}\operatorname{Im}%
G_{1,\zeta}(x_{0}-0,x_{0}+0;E+i0),
\]
we find%
\[
\sigma_{1,\zeta}^{\prime}(E)=-\frac{1}{\pi\kappa_{0}}\operatorname{Im}%
\Omega_{1,\zeta}(E+i0).
\]

\subsection{Spectrum}

\subsubsection{$\zeta=\pi/2$}

First we consider the case $\zeta=\pi/2$.

In this case, we have $U_{1,\pi/2}(x;\mathcal{E})=C_{1,1}(x;\mathcal{E})$ and%
\[
\sigma_{1,\pi/2}^{\prime}(E)=\frac{1}{\pi\kappa_{0}}\operatorname{Im}%
f_{1}(E+i0)=\frac{g}{2\pi\kappa_{0}^{2}}\operatorname{Im}[\psi(\alpha
)+\psi(\alpha_{-})+2\ln(2K/\kappa_{0})],
\]

\paragraph{$E=p^{2}\geq0$, $p\geq0$, $K=-ip=e^{-i\pi/2}p$}

We have%
\begin{align*}
&  \alpha=1-i\tilde{w},\;\alpha_{-}=-i\tilde{w},\;\tilde{w}=-g/2p,\\
&  \sigma_{1,\pi/2}^{\prime}(E)=\frac{g}{2\pi\kappa_{0}^{2}}[\operatorname{Im}%
(\psi(\alpha)+\psi(\alpha_{-}))-\pi]=\\
&  \,=\frac{g}{2\kappa_{0}^{2}}\left(  \coth(\pi g/2p)-1\right)  \equiv
\rho_{1,\pi/2}^{2}(E)
\end{align*}

\paragraph{$E=-\tau^{2}<0$, $\tau>0$, $K=\tau$}

We have%
\begin{align*}
&  \alpha=1+g/2\tau,\;\alpha_{-}=g/2\tau,\\
&  \sigma_{1,\pi/2}^{\prime}(E)=\frac{g}{\pi\kappa_{0}^{2}}\operatorname{Im}%
\psi(\alpha)_{\mathcal{E}=E+i0}.
\end{align*}
The function $\psi(\alpha)$ is real for $\mathcal{E}=E$ where $|\psi
(\alpha)|<\infty$. Therefore, $\sigma_{1,\pi/2}^{\prime}(E)$ can be not equal
to zero only in the points $\psi(\alpha)=\pm\infty$, i. e., in the points
$\alpha=-n$, $n\in\mathbb{Z}_{+}$.

\subparagraph{$g\geq0$}

In this case, we have $\alpha>0$ and the equation $\alpha=-n$ has no
solutions, i.e., $\sigma_{m}^{\prime}(E)=0$.

\subparagraph{$g<0$}

In this case, the equation $\alpha=-n$ has solutions,%
\begin{align*}
&  \tau_{n}=\frac{|g|}{2(1+n)},\;\mathcal{E}_{1,n}=-\frac{g^{2}}{4(1+n)^{2}%
},\;n\in\mathbb{Z}_{+},\\
&  \operatorname{Im}\psi(\alpha)=-\sum_{n\in\mathbb{Z}_{+}}\frac{4\pi\tau
_{n}^{3}}{|g|}\delta(E-\mathcal{E}_{1,n}).
\end{align*}

We thus find%
\begin{align*}
&  \sigma_{1,\pi/2}^{\prime}(E)=\sum_{n\in\mathbb{Z}_{+}}Q_{1,\pi/2,n}%
^{2}\delta(E-\mathcal{E}_{1,n}),\;Q_{1,\pi/2,n}=\frac{2\tau_{n}^{3/2}}%
{\kappa_{0}},\\
&  \mathrm{spec}\hat{h}_{1,\pi/2}=\{\mathcal{E}_{1,n},\;n\in\mathbb{Z}_{+}\}.
\end{align*}

Finally, we find:

i) $g\geq0$. The spectrum of $\hat{h}_{1,\pi/2}$ is simple and continuous,
$\mathrm{spec}\hat{h}_{1,\pi/2}=\mathbb{R}_{+}.$The set of the generalized
eigenfunctions $\{U_{1,\pi/2,E}(x)=\rho_{1,\pi/2}^{2}(E)C_{1,1}(x;E),\;E\geq
0\}$ forms a complete orthonormalized system in $L^{2}(\mathbb{R}_{+})$.

ii) $g<0$. The spectrum of $\hat{h}_{1,\pi/2}$ is simple and has additionally
the discrete part, $\mathrm{spec}\hat{h}_{1,\pi/2}=\mathbb{R}_{+}%
\cup\{\mathcal{E}_{1,n}<0,\;n\in\mathbb{Z}_{+}\}$. The set of the generalized
eigenfunctions $\{U_{1,\pi/2,E}(x)=\rho_{1,\pi/2}^{2}(E)C_{1,1}(x;E),\;E\geq
0\}$ and the eigenfunctions $\{U_{1,\pi/2,n}(x)=Q_{1,\pi/2,n}C_{1,1}%
(x;\mathcal{E}_{1,n}),\;n\in\mathbb{Z}_{+}\}$forms a complete orthonormalized
system in $L^{2}(\mathbb{R}_{+})$.

The same results for spectrum and eigenfunctions we obtain for the case
$\zeta=-\pi/2$.

Note that all results for spectrum and eigenfunctions can be obtained from
corresponding formulas of the sec \ 2 \ setting there $m=1$.

\subsubsection{$|\zeta|<\pi/2$}

Now we consider the case $|\zeta|<\pi/2$.

In this case, we can represent $\sigma_{1,\zeta}^{\prime}(E)$ in the form%
\[
\sigma_{1,\zeta}^{\prime}(E)=-\frac{1}{\pi\kappa_{0}\cos^{2}\zeta
}\operatorname{Im}\frac{1}{f_{1,\zeta}(E+i0)},\;f_{1,\zeta}(\mathcal{E}%
)=f_{1}(\mathcal{E})-\tan\zeta.
\]

\paragraph{$E=p^{2}\geq0$, $p\geq0$, $K=e^{-i\pi/2}p$}

In this case, we find%
\begin{align}
&  \sigma_{1,\zeta}^{\prime}(E)=\frac{1}{\pi\kappa_{0}}\frac{B_{1}(E)}%
{(A_{1}(E)\cos\zeta-\sin\zeta)^{2}+B_{1}^{2}(E)\cos^{2}\zeta}%
,\label{Coul2.3.7.2.1}\\
&  A_{1}(E)=\operatorname{Re}f_{1}(E),\;B_{1}(E)=\operatorname{Im}%
f_{1}(E)=\frac{\pi g}{2\kappa_{0}}[\coth(\pi g/2p)-1].\nonumber
\end{align}

\subparagraph{$g\leq0$}

In this case, $\sigma_{1,\zeta}^{\prime}(E)>0$ and is finite for $E>0$,
$\sigma_{1,\zeta}^{\prime}(0)\geq0$ and is finite for $|g|+|\zeta|\neq0$ , and
$\sigma_{1,\zeta}^{\prime}(E)$ has a behaviour $O(E^{-1/2})$ as $E\rightarrow
0$ for $g=\zeta=0$, such that the spectrum of $\hat{h}_{\zeta}$ is simple and
\ continuous, $\mathrm{spec}\hat{h}_{1,\zeta}=\mathbb{R}_{+}$.

\subparagraph{$g>0$}

In this case, $\sigma_{1,\zeta}^{\prime}(E)$ is given by eq.
(\ref{Coul2.3.7.2.1}) for $E>0$. But now, $B_{1}(0)=0$ and the denominator of
expr. (\ref{Coul2.3.7.2.1}) is equal to zero for $E=0$ and $\zeta=\zeta_{1}$,
$\tan\zeta_{1}=A_{1}(0)=(g/\kappa_{0})\ln(g/\kappa_{0})$, such that we should
study the behaviour of $\sigma_{1,\zeta}^{\prime}(E)$ for small $E$ in more
details. We have%
\[
f_{1,\zeta}(\mathcal{E})=(g/\kappa_{0})\ln(g/\kappa_{0})-\tan\zeta
+\mathcal{E}/(3g^{2})+O(\mathcal{E}^{2}).
\]
We see that if $\zeta\neq\zeta_{1}$ then $\sigma_{1,\zeta}^{\prime}(0)=0$. But
if $\zeta=\zeta_{1}$, we find%
\[
\sigma_{1,\zeta}^{\prime}(E)=-\frac{3g^{2}}{\pi\kappa_{0}\cos^{2}\zeta_{1}%
}\operatorname{Im}\frac{1}{E+i0}+O(1)=\frac{3g^{2}}{\kappa_{0}\cos^{2}%
\zeta_{1}}\delta(E)+O(1),
\]
and $\mathrm{spec}\hat{h}_{1,\zeta_{1}}=\mathbb{R}_{+}\cup\{E_{(-)}(\zeta
_{1})=0\}$.

\paragraph{$E=-\tau^{2}<0$, $\tau>0$, $K=\tau$}

In this case, we have%
\begin{align*}
&  f_{1}(E)=\frac{g}{2\kappa_{0}}[\psi(1+g/2\tau)+\psi(g/2\tau)+2\ln
(2\tau/\kappa_{0})]=\\
&  \,=\frac{g}{\kappa_{0}}[\psi(g/2\tau)+2\ln(2\tau/\kappa_{0})+\tau/g],
\end{align*}
so that .$f_{1}(E)$ is real and $\sigma_{1,\zeta}^{\prime}(E)$ can be not
equal to zero only in the points $E_{1,n}(\zeta)$ which are solutions of an
equation%
\begin{equation}
f_{1,\zeta}(E_{1,n}(\zeta))=0\;\mathrm{or}\;f_{1}(E_{1,n}(\zeta))=\tan\zeta.
\label{Coul2.3.7.3.1}%
\end{equation}

We thus obtain%
\[
\sigma_{1,\zeta}^{\prime}(E)=\sum_{n\in\mathcal{N}_{1}}Q_{1,\zeta,n}^{2}%
\delta(E-E_{1,n}(\zeta)),\;Q_{1,\zeta,n}=\frac{1}{|\cos\zeta|}\sqrt{\frac
{2}{\kappa_{0}f_{1}^{\prime}(E_{1,n}(\zeta))}},
\]
where $\mathcal{N}_{1}$ is some subset of integers, $\mathcal{N}_{1}%
\in\mathbb{Z}$, and $f^{\prime}(E_{1,n}(\zeta))>0$. Because any $E<0$ is
solution of eq. (\ref{Coul2.3.7.3.1}) for some $\zeta$, $f_{1}^{\prime}(E)>0$
for all $E$. Futhermore, we find:%
\[
\partial_{\zeta}E_{1,n}(\zeta)=[f_{1}^{\prime}(E_{1,n}(\zeta))\cos^{2}%
\zeta]^{-1}>0.
\]

\subparagraph{$g>0$}

In this case, we have: $f_{1}(E)$ is a smooth function on the interval
$-\infty<E<0;$ $f_{1}(E)$ increases monotonically from $-\infty$ to
$A(0)=(g/\kappa_{0})\ln(g/\kappa_{0})$ as $E$ run from $-\infty$ to $-0$. \ We
thus obtain: for $\zeta\in(\zeta_{1},\pi/2)$, eq. (\ref{Coul2.3.7.3.1}) has no
solutions; and for any fixed $\zeta\in(-\pi/2,\zeta_{1})$, eq.
(\ref{Coul2.5.7.3.1}) has one solution $E_{1,(-)}(\zeta)\in(-\infty,0)$
monotonically increasing from $-\infty$ to $-0$ as $\zeta$ run from $-\pi/2+0$
t0 $\zeta_{0}-0$.

\subparagraph{$g=0$}

In this case, we have $f_{1}(E)=-\tau/\kappa_{0}$ and eq. (\ref{Coul2.3.7.3.1}%
) has no solutions for $\zeta\in\lbrack0,\pi/2)$ and one solution
$E_{1,(-)}(\zeta)=-\kappa_{0}^{2}\tan^{2}\zeta$ for any $\zeta\in(-\pi/2,0)$.

\subparagraph{$g<0$}

In this case, we have: $f_{1}(E)=-\frac{|g|}{\kappa_{0}}[\psi(-|g|/2\tau
)-\tau/|g|+\ln(2\tau/\kappa_{0})]$; $f(\mathcal{E}_{1,n}\pm0)=\mp\infty$,
$n\in\mathbb{Z}_{+}$; in each interval $(\mathcal{E}_{0,n-1},\mathcal{E}%
_{0,n})$, $n\in\mathbb{Z}_{+}$, $f_{1}(E)$ increases monotonically from
$-\infty$ to $\infty$ as $E$ run from $\mathcal{E}_{1,n-1}+0$ to
$\mathcal{E}_{1,n}-0$; in each interval $(\mathcal{E}_{1,n-1},\mathcal{E}%
_{1,n})$, $n\in\mathbb{Z}_{+}$, for any fixed $\zeta\in(-\pi/2,\pi/2)$, eq.
(\ref{Coul2.3.7.3.1}) has one solution $E_{1,n}(\zeta)$ monotonically
increasing from $\mathcal{E}_{1,n-1}+0$ to $\mathcal{E}_{1,n}-0$ as $\zeta$
run from $-\pi/2+0$ to $\pi/2-0$. Here we set $\mathcal{E}_{1,-1}=-\infty$. We
find%
\[
\mathcal{N}_{1}=\left\{
\begin{array}
[c]{l}%
\varnothing,\;g>0,\;\zeta\in(\zeta_{1},\pi/2),\;\mathrm{or}\;g\geq
0,\;\zeta=\pm\pi/2,\;g=0,\;\zeta\in\lbrack0,\pi/2)\\
(-),\;g>0,\;\zeta\in(-\pi/2,\zeta_{1}]\\
(-),\;g=0,\;\zeta\in(-\pi/2,0)\\
\mathbb{Z}_{+},\;g<0,\;\zeta\in\lbrack-\pi/2,\pi/2]
\end{array}
\right.  ,
\]
where, for completeness, the cases $\zeta=\pm\pi/2$ and $E=0$ are included.

Note the relation%
\[
\lim_{\zeta\rightarrow\pi/2}E_{1,n}(\zeta)=\lim_{\zeta\rightarrow-\pi
/2}E_{1,n+1}(\zeta)=\mathcal{E}_{1,n},\;n\in\mathbb{Z}_{+}.
\]

Finally we obtain: the spectra of $\hat{h}_{1,\zeta}$ are simple,
$\mathrm{spec}\hat{h}_{1,\zeta}=\mathbb{R}_{+}\cup\{E_{1,n}(\zeta)\leq
0,\;n\in\mathcal{N}_{1}\}$, the set $\{U_{1,\zeta,E}(x)=\rho_{1,\zeta
}(E)U_{1,\zeta}(x;E),\;E\in\mathbb{R}_{+};\;U_{1,\zeta,n}(x)=Q_{1,\zeta
,n}U_{1,\zeta}(x;E_{1,n}(\zeta)\},\;n\in\mathcal{N}_{1}\}$ of (generalized)
eigenfunctions of $\hat{h}_{1,\zeta}$ forms the complete orthohonalized system
in $L^{2}(\mathbb{R}_{+})$, where%
\[
\rho_{1,\zeta}(E)=\left\{
\begin{array}
[c]{l}%
\mathrm{eq}.\;\mathrm{(\ref{Coul2.3.7.2.1})}\;\mathrm{for}\;g\leq0,\;E\geq0;\\
\mathrm{and}\;g>0,\;E>0;\\
\lim_{E\rightarrow+0}\rho_{1,\zeta_{1}}(E)\;\mathrm{for}\;g>0,\;E=0
\end{array}
\right.  .
\]

\subsection{$m=-1$}

Only modification which we must do is the following: the extension parameter
for the case $m=-1$ should be considered as indendent of the extension
parameter for the case $m=1$. It is convenient to denote the extension
parameter for the case $m=1$ as $\zeta_{(1)}$ and \ for the case $m=-1$ as
$\zeta_{(-1)}$.

\subsection{$m=0$}

\subsection{Useful solutions}

For $m=0$, eqs. (\ref{Coul2.1.1}) and\ref{Coul2.1.1a}) are redused
respectively to%
\begin{equation}
\partial_{x}^{2}\psi_{m}(x)+(\frac{1}{4x^{2}}-\frac{g}{x}+\mathcal{E})\psi
_{m}(x)=0, \label{Coul2.5.1.1}%
\end{equation}
and%
\[
\partial_{x}^{2}\psi_{m}(x)+(-\frac{\delta^{2}-1}{4x^{2}}-\frac{g}%
{x}+\mathcal{E})\psi_{m}(x)=0,\;|\delta|<1.
\]

We will use the following solutions of eq. (\ref{Coul2.5.1.1})%
\begin{align*}
&  C_{1,0}(x;\mathcal{E})=C_{1,0,0}(x;\mathcal{E})=(\kappa_{0}x)^{1/2}%
e^{-z/2}\Phi(\alpha,1;z),\\
&  C_{2,0}(x;\mathcal{E})=\left.  \partial_{\delta}C_{1,0,\delta
}(x;\mathcal{E})\right|  _{\delta=+0}=\\
&  \,=(\kappa_{0}x)^{1/2}e^{-z/2}\left.  \partial_{\delta}\Phi(\alpha_{\delta
},\beta_{\delta};z)\right|  _{\delta=+0}+(1/2)C_{1,0}(x;\mathcal{E})\ln
(\kappa_{0}x),\\
&  C_{3,0}(x;\mathcal{E})=C_{3,0,0}(x;\mathcal{E})=(\kappa_{0}x)^{1/2}%
e^{-z/2}\Psi(\alpha,1;z)=\\
&  \,=\frac{\omega_{0}(\mathcal{E})}{\Gamma(\alpha)}C_{1,0}(x;\mathcal{E}%
)-\frac{2}{\Gamma(\alpha)}C_{2,0}(x;\mathcal{E}),\\
&  \alpha=1/2-w,\;\omega_{0}(\mathcal{E})=2\psi(1)-\psi(\alpha)-\ln
(2K/\kappa_{0}).
\end{align*}

\subsubsection{Asymptotics}

\paragraph{$x\rightarrow0$}

We have%
\begin{align*}
C_{1,0}(u;\mathcal{E})  &  =(\kappa_{0}x)^{1/2}(1+O(x)),\;C_{2,0}%
(x;\mathcal{E})=(1/2)(\kappa_{0}x)^{1/2}\ln(\kappa_{0}x)(1+O(x)),\\
C_{3,0}(x;\mathcal{E})  &  =\left[  \frac{\omega_{0}(\mathcal{E})}%
{\Gamma(\alpha)}(\kappa_{0}x)^{1/2}-\frac{1}{\Gamma(\alpha)}(\kappa
_{0}x)^{1/2}\ln(\kappa_{0}x)\right]  (1+O(x)),\;\operatorname{Im}\emph{E}>0.
\end{align*}

\paragraph{$x\rightarrow\infty$, $\operatorname{Im}\mathcal{E}>0$}

We have%

\begin{align*}
C_{1,0}(x;\mathcal{E})  &  =\frac{\kappa_{0}^{1/2}(2K)^{\alpha-1}}%
{\Gamma(\alpha)}x^{-w}e^{Kx}(1+O(x^{-1}))=O(x^{-w}e^{x\operatorname{Re}K}),\\
C_{3,0}(x;\mathcal{E})  &  =\kappa_{0}^{1/2}(2K)^{-\alpha}x^{w}e^{-Kx}%
(1+O(x^{-1})).=O(x^{w}e^{-x\operatorname{Re}K})
\end{align*}

\subsubsection{Wronskian}%

\[
\mathrm{Wr}(C_{1,0},C_{2,0})=\kappa_{0}/2,\;\mathrm{Wr}(C_{1,0},C_{3,0}%
)=-\frac{\kappa_{0}}{\Gamma(\alpha)}%
\]

\subsection{Symmetric operator $\hat{h}_{0}$}

For given a differential operation $\check{h}_{0}=-\partial_{x}^{2}-\frac
{1}{4x^{2}}+\frac{g}{x}$ we determine the following symmetric operator
$\hat{h}_{0}$,%

\[
\hat{h}_{0}:\left\{
\begin{array}
[c]{l}%
D_{h_{0}}=\mathcal{D}(\mathbb{R}_{+}),\\
\hat{h}_{0}\psi(u)=\check{h}_{0}\psi(u),\;\forall\psi\in D_{h_{0}}%
\end{array}
\right.  .
\]

\subsection{Adjoint operator $\hat{h}_{0}^{+}=\hat{h}_{0}^{\ast}$}%

\[
\hat{h}_{0}^{+}:\left\{
\begin{array}
[c]{l}%
D_{h_{0}^{+}}=D_{\check{h}_{0}}^{\ast}(\mathbb{R}_{+})=\{\psi_{\ast}%
,\psi_{\ast}^{\prime}\;\mathrm{are\;a.c.\;in}\mathcal{\;}\mathbb{R}_{+}%
,\;\psi_{\ast},\hat{h}_{0}^{+}\psi_{\ast}\in L^{2}(\mathbb{R}_{+})\}\\
\hat{h}_{0}^{+}\psi_{\ast}(u)=\check{h}_{0}\psi_{\ast}(u),\;\forall\psi_{\ast
}\in D_{h_{0}^{+}}%
\end{array}
\right.  .
\]

\subsubsection{Asymptotics}

I) $x\rightarrow\infty$

Because $V(x)\rightarrow0$ as $x\rightarrow\infty$, we have: $\psi_{\ast
}(x),\psi_{\ast}^{\prime}(x),[\psi_{\ast},\psi_{\ast}](x)\rightarrow0$ as
$x\rightarrow\infty$, $\forall\psi_{\ast}\in D_{h_{0}^{+}}$.

II) $x\rightarrow0$

By the standard way, we obtain%
\begin{align*}
&  \psi_{\ast}(x)=c_{1}u_{1\mathrm{as}}(x)+c_{2}u_{2\mathrm{as}}%
(x)+O(x^{3/2}\ln x),\\
&  \psi_{\ast}^{\prime}(x)=c_{1}u_{1\mathrm{as}}^{\prime}(x)+c_{2}%
u_{2\mathrm{as}}^{\prime}(x)+O(x^{1/2}\ln x),\\
&  u_{1\mathrm{as}}(x)=(\kappa_{0}x)^{1/2},\;u_{2\mathrm{as}}(x)=(1/2)(\kappa
_{0}x)^{1/2}\ln(\kappa_{0}x).
\end{align*}

For the asymmetry form $\Delta_{h_{0}^{+}}(\psi_{\ast})$, we find%
\begin{align*}
&  \Delta_{h_{0}^{+}}(\psi_{\ast})=(\kappa_{0}/2)(\overline{c_{2}}%
c_{1}-\overline{c_{1}}c_{2})=(i\kappa_{0}/2)(\overline{c_{+}}c_{+}%
-\overline{c_{-}}c_{-}),\\
&  c_{\pm}=\frac{1}{\sqrt{2}}(c_{1}\pm ic_{2}).
\end{align*}

\subsection{Self-adjoint hamiltonians $\hat{h}_{0,\zeta}$}

The condition $\Delta_{h_{0}^{+}}(\psi)=0$ gives%
\begin{align*}
&  c_{-}=e^{2i\theta}c_{+},\;0\leq\theta\leq\pi,\;\theta=0\sim\theta
=\pi,\;\Longrightarrow\;\\
&  c_{1}\cos\zeta=c_{2}\sin\zeta,\;\zeta=\theta-\pi/2,\;|\zeta|\leq
\pi/2,\;\zeta=-\pi/2\sim\zeta=\pi/2,
\end{align*}
or%
\begin{align}
&  \psi(x)=C\psi_{\mathrm{as}}(x)+O(x^{3/2}\ln x),\;\psi^{\prime}%
(x)=C\psi_{\mathrm{as}}^{\prime}(x)+O(x^{1/2}\ln x),\label{Coul2.5.4.1}\\
&  \psi_{\mathrm{as}}(x)=u_{1\mathrm{as}}(x)\sin\zeta+u_{2\mathrm{as}}%
(x)\cos\zeta.\nonumber
\end{align}
We thus have a family of s.a.hamiltonians $\hat{h}_{0,\zeta}$,%
\[
\hat{h}_{0,\zeta}:\left\{
\begin{array}
[c]{l}%
D_{h_{0,\zeta}}=\{\psi\in D_{h_{0}^{+}},\;\psi
\;\mathrm{satisfy\;the\;boundary\;condition\;(\ref{Coul2.5.4.1})}\\
\hat{h}_{0,\zeta}\psi=\check{h}_{0}\psi,\;\forall\psi\in D_{h_{0,\zeta}}%
\end{array}
\right.  .
\]

\subsection{The guiding functional $\Phi_{0,\zeta}(\xi;\mathcal{E})$}

As a guiding functional $\Phi_{0,\zeta}(\xi;\mathcal{E})$ we choose%
\begin{align*}
&  \Phi_{0,\zeta}(\xi;\mathcal{E})=\int_{0}^{\infty}U_{0,\zeta}(x;\mathcal{E}%
)\xi(x)dx,\;\xi\in\mathbb{D}_{0,\zeta}=D_{r}(\mathbb{R}_{+})\cap
D_{h_{0,\zeta}}.\\
&  .U_{0,\zeta}(x;\mathcal{E})=C_{1,0}(x;\mathcal{E})\sin\zeta+C_{2,0}%
(x;\mathcal{E})\cos\zeta,
\end{align*}
$U_{0,\zeta}(u;\mathcal{E})$ is real-entire solution of eq. (\ref{Coul2.1.1})
(with $m=0$) satisfying the boundary condition (\ref{Coul2.5.4.1}).

The guiding functional $\Phi_{0,\zeta}(\xi;\mathcal{E}W)$ is simple and the
spectrum of $\hat{h}_{0,\zeta}$ is simple.

\subsection{Green function $G_{0,\zeta}(x,y;\mathcal{E})$, spectral function
$\sigma_{0,\zeta}(E)$}

We find the Green function $G_{0,\zeta}(x,y;\mathcal{E})$ as the kernel of the
integral representation
\[
\psi(x)=\int_{0}^{\infty}G_{0,\zeta}(,y;\mathcal{E})\eta(y)dy,\;\eta\in
L^{2}(\mathbb{R}_{+}),
\]
of unique solution of an equation%
\begin{equation}
(\hat{h}_{0,\zeta}-\mathcal{E})\psi(x)=\eta(x),\;\operatorname{Im}%
\mathcal{E}>0, \label{Coul2.5.6.1}%
\end{equation}
for $\psi\in D_{h_{0,\zeta}}$. General solution of eq. (\ref{Coul2.5.6.1})
(under condition $\psi\in L^{2}(\mathbb{R}_{+})$) can be represented in the
form%
\begin{align*}
\psi(x)  &  =aC_{3,0}(x;\mathcal{E})+\frac{\Gamma(\alpha)}{\kappa_{0}}%
C_{1,0}(x;\mathcal{E})\eta_{3}(\mathcal{E})+\frac{\Gamma(\alpha)}{\kappa_{0}%
}I(x),\;\eta_{3}(\mathcal{E})=\int_{0}^{\infty}C_{3,0}(y;\mathcal{E}%
)\eta(y)dy\\
I(x)  &  =C_{3,0}(x;\mathcal{E})\int_{0}^{x}C_{1,0}(y;\mathcal{E}%
)\eta(y)dy-C_{1,0}(x;\mathcal{E})\int_{0}^{x}C_{3,0}(y;\mathcal{E}%
)\eta(y)dy,\\
I(x)  &  =O\left(  x^{3/2}\ln x\right)  ,\;x\rightarrow0.
\end{align*}
A condition $\psi\in D_{h_{0,\zeta}}$(i.e.$\psi$ satisfies the boundary
condition (\ref{Coul2.5.4.1})) gives%
\[
a=-\frac{\Gamma^{2}(\alpha)\cos\zeta}{2\kappa_{0}\omega_{0,\zeta}%
(\mathcal{E})}\eta_{3}(\mathcal{E}),\;\omega_{0,\zeta}(\mathcal{E})=\frac
{1}{2}\omega_{0}(\mathcal{E})\cos\zeta+\sin\zeta,
\]%
\begin{align}
&  G_{0,\zeta}(x,y;\mathcal{E})=\Omega_{0,\zeta}(\mathcal{E})U_{0,\zeta
}(x;\mathcal{E})U_{0,\zeta}(y;\mathcal{E})+\nonumber\\
&  \,+\frac{2}{\kappa_{0}}\left\{
\begin{array}
[c]{c}%
\tilde{U}_{0,\zeta}(x;\mathcal{E})U_{0,\zeta}(y;\mathcal{E}),\;x>y\\
U_{0,\zeta}(x;\mathcal{E})\tilde{U}_{0,\zeta}(y;\mathcal{E}),\;x<y
\end{array}
\right.  ,\label{Coul2.5.6.2}\\
&  \Omega_{0,\zeta}(\mathcal{E})\equiv\frac{2\tilde{\omega}_{0,\zeta
}(\mathcal{E})}{\kappa_{0}\omega_{0,\zeta}(\mathcal{E})},\;\tilde{\omega
}_{0,\zeta}(\mathcal{E})=\frac{1}{2}\omega_{0}(\mathcal{E})\sin\zeta-\cos
\zeta,\nonumber\\
&  \tilde{U}_{0,\zeta}(x;\mathcal{E})=C_{1,0}(x;\mathcal{E})\cos\zeta
-C_{2,0}(x;\mathcal{E})\sin\zeta,\nonumber
\end{align}
where we used an equality%
\[
\Gamma(\alpha)C_{3,0}(x;\mathcal{E})=2\tilde{\omega}_{0,\zeta}(\mathcal{E}%
)U_{0,\zeta}(x;\mathcal{E})+2\omega_{0,\zeta}(\mathcal{E})\tilde{U}_{0,\zeta
}(x;\mathcal{E}).
\]
Note that the function $\tilde{U}_{0,\zeta}(x;\mathcal{E})$ is real-entire in
$\mathcal{E}$ and the last term in the r.h.s. of eq. (\ref{Coul2.5.6.2}) is
real for $W=E$. For $\sigma_{0,\zeta}^{\prime}(E)$, we find%
\[
\sigma_{0,\zeta}^{\prime}(E)=\frac{1}{\pi}\operatorname{Im}\Omega_{0,\zeta
}(E+i0).
\]

\subsection{Spectrum}

\subsubsection{$\zeta=\pi/2$}

First we consider the case $\zeta=\pi/2$.

In this case, we have $U_{0,\pi/2}=C_{1,0}(x;\mathcal{E})$ and%
\[
\sigma_{0,\pi/2}^{\prime}(E)=-\frac{1}{\pi\kappa_{0}}\operatorname{Im}%
[\psi(\alpha)+\ln(2K/\kappa_{0})],
\]

\paragraph{$E=p^{2}\geq0$, $p\geq0$, $K=e^{-i\pi/2}p$}

In this case, we find%
\begin{align*}
&  \sigma_{0,\pi/2}^{\prime}(E)=\frac{1}{2\kappa_{0}}-\frac{1}{\pi\kappa_{0}%
}\operatorname{Im}[\psi(1/2+ig/2p)=\frac{1}{2\kappa_{0}}[1-\tanh(\pi
g/2p)]\equiv\\
&  \equiv\rho_{0,\pi/2}^{2}(E),\;\mathrm{spec}\hat{h}_{0,\pi/2}=\mathbb{R}%
_{+}.
\end{align*}

\paragraph{$E=-\tau^{2}<0$, $\tau>0$, $K=\tau$}

In this case, we have
\[
\sigma_{0,\pi/2}^{\prime}(E)=-\frac{1}{\pi\kappa_{0}}\operatorname{Im}%
\psi(\alpha),\;\alpha=1/2+g/2\tau.
\]

\subparagraph{$g\geq0$}

In this case, we have $\alpha>0$, and $\sigma_{0,\pi/2}^{\prime}(E)=0$,
spectrum points are absent.

\subparagraph{$g<0$, $\alpha=1/2-|g|/2\tau$}

In this case, we have%
\begin{align*}
&  \sigma_{0,\pi/2}^{\prime}(E)=\sum_{n=0}^{\infty}Q_{0,\pi/2,n}^{2}%
\delta(E-\mathcal{E}_{0,n}),\;Q_{0,\pi/2,n}=\frac{2\tau_{n}}{\sqrt{\kappa
_{0}(1+2n)}},\\
&  \tau_{n}=\frac{|g|}{1+2n},\;\mathcal{E}_{0,n}=-\frac{g^{2}}{(1+2n)^{2}}.
\end{align*}

Finally:we obtain:

for $g\geq0$, the spectrum $\hat{h}_{0,\pi/2}$ is simple and continuous,
$\mathrm{spec}\hat{h}_{0,\pi/2}=\mathbb{R}_{+}$, and the set of generalized
eigenfunctions $\{U_{0,\pi/2,E}(x)=\rho_{0,\pi/2}(E)C_{1,0}(x;E),\;E\in
\mathbb{R}_{+}\}$ forms a complete orthonormalized system in $L^{2}%
(\mathbb{R}_{+})$;

for $g<0$, the spectrum $\hat{h}_{0,\pi/2}$ is simple and contains continuous
and discrete parts, $\mathrm{spec}\hat{h}_{0,\pi/2}=\mathbb{R}_{+}%
\cup\{\mathcal{E}_{0,n}<0,\;n\in\mathbb{Z}_{+}\}$, and the set of
(generalized) eigenfunctions $$\{U_{0,\pi/2,E}(x)=\rho_{0,\pi/2}(E)C_{1,0}%
(x;E),\;E\in\mathbb{R}_{+};U_{0,\pi/2,n}(x)=Q_{0,\pi/2,n}C_{1,0}%
(x;\mathcal{E}_{0,n}),\;n\in\mathbb{Z}_{+}\}$$ forms a complete orthonormalized
system in $L^{2}(\mathbb{R}_{+})$.

Note that there results for spectrum and the set of eigenfunctions can be
obtained from the corresponding results of sec. 2 by formal substitution
$|m|\rightarrow0$.

The same results we obtain for the case $\zeta=-\pi/2$.

\subsubsection{$|\zeta|<\pi/2$}

Now we consider the case $|\zeta|<\pi/2$.

In this case, we can represent $\sigma_{0,\zeta}^{\prime}(E)$ in the form%
\begin{align*}
\sigma_{0,\zeta}^{\prime}(E)  &  =-\frac{2}{\pi\kappa_{0}\cos^{2}\zeta
}\operatorname{Im}\frac{1}{f_{0,\zeta}(E+i0)},\;f_{0,\zeta}(\mathcal{E}%
)=f_{0}(\mathcal{E})+\tan\zeta,\\
f_{0}(\mathcal{E})  &  =\omega_{0}(\mathcal{E})/2.
\end{align*}

\paragraph{$E=p^{2}\geq0$, $p\geq0$, $K=e^{-i\pi/2}p$}

In this case, we find%
\begin{align}
&  \sigma_{0,\zeta}^{\prime}(E)=\frac{8}{\kappa_{0}}\frac{B_{0}(E)}%
{16(A_{0}(E)\cos\zeta+\sin\zeta)^{2}+\pi^{2}B_{0}^{2}(E)\cos^{2}\zeta
},\label{Coul2.5.7.2.1}\\
&  A_{0}(E)=\operatorname{Re}f_{0}(E),\;B_{0}(E)=(4/\pi)\operatorname{Im}%
f_{0}(E)=1-\tanh(\pi g/2p).\nonumber
\end{align}

\subparagraph{$g\leq0$}

In this case, $B_{0}(E)>0$, $\forall E\geq0$, and $\sigma_{0,\zeta}^{\prime
}(E)>0$ and is finite, such that the spectrum of $\hat{h}_{0,\zeta}$ is simple
and \ continuous, $\mathrm{spec}\hat{h}_{0,\zeta}=\mathbb{R}_{+}$.

\subparagraph{$g>0$}

In this case, $\sigma_{0,\zeta}^{\prime}(E)$ is given by eq.
(\ref{Coul2.5.7.2.1}) for $E>0$. But now, $B_{0}(0)=0$ and the denominator of
expr. (\ref{Coul2.5.7.2.1}) is equal to zero for $E=0$ and $\zeta=\zeta_{0}$,
$\tan\zeta_{0}=-A_{0}(0)=(1/2)\ln(g/\kappa_{0})-\psi(1)$, such that we should
study the behaviour of $\sigma_{0,\zeta}^{\prime}(E)$ for small $E$ in more
details. We have%
\[
f_{0,\zeta}(\mathcal{E})=\tan\zeta-[(1/2)\ln(g/\kappa_{0})-\psi
(1)]+\mathcal{E}/(12g^{2})+O(\mathcal{E}^{2}).
\]
We see that if $\zeta\neq\zeta_{0}$ then $\sigma_{0,\zeta}^{\prime}(0)=0$. But
if $\zeta=\zeta_{0}$, we find%
\[
\sigma_{0,\zeta}^{\prime}(E)=-\frac{24g^{2}}{\pi\kappa_{0}\cos^{2}\zeta_{0}%
}\operatorname{Im}\frac{1}{E+i0}+O(1)=\frac{24g^{2}}{\kappa_{0}\cos^{2}%
\zeta_{0}}\delta(E)+O(1),
\]
and $\mathrm{spec}\hat{h}_{0,\zeta_{0}}=\mathbb{R}_{+}\cup\{E_{(-)}(\zeta
_{0})=0\}$.

\paragraph{$E=-\tau^{2}<0$, $\tau>0$, $K=\tau$}

In this case, we have%
\[
f_{0}(E)=\psi(1)-\frac{1}{2}\psi(1/2+g/2\tau)-\frac{1}{2}\ln(2\tau/\kappa
_{0}),
\]
so that .$f_{0}(E)$ is real and $\sigma_{0,\zeta}^{\prime}(E)$ can be not
equal to zero only in the points $E_{0,n}(\zeta)$ which are solutions of an
equation%
\begin{equation}
f_{0,\zeta}(E_{0,n}(\zeta))=0\;\mathrm{or}\;f_{0}(E_{0,n}(\zeta))=-\tan\zeta.
\label{Coul2.5.7.3.1}%
\end{equation}

We thus obtain%
\[
\sigma_{0,\zeta}^{\prime}(E)=\sum_{n\in\mathcal{N}_{0}}Q_{0,\zeta,n}^{2}%
\delta(E-E_{0,n}(\zeta)),\;Q_{0,\zeta,n}=\frac{1}{|\cos\zeta|}\sqrt{\frac
{2}{\kappa_{0}f_{0}^{\prime}(E_{0,n}(\zeta))}},
\]
where $\mathcal{N}_{0}$ is some subset of integers, $\mathcal{N}_{0}%
\in\mathbb{Z}$, and $f^{\prime}(E_{0,n}(\zeta))>0$. Because any $E<0$ is
solution of eq. (\ref{Coul2.5.7.3.1}) for some $\zeta$, $f_{0}^{\prime}(E)>0$
for all $E$. Futhermore, we find:%
\[
\partial_{\zeta}E_{0,n}(\zeta)=-[f_{o}^{\prime}(E_{0,n}(\zeta))\cos^{2}%
\zeta]^{-1}<0.
\]

\subparagraph{$g>0$}

In this case, we have: $f_{0}(E)$ is a smooth function on the interval
$-\infty<E<0;$ $f_{0}(E)$ increases monotonically from $-\infty$ to
$A(0)=\psi(1)-\frac{1}{2}\ln(g/\kappa_{0})$ as $E$ run from $-\infty$ to $-0$.
\ We thus obtain: for $\zeta\in(-\pi/2,\zeta_{0})$, eq. (\ref{Coul2.5.7.3.1})
has no solutions; and for any fixed $\zeta\in(\zeta_{0},\pi/2)$, eq.
(\ref{Coul2.5.7.3.1}) has one solution $E_{0,(-)}(\zeta)\in(-\infty,0)$
monotonically increasing from $-\infty$ to $-0$ as $\zeta$ run from $\pi/2-0$
t0 $\zeta_{0}+0$.

\subparagraph{$g=0$}

In this case, we have $f_{0}(E)=\psi(1)-\frac{1}{2}\psi(1/2)-\frac{1}{2}%
\ln(2\tau/\kappa_{0})$ and eq. (\ref{Coul2.5.7.3.1}) has one solution
$E_{0,(-)}(\zeta)=(\kappa_{0}^{2}/4)e^{4\psi(1)-2\psi(1/2)+4\tan\zeta}$.

\subparagraph{$g<0$}

In this case, we have: $f_{0}(E)=\psi(1)-\frac{1}{2}\psi(1/2-|g|/2\tau
)-\frac{1}{2}\ln(2\tau/\kappa_{0})$; $f(\mathcal{E}_{0,n}\pm0)=\mp\infty$,
$n\in\mathbb{Z}_{+}$; in each interval $(\mathcal{E}_{0,n-1},\mathcal{E}%
_{0,n})$, $n\in\mathbb{Z}_{+}$, $f_{0}(E)$ increases monotonically from
$-\infty$ to $\infty$ as $E$ run from $\mathcal{E}_{0,n-1}+0$ to
$\mathcal{E}_{0,n}-0$; in each interval $(\mathcal{E}_{0,n-1},\mathcal{E}%
_{0,n})$, $n\in\mathbb{Z}_{+}$, for any fixed $\zeta\in(-\pi/2,\pi/2)$, eq.
(\ref{Coul2.5.7.3.1}) has one solution $E_{0,n}(\zeta)$ monotonically
increasing from $\mathcal{E}_{0,n-1}+0$ to $\mathcal{E}_{0,n}-0$ as $\zeta$
run from $\pi/2-0$ t0 -$\pi/2+0$. Here we set $\mathcal{E}_{0,-1}=-\infty$. We
find%
\[
\mathcal{N}_{0}=\left\{
\begin{array}
[c]{l}%
\varnothing,\;g>0,\;\zeta\in(-\pi/2,\zeta_{0})\;\mathrm{or}\;g\geq
0,\;\zeta=\pm\pi/2\\
(-),\;g>0,\;\zeta\in\lbrack\zeta_{0},\pi/2)\\
(-),\;g=0,\;\zeta\in(-\pi/2,\pi/2)\\
\mathbb{Z}_{+},\;g<0,\;\zeta\in\lbrack-\pi/2,\pi/2]
\end{array}
\right.  ,
\]
where, for completeness, the cases $\zeta=\pm\pi/2$ and $E=0$ are included.

Note the relation%
\[
\lim_{\zeta\rightarrow-\pi/2}E_{0,n}(\zeta)=\lim_{\zeta\rightarrow\pi
/2}E_{0,n+1}(\zeta)=\mathcal{E}_{0,n},\;n\in\mathbb{Z}_{+}.
\]

Note also that all results for spectrum (and for eigenfunctions) in the case
$g=0$ can be obtained by the formal limit $g\rightarrow0$ in the cases $g>0$
or $g<0$.

Finally we obtain: the spectra of $\hat{h}_{0,\zeta}$ are simple,
$\mathrm{spec}\hat{h}_{0,\zeta}=\mathbb{R}_{+}\cup\{E_{0,n}(\zeta)\leq
0,\;n\in\mathcal{N}_{0}\}$, the set $\{U_{0,\zeta,E}(x)=\rho_{0,\zeta
}(E)U_{0,\zeta}(x;E),\;E\in\mathbb{R}_{+};\;U_{0,\zeta,n}(x)=Q_{0,\zeta
,n}U_{0,\zeta}(x;E_{0,n}(\zeta)\},\;n\in\mathcal{N}_{0}\}$ of (generalized)
eigenfunctions of $\hat{h}_{0,\zeta}$ forms the complete orthohonalized system
in $L^{2}(\mathbb{R}_{+})$, where%
\[
\rho_{0,\zeta}(E)=\left\{
\begin{array}
[c]{l}%
\mathrm{eq}.\;\mathrm{(\ref{Coul2.5.7.2.1})}\;\mathrm{for}\;g\leq0,\;E\geq0\\
\mathrm{and\;}g>0,\;E>0,\\
\lim_{E\rightarrow+0}\rho_{0,\zeta}(E),\;\mathrm{for}\;g>0,\;E=0
\end{array}
\right.  .
\]

\subsection{List of coincidences}

Make the following identifications%
\begin{align*}
&  u=\sqrt{x/\kappa_{0}},\;W=-4\kappa_{0}g,\;\lambda=-4\kappa_{0}%
^{2}\mathcal{E},\\
&  x=\kappa_{0}u^{2},\;\mathcal{E}=-\lambda/4\kappa_{0}^{2},\;g=-W/4\kappa
_{0},\;\Rightarrow
\end{align*}%
\begin{align*}
&  K=\sqrt{\lambda}/2\kappa_{0}=\varkappa^{2}/2\kappa_{0},\;\sqrt{K}%
=\varkappa/\sqrt{2\kappa_{0}},\;\kappa_{0}^{2}/\varkappa^{2}=\kappa_{0}/2K,\\
&  z=\rho,\;\sqrt{g}=\sqrt{-W}/2\sqrt{\kappa_{0}},\\
&  w_{C}=-\frac{g}{2K}=\frac{W}{4\sqrt{\lambda}}=w_{O},\;\alpha_{C}=\alpha
_{O},\;\alpha_{C-}=\alpha_{O-},\;\beta_{C}=\beta_{O},
\end{align*}%
\begin{align*}
&  E_{C}>0,\;K=-i\sqrt{E_{C}},\;\lambda<0;\;\sqrt{\lambda}=-i\sqrt{|\lambda
|},\\
&  \tilde{w}_{C}=-iw_{C}=i\frac{g}{2K}=-\frac{g}{2\sqrt{E_{C}}}=\frac{E_{O}%
}{4\sqrt{|\lambda|}}=-i\frac{E_{O}}{4\sqrt{\lambda}}=\tilde{w}_{O}%
\end{align*}

\subsection{$|m|\geq1$}

We find%
\[
C_{k,m}(x;\mathcal{E})=(k_{0}u)^{1/2}O_{k,m}(u;W),\;k=1,4,3.
\]%
\begin{align*}
&  C_{Om}(W)=\frac{(\kappa_{0}^{2}/\varkappa^{2})^{|m|}\Gamma(|m|)}%
{\Gamma(\alpha_{O})}=\frac{(\kappa_{0}/2K)^{|m|}\Gamma(|m|)}{\Gamma(\alpha
_{C})}=C_{Cm}(\mathcal{E}),\\
&  \omega_{Om}(W)=2\kappa_{0}|m|C_{Om}(W)=2\kappa_{0}|m|C_{Cm}(\mathcal{E}%
)=2\omega_{Cm}(\mathcal{E}),\\
&  B_{Om}(W)=\frac{(-1)^{|m|+1}}{2\Gamma(\beta_{O})\Gamma(\alpha_{O-})}\left[
\psi(\alpha_{O-})+\psi(\alpha_{O})-4\ln(\kappa_{0}/\varkappa)\right]  =\\
&  \,=\frac{(-1)^{|m+1|}}{2\Gamma(\beta_{C})\Gamma(\alpha_{C-})}\left[
\psi(\alpha_{C-})+\psi(\alpha_{C})+2\ln(2K/\kappa_{0})\right]  =B_{Cm}%
(\mathcal{E}).
\end{align*}

\[
\Omega_{Cm}(\mathcal{E})\equiv\frac{B_{Cm}(\mathcal{E})}{\omega_{Cm}%
(\mathcal{E})}=2\frac{B_{Om}(W)}{\omega_{Om}(W)}=2\Omega_{Om}(W).
\]

\subsection{$m=0$}%

\begin{align*}
&  C_{k,0}(x;\mathcal{E})=(k_{0}u)^{1/2}O_{k,0}(u;W),\;k=1,2,3,\\
&  U_{C\zeta}(x;\mathcal{E})=(k_{0}u)^{1/2}U_{O\zeta}(u;W),\;\tilde{U}%
_{C\zeta}(x;\mathcal{E})=(k_{0}u)^{1/2}\tilde{U}_{O\zeta}(u;W).
\end{align*}

\begin{align*}
&  \omega_{C0}(\mathcal{E})=2\psi(1)-\psi(\alpha_{C})-\ln(2K/\kappa_{0})=\\
&  \,=2\psi(1)-\psi(\alpha_{O})+2\ln(\kappa_{0}/\varkappa)=\omega_{O0}(W),\\
&  \omega_{C\zeta}(\mathcal{E})=\omega_{O\zeta}(W),\;\tilde{\omega}_{C\zeta
}(\mathcal{E})=\tilde{\omega}_{O\zeta}(W).
\end{align*}

\section{Conclusions} 

As we found, two dimensional oscillator and coulomb problems on pseudoshpere
are described by the same equations in terms of the variables $\alpha$ and $\beta$ 
This means that each point of the spectra of one of these theories corresponds a point 
of the spectra of the other theory, i.e. there is one-to-one correspondence between 
points of the the planes $E_{O}, \lambda$ and $E_{C}, g$. 

\section{Acknowledgement}I.Tyutin  thanks RFBR Grant 11-01-00830 for partial support.

\end{document}